\documentclass[useAMS,usenatbib]{mn2e}

\usepackage{amssymb,amsmath}
\usepackage{graphicx}
\usepackage{natbib}
\usepackage{aas_macros}
\usepackage{textcomp}

\newcommand{\hi}{H~{\sc i}}

\title[The turbulent life of dust grains]{The turbulent life of dust grains in the supernova-driven, multi-phase interstellar medium}

\author[Thomas Peters et al.]{
\parbox[h]{\textwidth}{
Thomas Peters$^1$\thanks{E-Mail: tpeters@mpa-garching.mpg.de},
Svitlana Zhukovska$^1$,
Thorsten Naab$^1$,
Philipp Girichidis$^1$,
Stefanie Walch$^2$,
Simon C. O. Glover$^3$,
Ralf S. Klessen$^{3,4}$,
Paul C. Clark$^5$
and Daniel Seifried$^2$
}\vspace{0.4cm}\\
\parbox{\textwidth}{$^1$Max-Planck-Institut f\"{u}r Astrophysik, Karl-Schwarzschild-Str. 1, D-85748 Garching, Germany\\
$^2$I. Physikalisches Institut, Universit\"{a}t zu K\"{o}ln, Z\"{u}lpicher Strasse 77, D-50937 K\"{o}ln, Germany\\
$^3$Universit\"{a}t Heidelberg, Zentrum f\"{u}r Astronomie, Institut f\"{u}r Theoretische Astrophysik, Albert-Ueberle-Str. 2, D-69120 Heidelberg, Germany\\
$^4$Universit\"at Heidelberg, Interdisziplin\"{a}res Zentrum f\"{u}r Wissenschaftliches Rechnen (IWR), D-69120 Heidelberg, Germany\\
$^5$School of Physics \& Astronomy, Cardiff University, 5 The Parade, Cardiff CF24 3AA, Wales, UK
}}

\begin{document}

\maketitle

\begin{abstract}
Dust grains are an important component of the interstellar medium (ISM) of galaxies.
We present the first direct measurement of the residence times
of interstellar dust in the different ISM phases, and of the transition rates between these
phases, in realistic hydrodynamical simulations of the multi-phase ISM.
Our simulations include a time-dependent chemical network that follows the abundances of H$^+$, H, H$_2$, C$^+$ and CO
and take into account self-shielding by gas and dust using a tree-based radiation transfer method.
Supernova explosions are injected either at random locations, at density peaks, or as a mixture of the two.
For each simulation, we investigate how matter circulates between the ISM
phases and find more sizeable transitions than considered in simple mass exchange
schemes in the literature. The derived residence times in the ISM phases are
characterised by broad distributions, in particular for the molecular, warm and hot medium.
The most realistic simulations with random and mixed driving have median residence times in the molecular,
cold, warm and hot phase around $17$, $7$, $44$ and $1\,$Myr, respectively. The transition rates measured in the random driving run
are in good agreement with observations of Ti gas-phase depletion in the warm and cold phases in a simple depletion model, although
the depletion in the molecular phase is under-predicted. ISM phase definitions based on chemical abundance
rather than temperature cuts are physically more meaningful, but lead to significantly different transition rates
and residence times because there is no direct correspondence between the two definitions.
\end{abstract}

\section{Introduction}
\label{sec:intro}

Refractory sub-\textmu m particles or dust grains are an integral component of the
interstellar medium (ISM) of galaxies. Interstellar dust consists of silicate and carbonaceous
grains with a mass fraction of less than 1\% of the gas mass. Nevertheless, these dust grains influence the
physics and chemistry of the ISM in multiple ways \citep{tielensbook,Draine:2011tr}. One of the most important effects of interstellar grains
is their absorption of the ultraviolet (UV) interstellar radiation field. At least 30\% of the stellar light at UV wavelengths
is reprocessed by grains and re-emitted in the infrared \citep{Soifer:1987jk}. A small fraction of the absorbed energy is
returned to the ISM through photoelectric emission, which is the dominant heating source of the diffuse ISM \citep{baktie94}.
Dust also affects the thermal balance of the ISM by locking away important coolants such as C$^+$ and Si$^+$ \citep{Bekki:2015hn, McKinnon:2016ft}
and by causing the freeze-out of CO in the dense phase \citep{goldsmith01,bertaf07,hollenbach09,caselli11,fontani12,hocuk14}.
Grains have a two-fold role in astrochemistry, namely shielding the molecules from the dissociating UV photons and providing
the surfaces for the formation of complex organic molecules and H$_2$ molecules, the main constituent of molecular clouds.
The size distribution and chemical composition are the key dust properties determining grain interactions with matter and radiation.

Numerical simulations of galactic evolution usually assume that interstellar dust constitutes a fixed fraction of the metals
and that its properties have the average characteristics of dust in the local Galaxy \cite[][and references therein]{walchetal15}.
However, there is strong observational evidence that grain properties depend on the local environment as indicated by variations of
interstellar element depletion \citep{Jenkins:2009p2144}, extinction curves \citep{Fitzpatrick:2007p6352},
dust-to-gas ratios \citep{RomanDuval:2014gu, Reach:2015gx}, opacities \citep{Roy:2013hm}, and spectral characteristics \citep{Dartois:2004p466},
among many others \citep[see][for a review]{Dorschner:1995p7228}. These observations demonstrate that both the composition and the size
distribution of grains are not constant even in our nearest environment.

The observed changes in dust properties are attributed to grain evolution driven by local conditions in the ISM. In a simplified way,
the ISM can be described by a three-phase model proposed by \cite{McKee:1977p7211}, which consists of cold clouds (cold neutral medium, CNM),
surrounded by the warm neutral envelopes (warm neutral medium, WNM) and warm ionised matter (WIM); the hot ionised medium (HIM) fills most
of the volume. Additionally, molecular clouds represent dense regions of the ISM that are opaque to UV radiation and
where star formation takes place. Due to their coupling to the interstellar gas, interstellar grains follow the cycle of matter between the ISM
phases, which is regulated by the formation of molecular clouds and their disruption by stellar feedback processes
\citep[e.g.][]{Wooden:2004p1005, Dobbs:2012in}.

During their time in the warm medium, grains are altered and partially or completely
destroyed by a number of processes such as vaporisation in grain-grain collisions and erosion by ion sputtering in interstellar shocks,
and by UV irradiation by the interstellar radiation field
\citep{McKee:1989p1030, Tielens:1994p6577, Jones:1994p1037, Jones:1996p6593, Jones:2013gg, Slavin:2015in}. In clouds, grains are protected
from UV irradiation and sputtering in shocks, and the dust mass can grow by accretion of gas-phase species
\citep{Greenberg:tw, Draine:2009p6616}. Coagulation becomes the dominant outcome of grain-grain collisions in dense clouds, resulting
in the removal of small grains and the build-up of large grains \citep{Hirashita:2009p663, Ormel:2011cj}. 

Dust abundances in the interstellar gas are therefore closely related to the evolution of the ISM, in particular to its structure and the
mass transfer between the phases.  To explain the large differences in element depletion between the WNM and CNM \citep{Savage:1996p486},
simple models of dust evolution based on two- and three-phase ISM models with  various schemes for phase transitions were proposed in
the literature \citep{draine90, ODonnell:1997p683, Tielens:1998p7054, Weingartner:1999p6573}. \cite{draine90} demonstrated
that the relative depletion in the CNM and WNM depend on the adopted scheme of the phase transition: whether gas from the WNM is
transferred directly to molecular clouds or through the CNM. Moreover, the observed scatter in the dust-to-gas ratio in spiral galaxies
can also be explained by the cycling of matter in the multi-phase ISM, with the timescales of the transitions determining the amplitude
of the dust-to-gas ratio variations \citep{Hirashita:2000p6583}.

The residence times that the grains spend in the different ISM phases are important quantities that regulate the processing of the
grains in these environments. \cite{Hirashita:2009p663} demonstrated that the grain size distribution is very sensitive to the residence
times in the ISM phases: grain shattering by turbulence can overproduce the number of small grains in the WIM and limit the sizes of
large grains in the WNM. Coagulation dominating in dense cloud cores can completely remove small grains when the residence time exceeds $10$~Myr.

The residence times in the multi-phase ISM also determine the exposure of interstellar grains to UV radiation and the susceptibility to collisions with ions.
Information on the exposure duration is useful to set up realistic conditions for laboratory experiments on dust analogues simulating
the formation and evolution of organic refractory matter \citep{Jenniskens:1993p2249}. Moreover, some stardust grains present
in the Solar System during its formation process can be identified in meteorites and studied in the laboratory.
Analysis of the surfaces of these grains provides insights into their processing in the ISM and can be done more accurately if
the exposure times are known \citep{Gyngard:2009ch, Heck:2009eg, Amari:2014iy}. 

Models of interstellar dust typically use a highly idealised description of the ISM phases, where each phase is
represented by a single characteristic temperature and density. Recent, more realistic hydrodynamical simulations of the ISM
\citep{walchetal15,girietal16} present an opportunity
to move away from this simplified picture.
They allow us to probe dust evolution in the context of an inhomogeneous ISM under a wide
range of physical conditions and a complex evolutionary history. For example, in such simulations molecular clouds form dynamically out of the diffuse phase and
get destroyed by stellar feedback. The physical conditions within the molecular cloud are then time-dependent, and there will be a distribution
of molecular cloud lifetimes rather than a single unique value.

In this paper, we investigate the lifecycle of interstellar grains in the multi-phase ISM
with simulations from the SILCC (SImulating the LifeCycle of molecular Clouds)
project \citep{walchetal15,girietal16}. The SILCC simulations include a time-dependent chemical network and therefore provide
detailed descriptions of the relevant ISM structure and phases. 
We present a pilot study in which we use Lagrangian tracer particles to measure
residence times in and transition rates between the different ISM phases in local galactic-scale simulations of a supernova-driven
ISM under solar neighbourhood conditions. Details of our numerical scheme and the simulation setup are given in Section~\ref{sec:sim}.
We describe the evolution of the simulations in Section~\ref{sec:adv} and of individual tracer particle trajectories in Section~\ref{sec:traj}.
We then discuss how the tracer particles sample the multi-phase ISM (Section~\ref{sec:sam}) and present our measurements
of mass circulation and transition rates between the phases (Section~\ref{sec:trans}), residence times within the phases (Section~\ref{sec:res})
and shielding from UV radiation
(Section~\ref{sec:shield}). We
point out some caveats in Section~\ref{sec:cav} and conclude in Section~\ref{sec:conc}.

\section{Simulations}
\label{sec:sim}

We present kpc-scale stratified box simulations run with the adaptive mesh refinement code \texttt{FLASH}~4 \citep{fryxell00,dubey09}
with a stable, positivity-preserving magnetohydrodynamics solver \citep{bouchut07,waagan09,waagan11}
and a method based on a Barnes-Hut tree \citep{barhut86} to incorporate self-gravity (R.~W\"{u}nsch et al. in prep.).
Our simulation box has dimensions $0.5\,$kpc$\,\times\, 0.5\,$kpc$\,\times\, 2.5\,$kpc and a maximum grid resolution of $3.9\,$pc.
We apply periodic boundary conditions in the plane of the disc ($x$ and $y$ directions) and outflow boundary conditions in the
vertical ($z$) direction, so that gas can leave but not enter the simulation box.
The simulation domain has lower heights above and below the disc plane than previous SILCC simulations \citep{walchetal15,girietal16},
but otherwise the code and setup are identical.

We impose an external potential to model the gravitational force of a stellar disc on the gas.
We choose an isothermal sheet potential with parameters that fit solar neighbourhood values, namely a stellar surface density
$\Sigma_{*} = 30\,$M$_\odot\,$pc$^{-2}$ and a vertical scale height $z_\mathrm{d} = 100\,$pc.
We set up the gas with a Gaussian distribution in $z$-direction, with a scale height of $60\,$pc,
and a gas surface density $\Sigma_\mathrm{gas} = 10\,$M$_\odot\,$pc$^{-2}$.
The simulations presented in this paper do not include magnetic fields.
For more information on the initial conditions and simulation setup see \citet{walchetal15} and \citet{girietal16}.

We use a time-dependent chemical network \citep{nellan97,glovmcl07,glovmcl07b,glovclar12}
that follows the abundances of free electrons, H$^{+}$, H, H$_{2}$, C$^{+}$, O and CO. 
We employ the \texttt{TreeCol} algorithm (\citealt{clarketal12}, R.~W\"{u}nsch et al. in prep.) to take into account
dust shielding and molecular self-shielding.
In the warm and cold gas, the primary cooling
processes are Lyman-$\alpha$ cooling, H$_{2}$ ro-vibrational line cooling, fine-structure emission from C$^{+}$ and O, and
rotational emission from CO \citep{glovetal10,glovclar12}. In hot gas, electronic excitation of helium and of partially  
ionised metals must also be taken into account, which is done using the \cite{gnafer12} cooling rates
assuming collisional ionisation equilibrium.
Furthermore, we include diffuse heating from the photoelectric effect, cosmic rays and X-rays
following the prescriptions of \citet{baktie94}, \citet{gollan78} and \citet{woletal95}, respectively. More information
on the chemical network and the various heating and cooling processes can be found in \citet{gattoetal15} and \citet{walchetal15}.

For our gas surface density the Kennicutt-Schmidt relation \citep{schmidt59,kenn98} yields a star formation rate
surface density $\Sigma_\mathrm{SFR} = 6 \times 10^{-3}\,$M$_\odot\,$yr$^{-1}\,$kpc$^{-2}$. Assuming one supernova event
per $100\,$M$_\odot$ in stars, this corresponds to a supernova rate surface density $\Sigma_\mathrm{SN} = 60\,$Myr$^{-1}\,$kpc$^{-2}$,
or 15 supernova explosions per Myr in our simulation volume.
We inject a thermal energy $E_\mathrm{SN} = 10^{51}\,$erg with each such explosion at this constant rate.

We consider three different ways to position the sites of the supernova explosions (see \citealt{girietal16} for a detailed justification). 
For random driving, we distribute the explosion sites according to a probability distribution that is uniform in the $x$-$y$ plane
and Gaussian in $z$ direction with a scale height of $50\,$pc \citep{tamloesch94}. For peak driving, supernovae always
explode at the current location of the maximum density peak. For mixed driving, 50\% of all supernovae are randomly distributed
and 50\% explode at density peaks.
We refer to \citet{gattoetal15} for details on the implementation of supernova positioning and energy injection and to \citet{walchetal15}
and \citet{girietal16} for a systematic investigation of the effects of supernova positioning and supernova rates on
the chemical and dynamical properties of the resulting ISM, respectively.
In their notation,
our simulations with random, mixed and peak driving correspond to the runs S10-KS-rand, S10-KS-mix and S10-KS-peak, respectively.
We have run all three simulations for a duration of $80\,$Myr.

Our simulations represent an idealisation of star formation feedback in the ISM. Because we do not model the formation of star
clusters self-consistently (but see \citealt{gatto16,peters16b}), we systematically explore how the location
of supernova explosions affects the ISM. Peak driving is motivated by the idea that star formation happens in the densest regions
of the ISM. Random driving takes the finite time between the formation of stars and their explosions as supernovae
as well as the existence of field and runaway OB stars into account, which leads to the expectation that a large fraction of supernovae
are located in underdense gas at random positions. Mixed driving is a compromise between the two extremes. The ISM is most realistic for simulations
with random and mixed driving is terms of mass fractions and volume-filling fractions of the different ISM phases, while with
peak driving the formation of H$_2$ and the hot phase is suppressed \citep{walchetal15}.
Although some dynamical quantities like velocity dispersions and outflow rates converge,
the simulations do not reach a steady state because they do not include a self-consistent treatment of star formation and supernova explosions \citep{girietal16}.
They should be understood as controlled numerical experiments that allow us to study the properties of interstellar dust
in the multi-phase ISM produced by the different forms of supernova driving.

At the beginning of each simulation, we distribute $N_\mathrm{part} = 10^6$ tracer particles over the simulation volume such that the local
tracer particle number density is proportional to the local gas density. The equation of motion for the tracer particles is integrated
using Heun's method. We interpolate the data stored on the grid (gas density and temperature, chemical abundances, and the
local radiation field) to the location of the particle within the grid tri-linearly and save it together with the particle positions and velocities
to a file every 10\,kyr, which is a factor of a few larger than the typical simulation timestep.
The tracer particles are passive and do not affect the dust abundance used in the chemical network, which we assume to be constant.
Furthermore, we do not inject any additional particles during the simulation since the timescale of dust production is long
compared to the total runtime.

The tracer particles are interpreted as representative ensembles of dust grains. We can assign a mass to each tracer particle by assuming
that the total mass of dust is equally distributed among all tracer particles. For simplicity, we adopt the constant gas-to-dust mass ratio of $162$
derived for the interstellar dust model BARE-GR-S, which consists of bare silicate and graphitic grains \citep{Zubko:2004p4116}.
A higher gas-to-dust ratio than a commonly used value of 100 is required by interstellar dust models that simultaneously fit dust extinction,
infrared diffuse emission and element abundance constraints \citep{Zubko:2004p4116, Draine:2007p6602}.
For a gas-to-dust
ratio of $162$ and a total gas mass in the simulation volume $M_\mathrm{gas} = 2.5 \times 10^6\,$M$_\odot$, the tracer particle mass becomes
\begin{equation}
m_\mathrm{part} = \frac{M_\mathrm{gas}}{162\,N_\mathrm{part}} = 1.5 \times 10^{-2}\,\mathrm{M}_\odot.
\end{equation}
This definition allows us to convert between tracer particle numbers and dust masses.

The BARE-GR-S dust model is designed to match various observational constraints in the local diffuse ISM. It is known that the gas-to-dust ratio
in the dense gas can be a few times lower compared to the diffuse ISM owing to accretion of gas-phase species onto grain surfaces. It is however
very difficult to infer the true variations of the gas-to-dust ratio from observed dust emission maps, because it requires one to disentangle
the effects of a number of
different physical processes which affect dust emission in translucent molecular clouds: changes of the far-infrared dust properties due to coagulation of grains,
the contribution of the CO-dark molecular gas to the gas-to-dust ratio and the true decrease in the gas-to-dust ratio due to the dust growth in the ISM \citep{RomanDuval:2014gu}.
Recent models of dust evolution in the inhomogeneous multi-phase ISM study the variations of silicate dust abundance with local conditions, but
are still uncertain in variations in the gas-to-dust ratio \citep{zhuk16}.

The difference in element depletion between the lines of sight with the lowest and the highest depletion levels in the large sample of data
compiled by \cite{Jenkins:2009p2144} implies a variation of the gas-to-dust ratio of a factor of $2$. With this factor, the dust mass
associated with a tracer particle $m_\mathrm{part}$ in the dense phase can be higher, which is presently neglected in this work.

\section{Distribution of tracer particles}
\label{sec:adv}


The gas flow in the simulations is created mainly by two dynamical processes: gravity, which pulls the gas towards the disc midplane
and leads to the formation of self-gravitating molecular clouds, and blast waves that are created by supernova explosions.
As the gas moves, the tracer particles are advected with the flow. In a compressible gas, the number density of tracer
particles is roughly proportional to the gas density. Therefore, tracer particles tend to accumulate in dense regions, and only
relatively few tracer particles are present in voids or get injected into the halo. A very small fraction of tracer particles
leaves the simulation box at $z = \pm1.25\,$kpc during the simulation runtime: 2.95\% for random driving, 0.04\% for mixed driving
and 0.02\% for peak driving. Since we do not know the fate of these particles, we only include tracer particles in our subsequent
analysis that stay within the simulation box until the end.

\begin{figure*}
\centerline{\includegraphics[width=\textwidth]{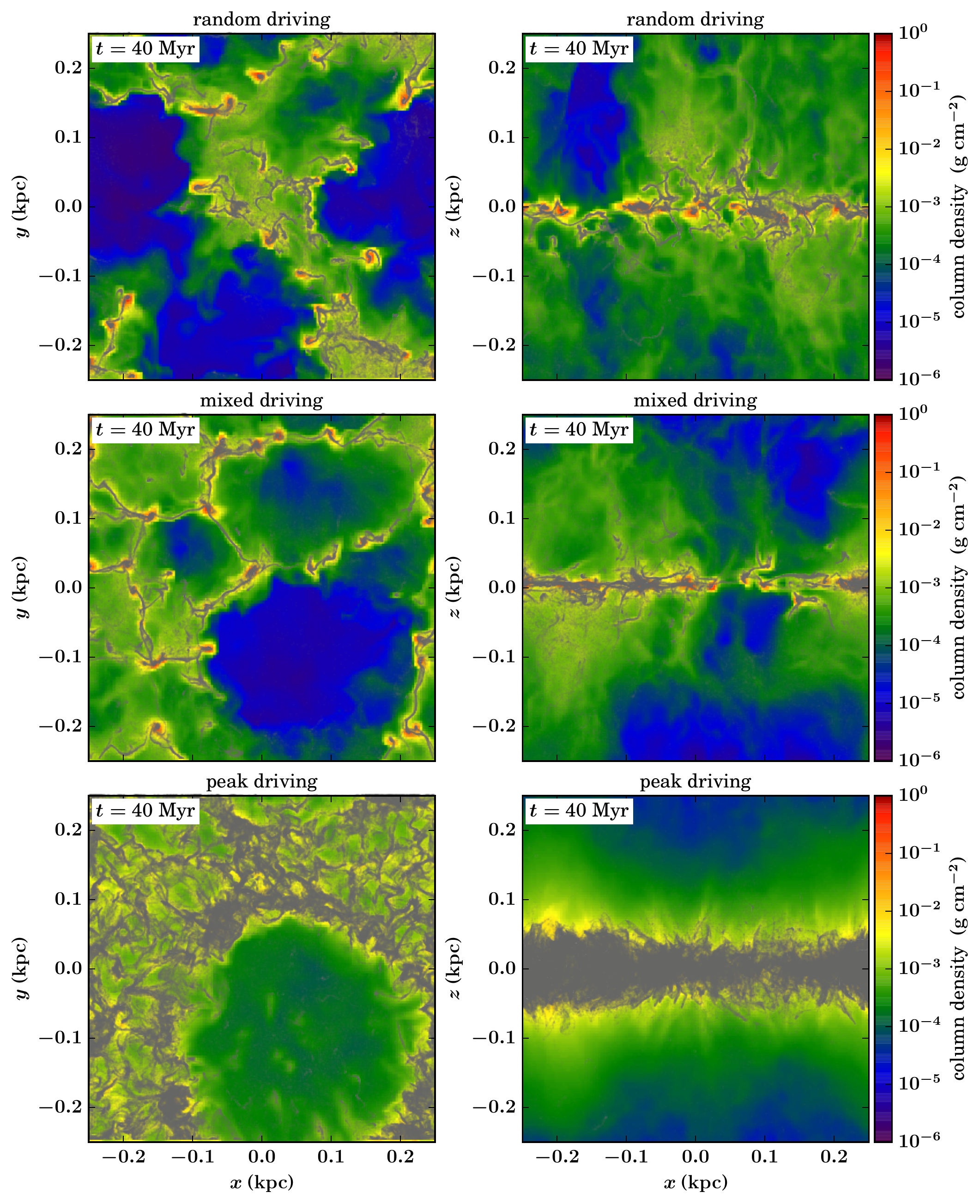}}
\caption{Projections of gas density (left: face-on, right: edge-on) for the simulation with random driving (top),
mixed driving (middle) and peak driving (bottom) after 40~Myr of evolution. In the $z$-direction,
only the inner $0.25$\,kpc of the whole box height ($1.25$\,kpc) is shown. Each tracer particle is represented by a grey dot.
Movies of the simulations are available on the SILCC website.}
\label{fig:proj}
\end{figure*}

Figure~\ref{fig:proj} shows
face-on and edge-on gas column densities together with the tracer particle locations for
random, mixed and peak driving after 40~Myr of evolution\footnote{An animated version of this figure
can be found on the SILCC project website, http://hera.ph1.uni-koeln.de/$\sim$silcc/.}.
At the start of the simulation, the gas collapses towards the disc midplane until the supernova explosions have
created enough thermal and kinetic pressure to support the gas against collapse.
After 10\,Myr, the random supernova driving has left a large fraction of the initial particle distribution
in the disc plane unaffected, whereas the tracer particles are completely mixed after 40\,Myr. 
As the simulations proceed, 
self-gravitating structures form that resist destruction by supernova blast waves.
At 80\,Myr, most of the tracer particles
reside in a few dense regions and within an outflow driven by supernova explosions from the diffuse medium.

For the case of mixed driving, some differences occur. Here, every second explosion is located
at density peaks. Injecting supernovae at the places of highest density can lead to a runaway process when the supernova injection
triggers the formation of a blast wave that creates a high-density shock.
The density peak at the moment of the next supernova injection will then likely be located somewhere at the density enhancement
created by this shock.
A repeated injection of supernovae in a relatively small
volume can then lead to an amplification of density contrasts that attracts subsequent supernova injections.
This is what happens here in the first 10\,Myr and what creates the spherical region visible in the face-on projection.
Because only half of all supernovae explode at random positions, a large fraction of the particles remain unperturbed
by supernova explosions at this time. However, after 40\,Myr also in this simulation the tracer particles are completely mixed,
and the disc scale height is similar to the simulation with random driving. Once the ISM has a sufficiently complex structure,
the influence of the density peak supernovae on the mixing is reduced because a certain region in the disc is no longer favoured
for supernova injections. The final snapshot at 80\,Myr also looks similar to the situation for random driving.

In the simulation with pure peak driving, the evolution is markedly different.
Now, all supernovae explode within a region of diameter $\sim\,0.2\,$kpc in the first 10\,Myr because of the described runaway
process. Supernova explosion sites propagate radially outwards from the location of the first explosion. At 40\,Myr,
the ISM has a complex structure and the tracer particles are fully mixed. However, the initially clustered supernova explosions have created
a cavity that remains for the rest of the simulation runtime. Since supernovae are injected at density peaks, much
less hot gas is created, and almost no galactic wind is driven \citep{girietal16}.
At 80\,Myr, the gas density in the disc is higher than for random and mixed driving,
and the tracer particles are more homogeneously distributed.

\section{Individual trajectories}
\label{sec:traj}

While ISM dust models typically represent the ISM phases with a single characteristic temperature,
a more accurate phase definition assigns a whole temperature range to each phase \citep[e.g.][]{ferri01}.
Guided by our previous study of ISM phases in the SILCC simulations by \citet{walchetal15},
we consider four different phases defined by temperature cuts in this paper:
\begin{enumerate}
\item molecular phase ($T < 50\,$K),
\item cold phase ($50 \leq T < 300\,$K),
\item warm phase ($300 \leq T < 3 \times 10^5\,$K),
\item hot phase ($T \geq 3 \times 10^5\,$K).
\end{enumerate}
There are two important differences to our analysis in \citet{walchetal15}.
First, we do not include a warm-hot phase because it is thermally unstable
and therefore short-lived. Hence, we subsume the warm-hot phase from \citet{walchetal15} in our warm phase.
Second, we choose a slightly higher temperature cut of $50\,$K to separate the molecular and the cold phase
instead of $30\,$K as in  \citet{walchetal15}.

\begin{figure*}
\centerline{\includegraphics[width=0.5\linewidth]{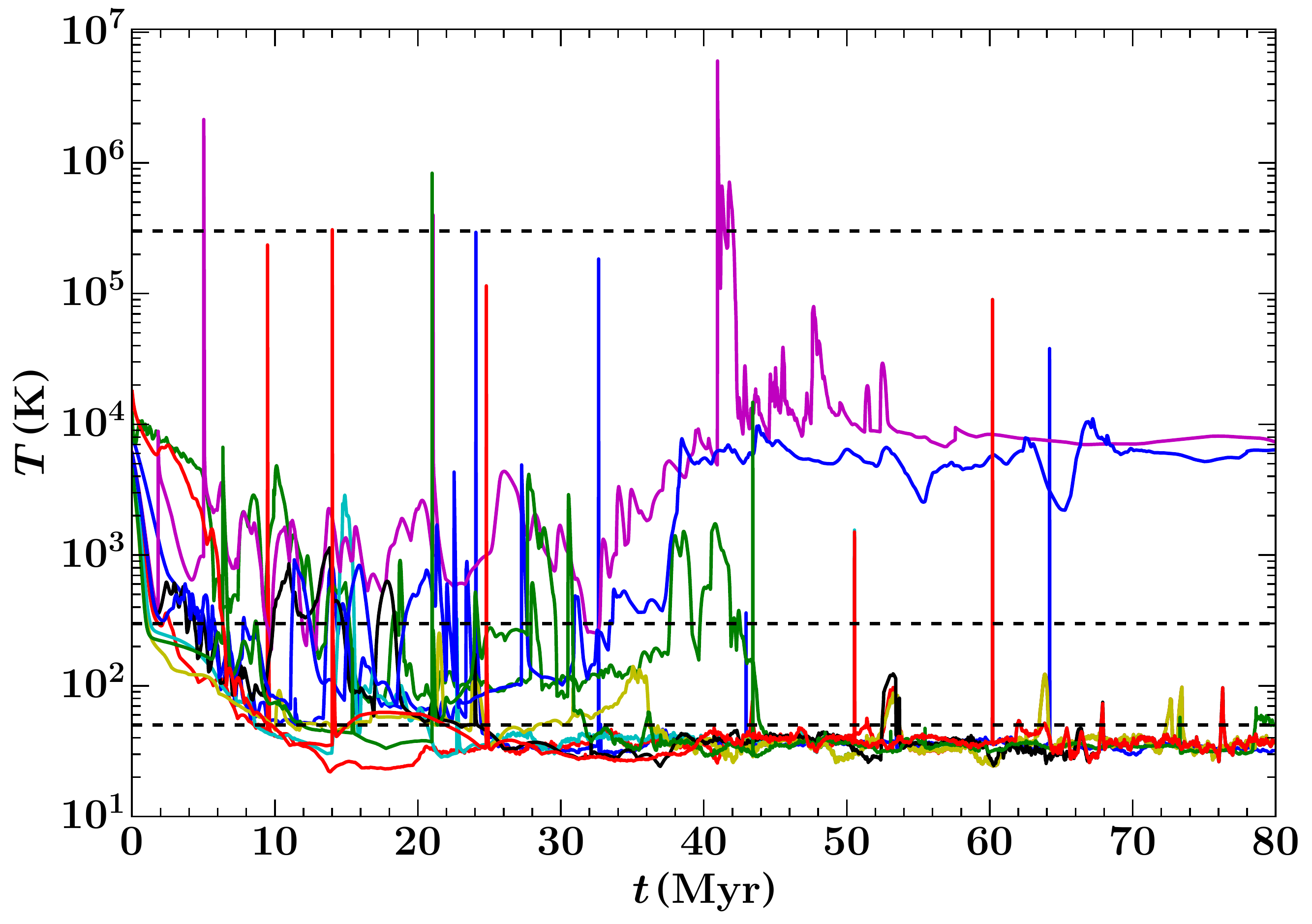}
\includegraphics[width=0.5\linewidth]{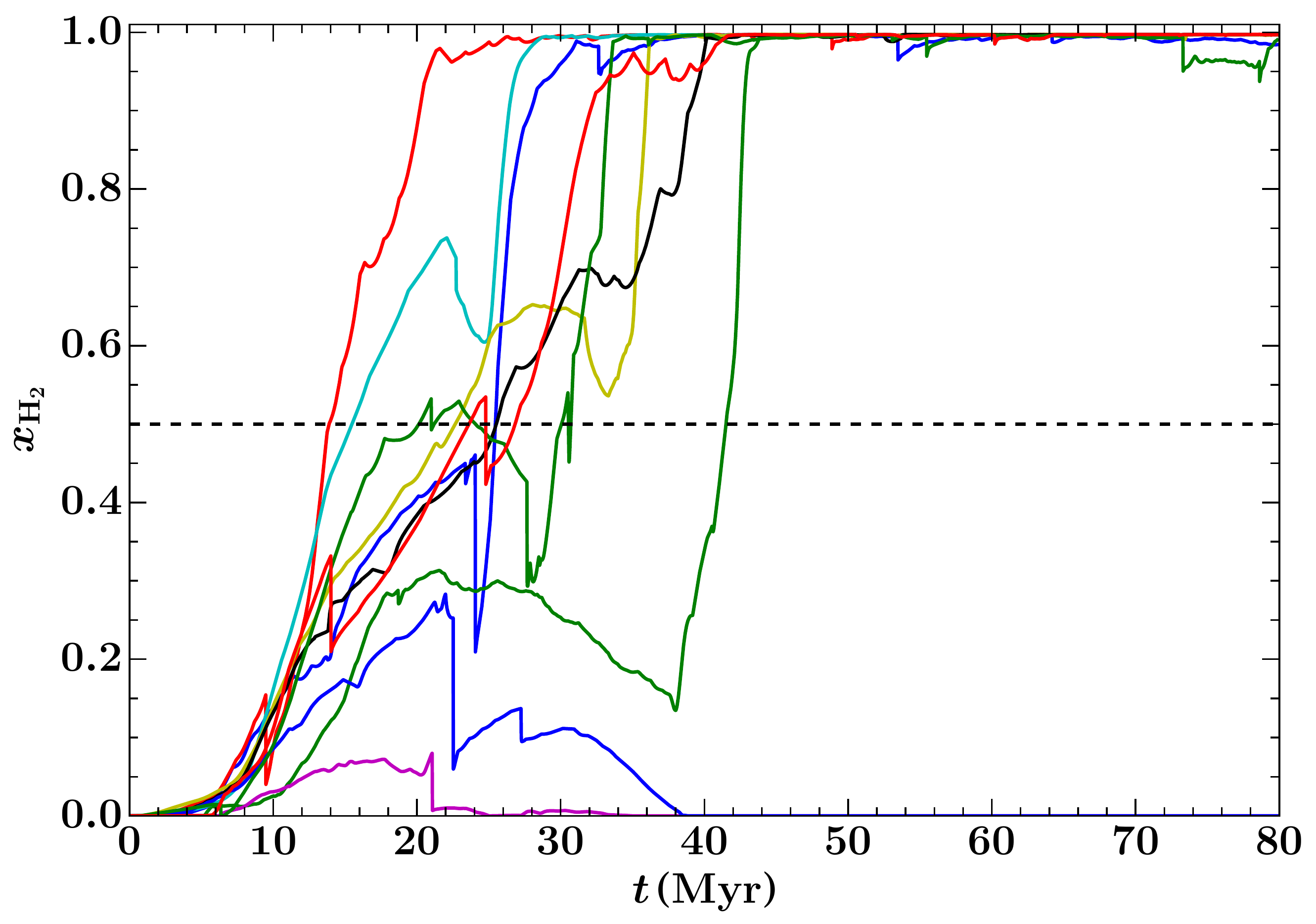}}
\centerline{\includegraphics[width=0.5\linewidth]{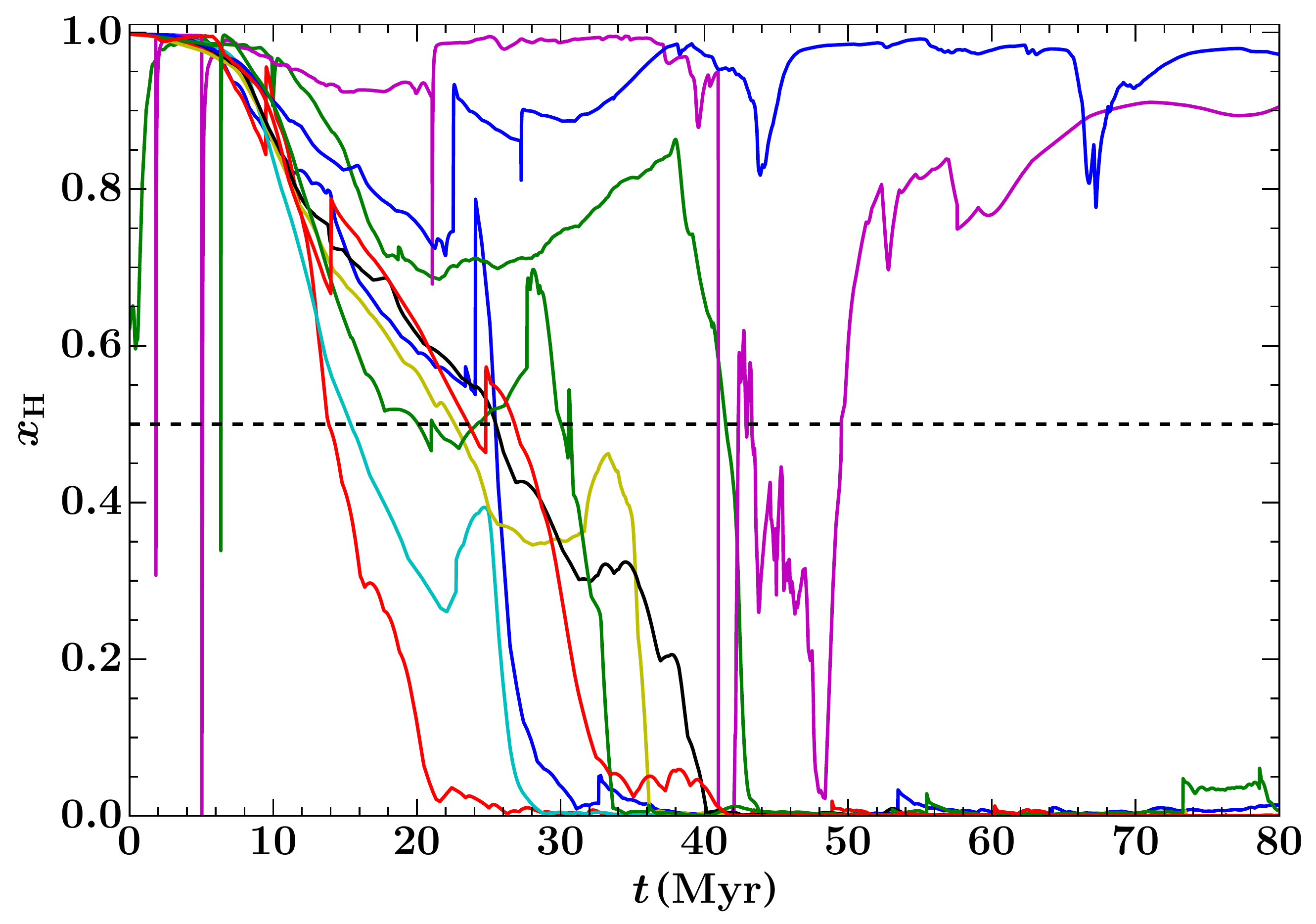}
\includegraphics[width=0.5\linewidth]{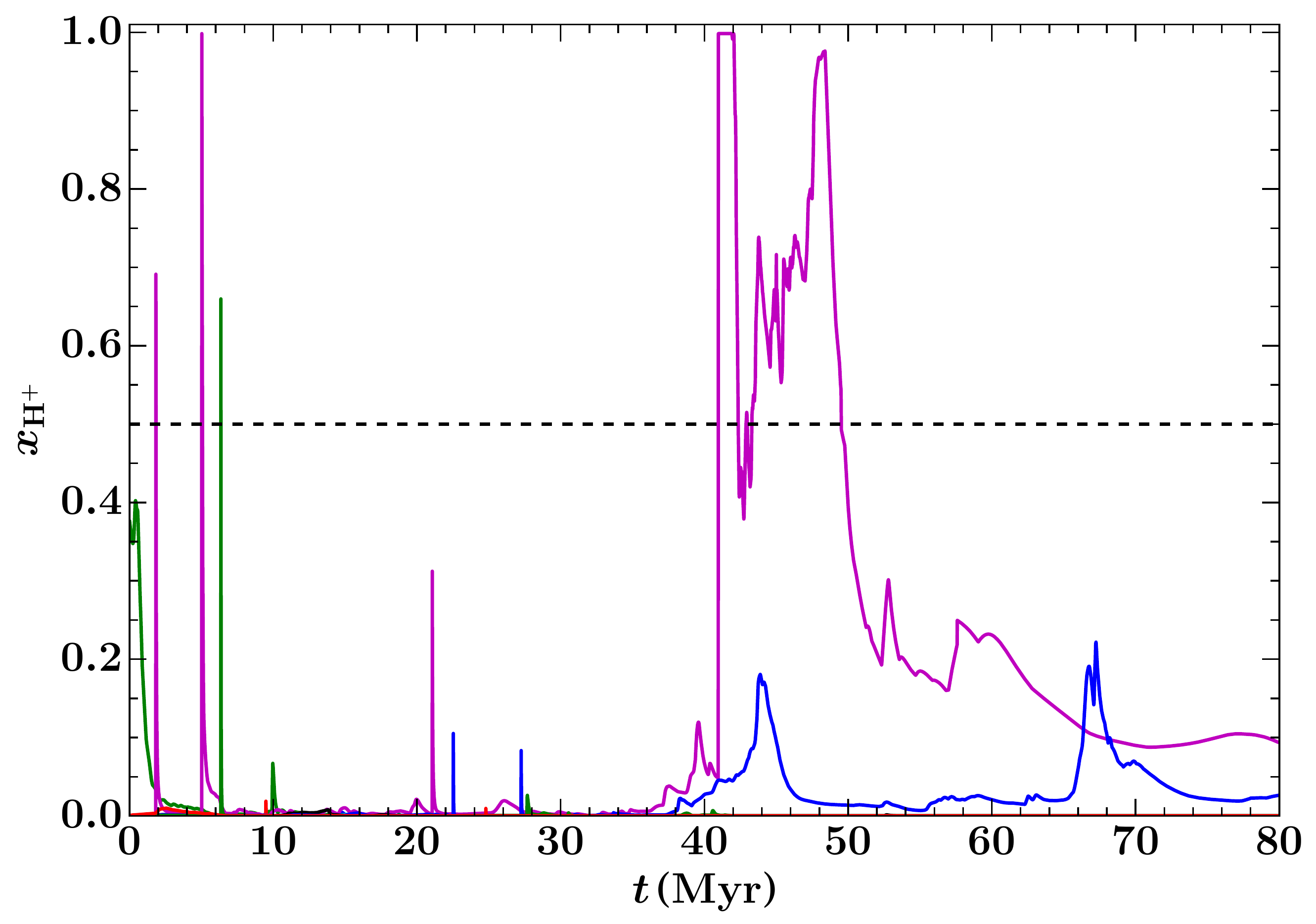}}
\caption{Histories of temperature and molecular, atomic and ionised hydrogen mass fractions for 10
randomly selected particles from the simulation with random driving. The dashed lines indicate the phase boundaries.
The individual particles are identically coloured in the four panels. The particles have complex evolutionary histories and typically change
their ISM phase multiple times during the simulation.}
\label{fig:traj}
\end{figure*}

The reason for this higher temperature cut becomes evident from Figure~\ref{fig:traj}, where we plot the temperature
histories of 10 randomly selected particles from the simulation with random driving. At our spatial grid resolution of
$3.9\,$pc, we barely resolve temperatures below $30\,$K, so that this threshold value would lead to artifically low
residence times in the molecular phase and artificially high transition rates between the molecular and the cold phase.

In \citet{walchetal15}, we compared our simulation results with the classical \citet{McKee:1977p7211} pressure equilibrium
model of a three-phase ISM. We found that the hot gas pressure in our simulations is in approximate equilibrium with the warm phase,
and that the volume filling fraction of the hot gas, which fills the intercloud volume, is in agreement with predictions of this model.

We stress that \citet{walchetal15} have shown that the temperature
cuts do not directly correspond to transitions in the chemical composition.
We therefore define a second,
independent, set of ISM phases using the chemical abundances of the different forms of hydrogen in our
chemical network, or more precisely their corresponding mass fractions $x$:
\begin{enumerate}
\item H$_2$ phase ($x_{\mathrm{H}_2} > 50\%$),
\item H phase ($x_{\mathrm{H}} > 50\%$),
\item H$^+$ phase ($x_{\mathrm{H}^+} > 50\%$).
\end{enumerate}
The histories of the mass fractions are also shown in Figure~\ref{fig:traj} for the same 10 particles.
The time evolution of the chemical abundances is much smoother than the temperature  evolution. In part, this is
because the chemical timescale is often much longer than the dynamical timescale or cooling timescale, meaning
that the chemical makeup of the gas frequently lags behind its thermal state. Consequently, if the duration of a
heating event and the subsequent cooling is short enough, the gas can move from one thermal phase to another and
back again without ever significantly changing its chemical state. In addition, the temperature cuts that we
use to define our different thermal phases are often only weakly correlated with chemical changes in the gas.
For example, although the temperature range we adopt for our ``molecular'' phase is a relatively
good match for the temperature inferred from CO observations, there is both observational
and theoretical evidence that CO-dark H$_{2}$ occupies a much broader range of temperatures
\citep{rachford09,glosmi16}. Therefore, gas moving from the CO-bright 
regime to the CO-dark regime will undergo a transition from our ``molecular'' to our
``cold'' phase, but may remain dominated by H$_{2}$ throughout. 

Since there is no one-to-one correspondence between the two sets of phase definitions (see also discussion in \citealt{walchetal15}),
the chemical abundance cuts provide a separate set of statistics for residence times and transition rates.
These data are more physically meaningful but harder to compare to values in the literature because they
are different from the traditional ISM phase definitions.

The displayed evolutionary tracks illustrate the complexity of interpreting the data in terms of grain properties. For example, in gas that is dense
enough
to shield itself from the interstellar radiation field, allowing molecular hydrogen to form,
ice mantles can grow on the surfaces
of dust grains. However, the tracks demonstrate that particles in the H$_2$ phase are not protected from high temperatures,
and even short periods of heating beyond the evaporation temperature (or exposure to UV radiation) can destroy
the ice mantles again. We conclude that a single physical quantity (density, column density, temperature, or chemical composition)
is not enough to characterise the grain properties, but all of them must be considered in concert.

\section{Sampling of the multi-phase ISM}
\label{sec:sam}

As the tracer particles get advected with the flow, they sample the entire phase space of the multi-phase ISM.
Initially, they are homogeneously distributed across the disc, with a tracer particle number density that scales with the gas density.
When the gas starts to move, regions of compressed gas will contain more tracer particles than voids.
Naturally, since the molecular and cold phase have a higher gas density than the warm and hot phase,
the number density of tracer particles residing in these phases will also be larger. Because the density contrast in
the ISM amounts to several orders of magnitude, we may inadequately sample the underdense gas
with our $N_\mathrm{part} = 10^6$ tracer particles.

\begin{figure*}
\centerline{\includegraphics[width=0.5\linewidth]{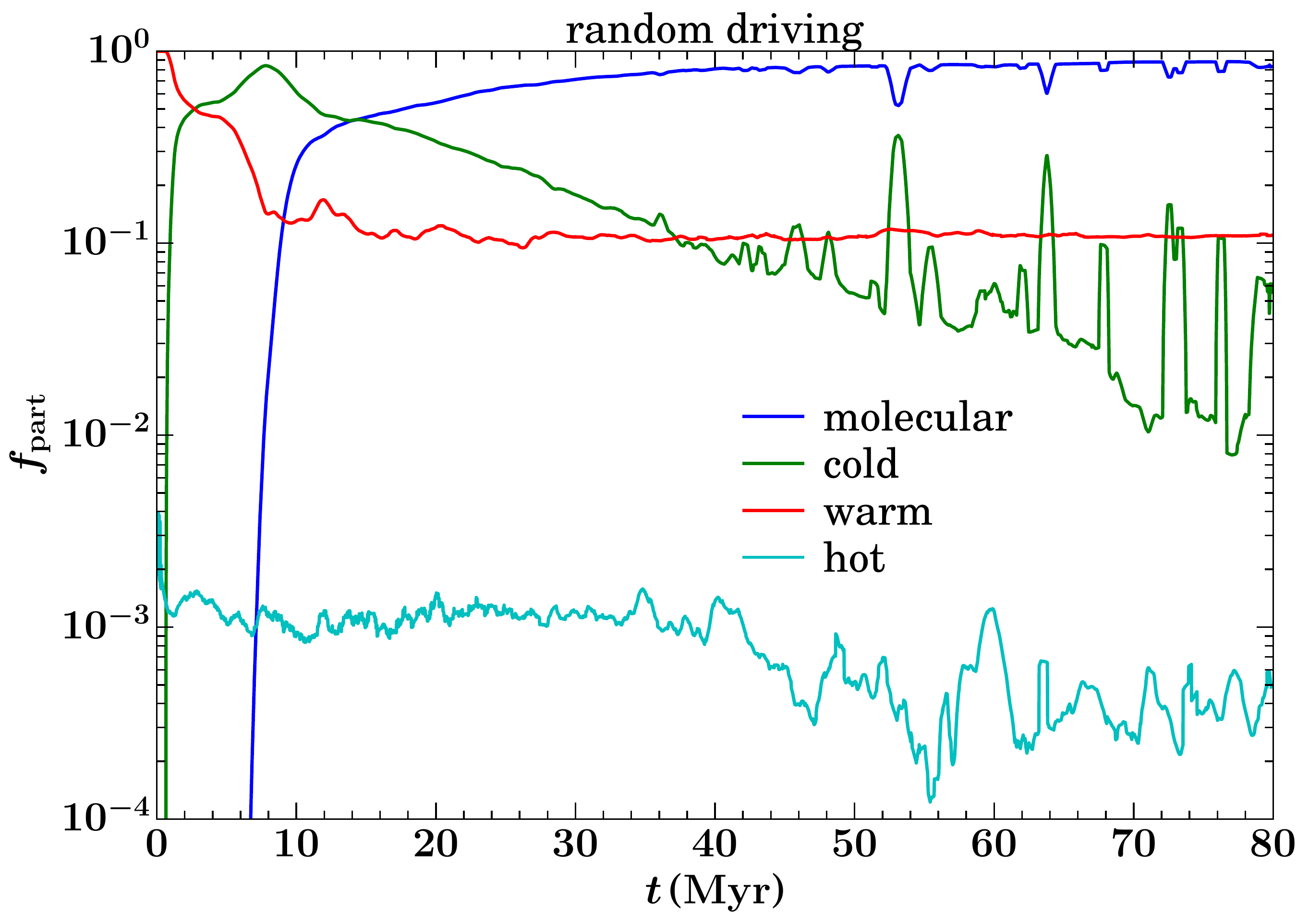}
\includegraphics[width=0.5\linewidth]{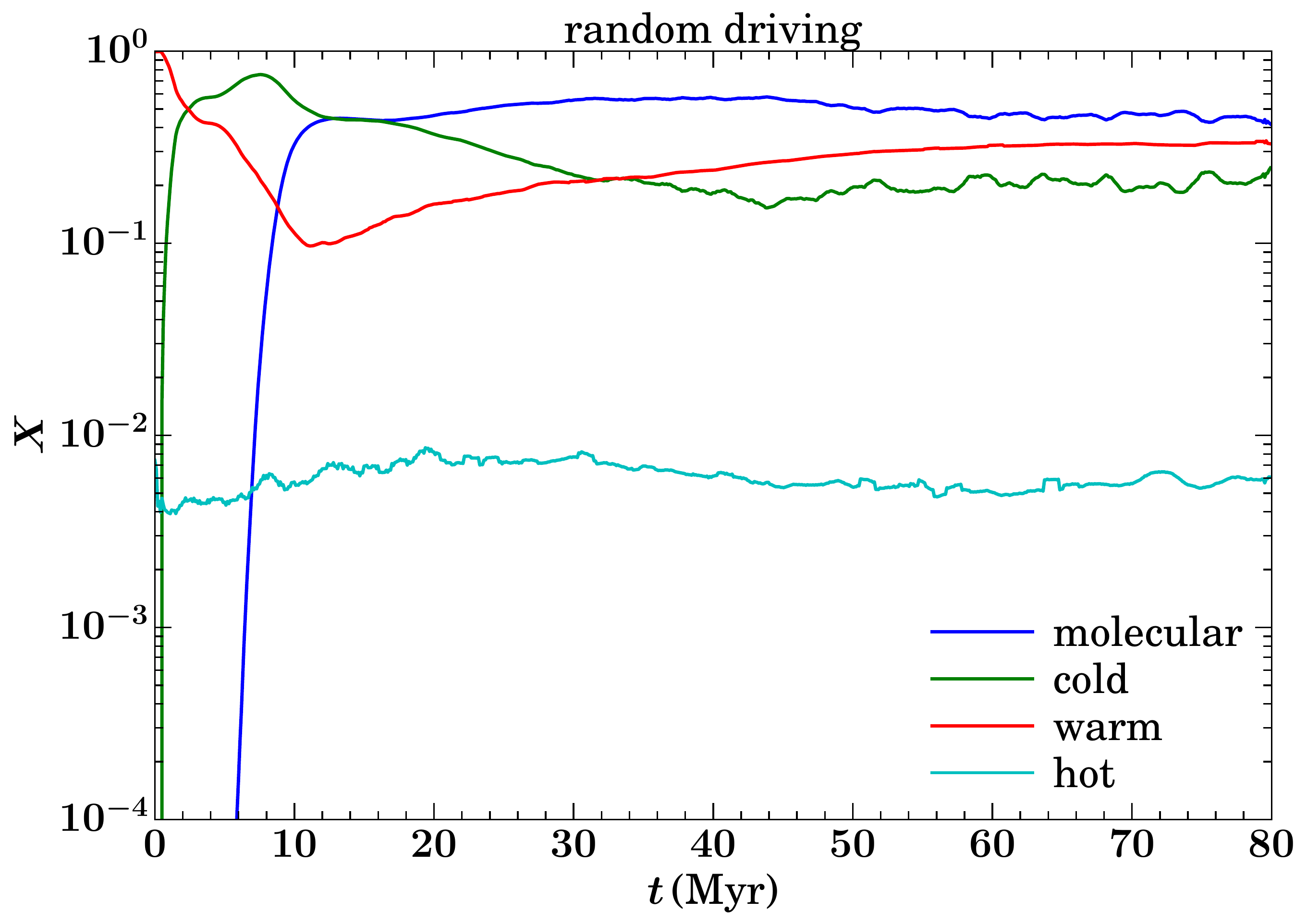}}
\centerline{\includegraphics[width=0.5\linewidth]{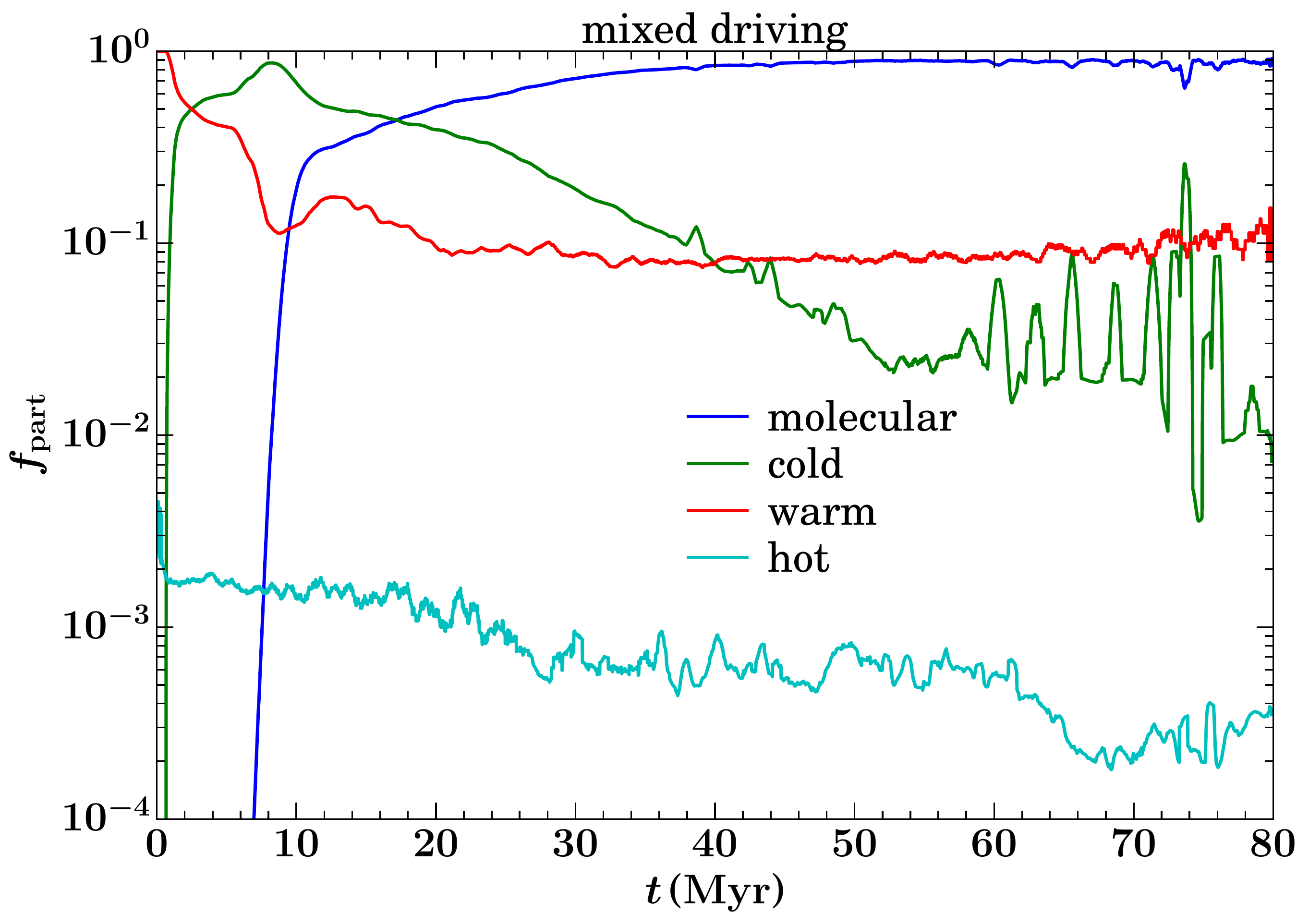}
\includegraphics[width=0.5\linewidth]{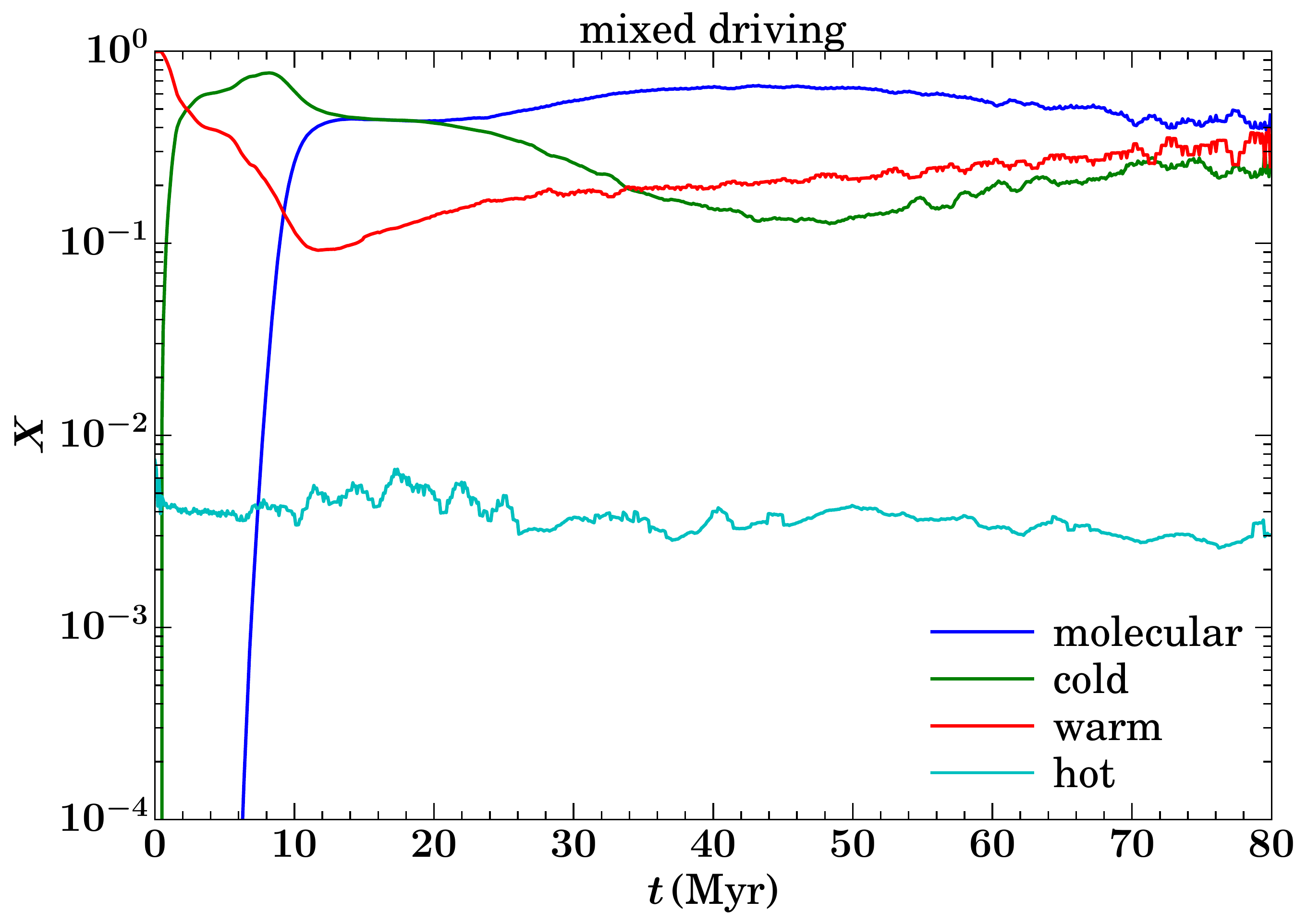}}
\centerline{\includegraphics[width=0.5\linewidth]{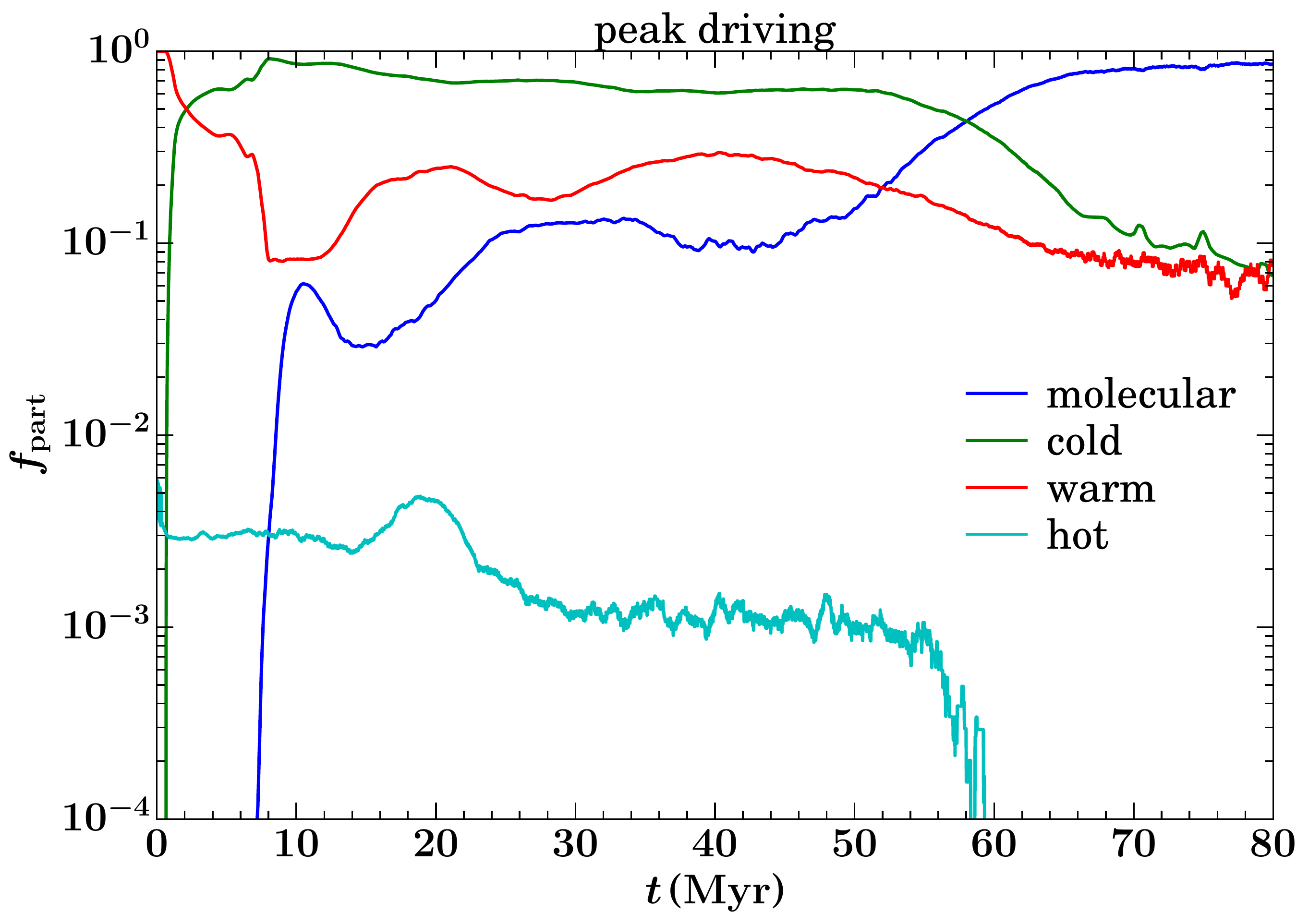}
\includegraphics[width=0.5\linewidth]{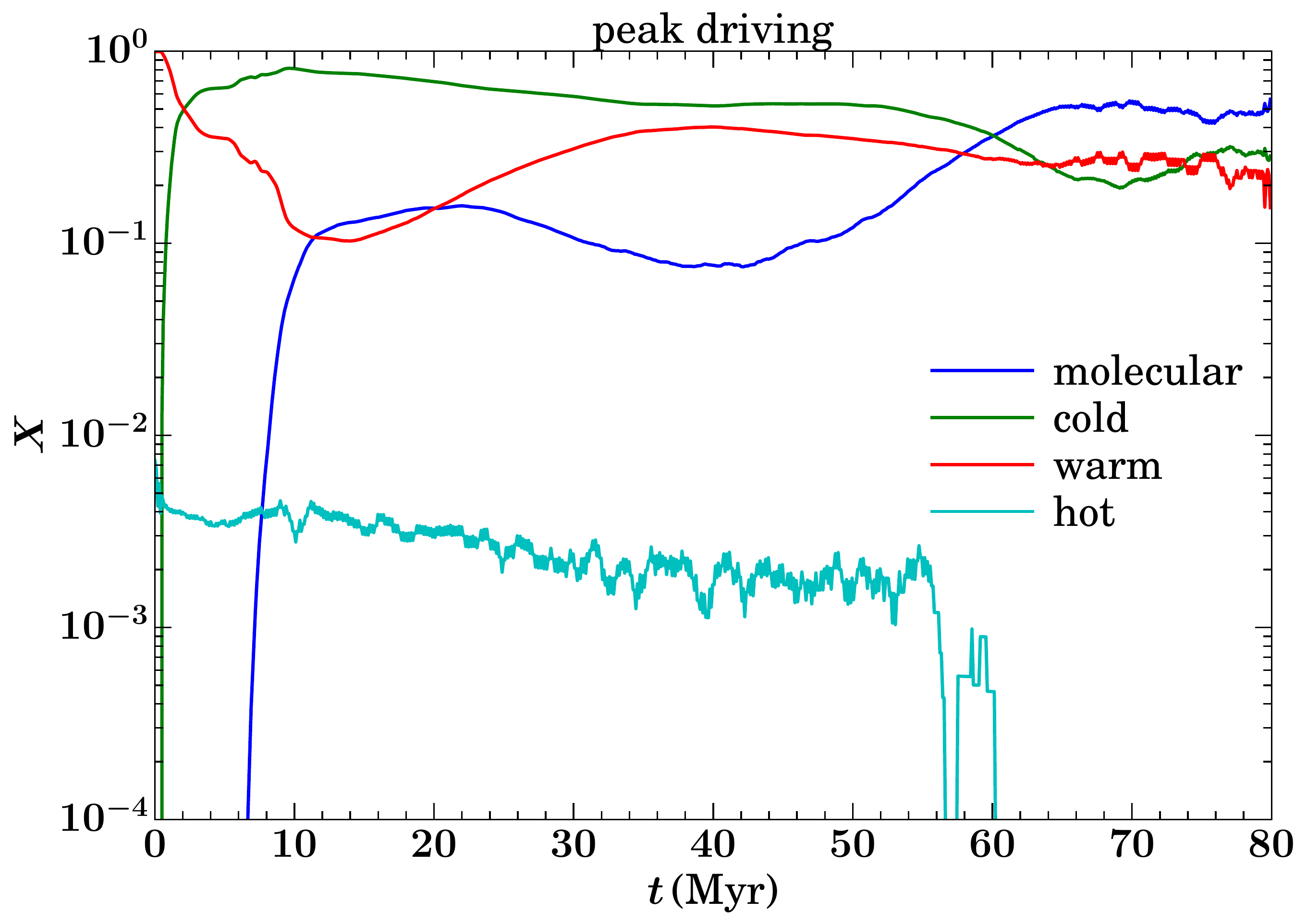}}
\caption{Particle fractions $f_\mathrm{part}$ (left) and total gas mass fractions $X$ (right) for the different ISM phases
defined by temperature cuts as function of time $t$ for random (top), mixed (middle) and peak (bottom) driving.}
\label{fig:pn}
\end{figure*}

\begin{figure*}
\centerline{\includegraphics[width=0.5\linewidth]{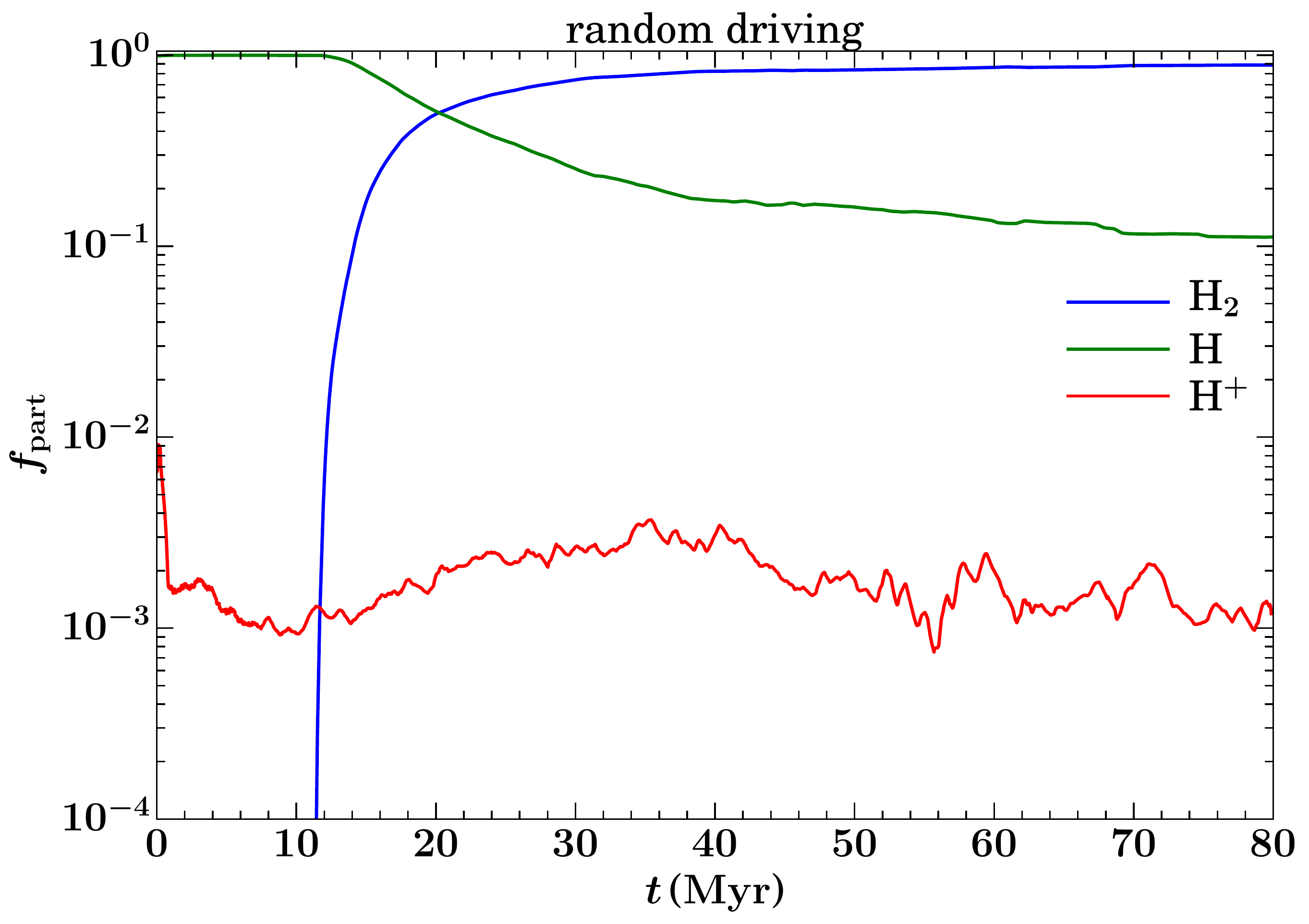}
\includegraphics[width=0.5\linewidth]{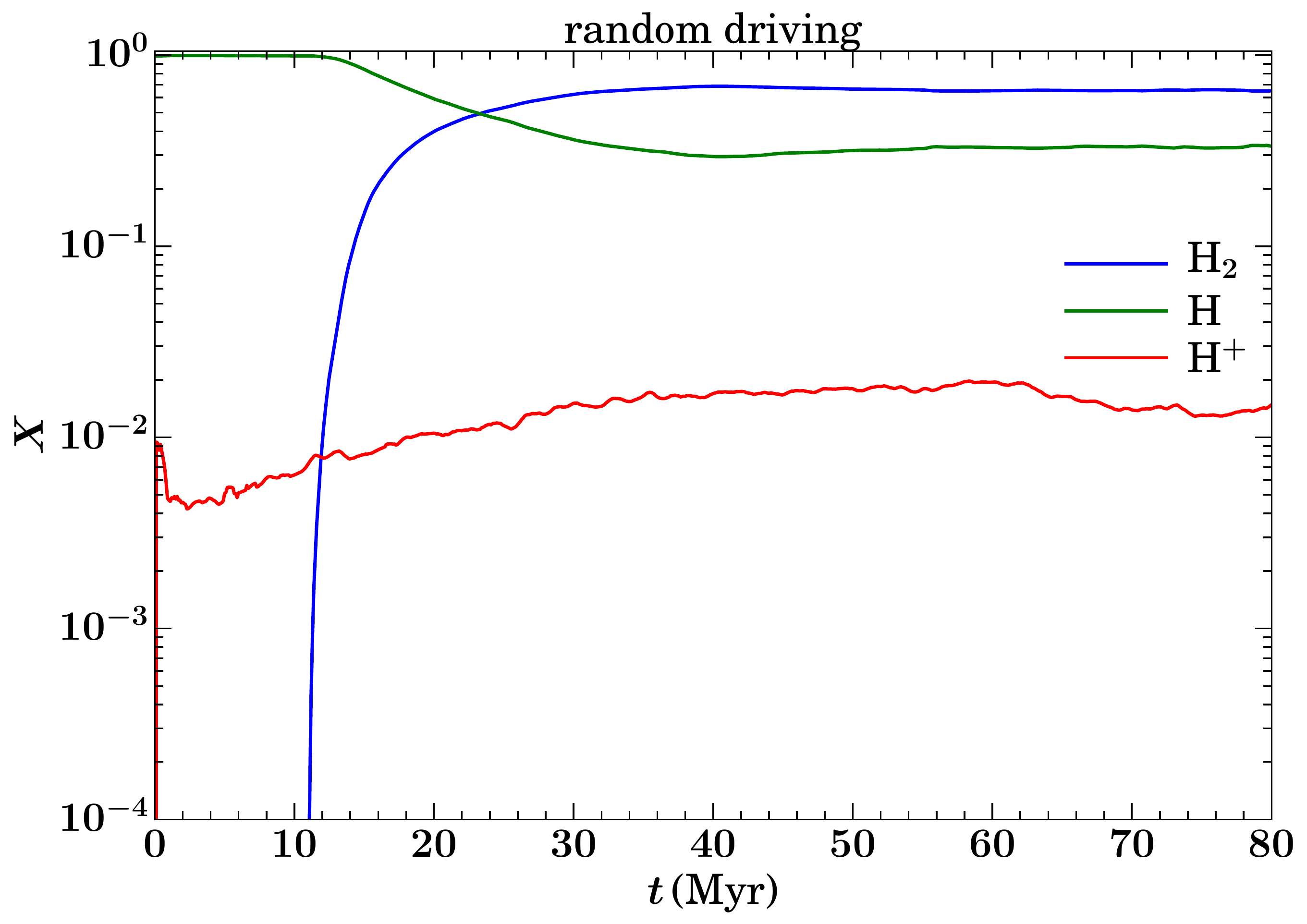}}
\centerline{\includegraphics[width=0.5\linewidth]{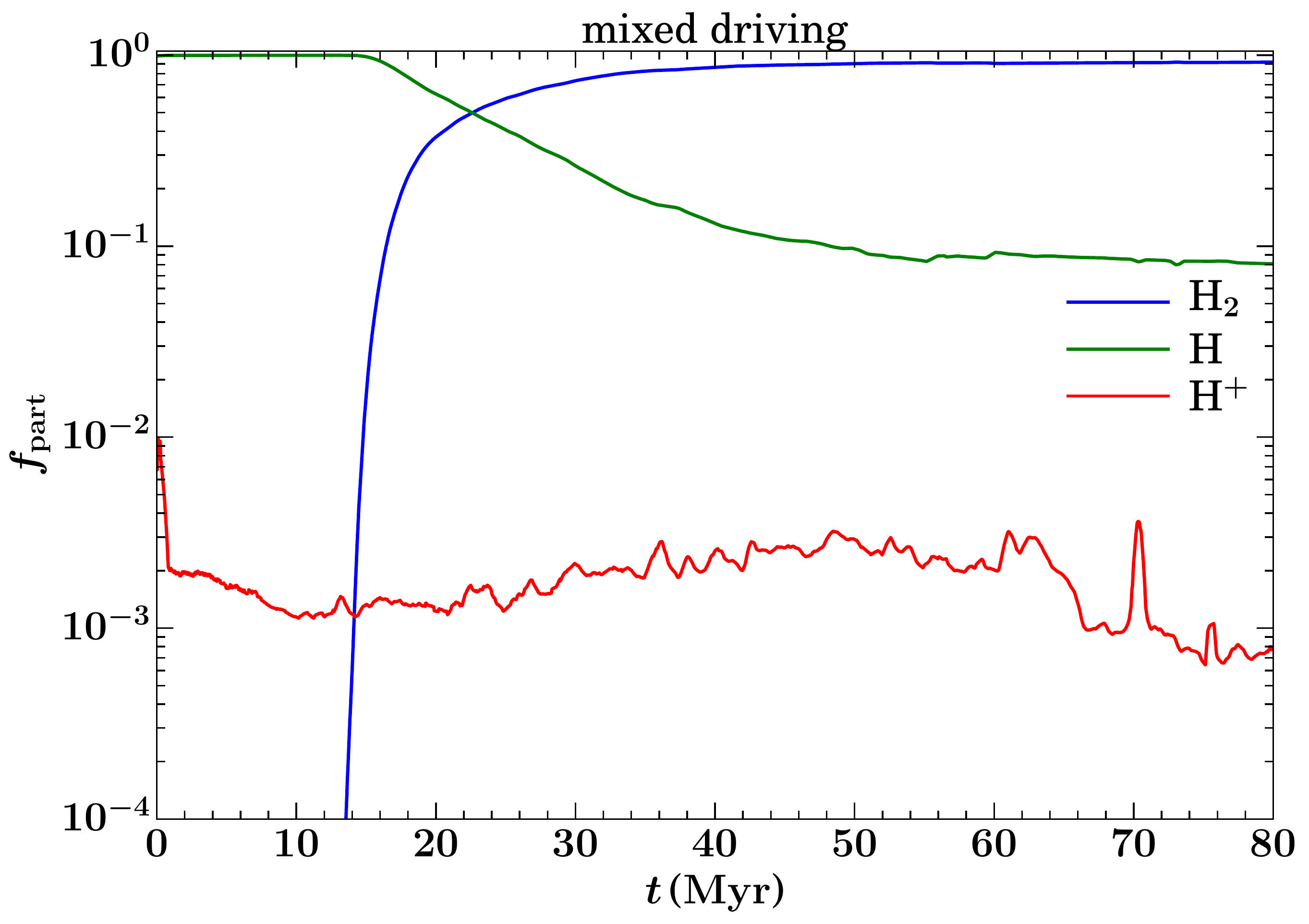}
\includegraphics[width=0.5\linewidth]{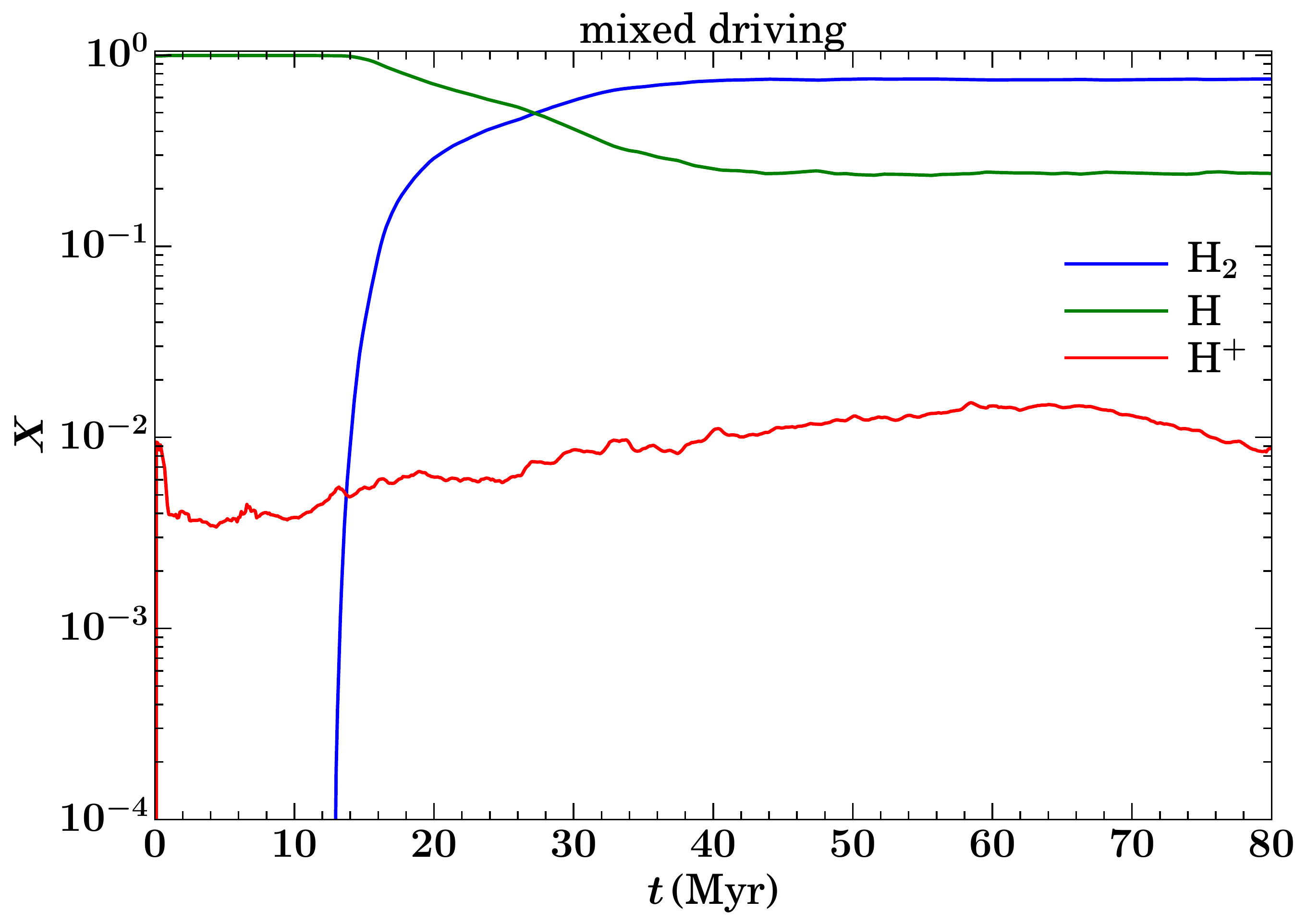}}
\centerline{\includegraphics[width=0.5\linewidth]{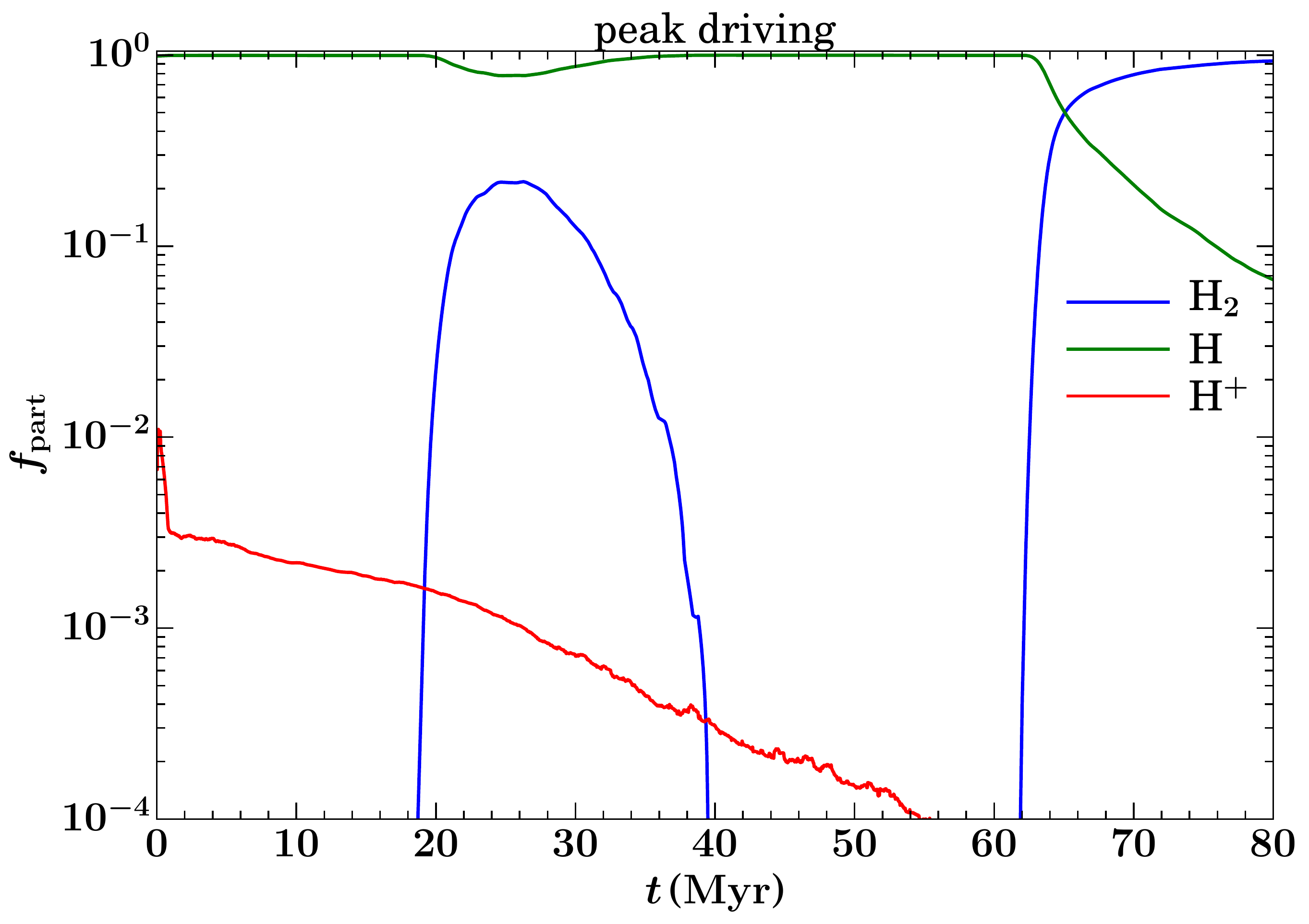}
\includegraphics[width=0.5\linewidth]{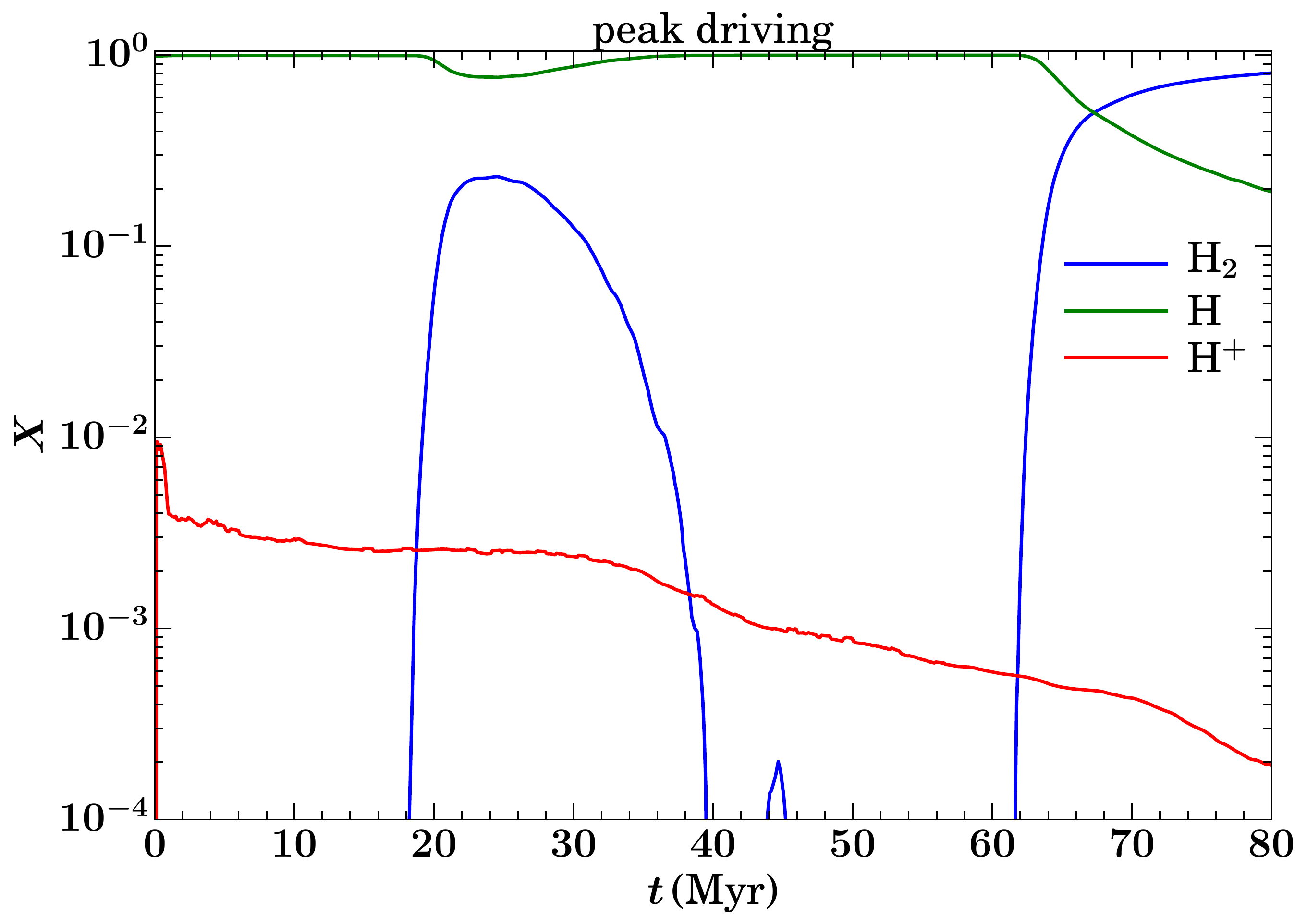}}
\caption{Particle fractions $f_\mathrm{part}$ (left) and total gas mass fractions $X$ (right) for the different ISM phases
defined by chemical abundances as function of time $t$ for random (top), mixed (middle) and peak (bottom) driving.}
\label{fig:pnabun}
\end{figure*}

To check how well the fraction of particles $f_\mathrm{part}$ in the different ISM phases
represents the corresponding total gas mass fractions $X$, we plot both quantities for the three simulations
as a function of time $t$ in Figure~\ref{fig:pn} for the ISM phases defined by temperature cuts and in
Figure~\ref{fig:pnabun} for the chemical abundance cuts.
We see that for random and mixed driving, we need about
$10\,$Myr to produce a molecular phase with a total mass fraction in excess of $10\%$.
After this time, the mass fractions of the four phases only change within a factor of a few.

However, the particle fractions evolve notably differently than the mass fractions.
The particle fraction in the molecular phase steadily increases over the simulation time,
while the particle fraction in the cold phase decreases.
A growing number of particles fall into the deep gravitational potential wells of molecular
clouds (compare Figure~\ref{fig:proj}) and enter the molecular phase from the cold phase, explaining these overall trends.
They can only escape from these regions when a supernova explodes in a nearby location.
During such an event, a large number of particles gets ejected instantaneously from the molecular into
the cold phase, from where they fall back into the molecular phase by gravitational attraction.
This is the reason for the series of spikes that are superimposed on the
general trend for the cold phase evolution after $40\,$Myr.
The particle fractions of the warm and hot phase remain roughly constant, although the hot
phase shows large fluctuations due to the small absolute particle numbers in this phase.

For peak driving, the situation is qualitatively different. Here, continuous supernova explosions
in the dense gas delay the formation of a molecular phase substantially. It takes 60\,Myr until
the mass fraction for the molecular phase reaches similar values as for random and mixed driving.
Simultaneously, the hot phase disappears completely. The regions in which the supernovae explode
are now so dense that no significant amount of hot gas is produced anymore.

In general, the particle fractions $f_\mathrm{part}$ and the total gas mass fractions $X$
agree within a factor of a few. In absolute values, the differences are largest in the molecular phase,
while the biggest relative error between $f_\mathrm{part}$ and $X$ occurs for the hot phase. Because of the discrepancies between
$f_\mathrm{part}$ and $X$, we must be aware that the tracer particles do not equally sample the full
simulation box. We note that this may be due to a fundamental problem of Lagrangian tracer particles in
highly compressible flows \citep{genel13}, and therefore the
mismatch is unlikely to be solved by simply increasing
the total number of tracer particles. Instead, early stellar feedback from winds \citep{gatto16}
and radiation \citep{peters16b} may prevent the tracer particles from being locked inside
clouds and facilitate a much better circulation of the particles through the different ISM phases.
We plan to test this hypothesis in future work.

The evolution of the chemical phases shown in Figure~\ref{fig:pnabun} is more regular compared to the temperature cuts,
which can be directly traced back to the less erratic individual trajectories.
The definition of the chemical phases is more robust with respect to perturbations.
In particular, supernova explosions near molecular clouds do not eject a large number of locked particles into
regions where the molecular mass fraction is less than 50\%, so that these particles do not change
their phase.
In general, the agreement between
particle fractions and total gas mass fractions is better than for the temperature cuts.
A substantial difference in the time evolution of the ISM phases between the simulations can be observed in the run
with peak driving, where an H$_2$ phase is beginning to build up at $20\,$Myr but then gets
completely destroyed again at $40\,$Myr. The H$_2$ phase can only persist after $60\,$Myr.
In this simulation, the H phase is by far dominant for most of the simulation runtime because
the supernova explosions at density peaks can very effectively delay the formation of significant
amounts of molecular hydrogen.

\section{Transition rates}
\label{sec:trans}

Mass exchange between the ISM phases plays an important role in the lifecycle of interstellar grains.
It controls the circulation of gas from the WNM, where dust abundances are reduced by destruction in interstellar shocks,
and matter enriched with dust due to accretion on grain surfaces at higher gas densities. 
Several models of dust evolution in the idealised two- and three-phase ISM have been proposed  
to explain the observed differences between element depletion in the warm medium and cold clouds
\citep{draine90, ODonnell:1997p683, Tielens:1998p7054, Weingartner:1999p6573}. 
Models with a two-phase ISM neglect the difference between the molecular and diffuse \hi\ clouds and consider 
mass circulation only between the ambient warm medium and clouds \citep{ODonnell:1997p683, Tielens:1998p7054}. 
\cite{ODonnell:1997p683} showed that a three-phase ISM model with an additional mass exchange with the molecular clouds 
reproduces the observations better. Assuming a steady state for the interchange between phases and timescales 
of dust destruction and accretion, one can estimate the rates of mass exchange required to reproduce the 
observed element depletion \citep{draine90, Weingartner:1999p6573}. The resulting rates, however, depend 
on the adopted scheme of mass transfer between phases that in the case of the three-phase ISM can occur via
different routes. For example, depending on whether the WNM is converted to molecular clouds directly or through the CNM, the mass
exchange rates between the CNM and molecular clouds can differ by a factor of $8$ \citep{draine90}.

In addition to the mass circulation scheme, the outcome of dust evolution models
depends on the implementation of dust destruction by supernova shocks and dust growth by
accretion in clouds, which introduce more uncertainties in the models. For example,
unknown details of the growth process, in particular, the efficiency of sticking of
incident species on the grain surfaces can strongly affect the value of the accretion
timescale and the resulting dust abundance distribution \citep{zhuk16}.

In this work, we clarify the uncertainties in dust evolution modelling with respect
to the matter cycle and mass exchange scheme by means of numerical simulations of
the turbulent ISM and provide a basis for studies of grain processing in the ISM.
The simulations allow us to directly measure the mass interchange rates between the
different ISM phases defined in Section~\ref{sec:traj} in a dynamically more realistic situation. In the following, we analyse the mass
exchange rates predicted by the three simulations with the different supernova
locations to determine the dominant transitions between phases.

\begin{figure*}
\centerline{\includegraphics[width=0.5\linewidth]{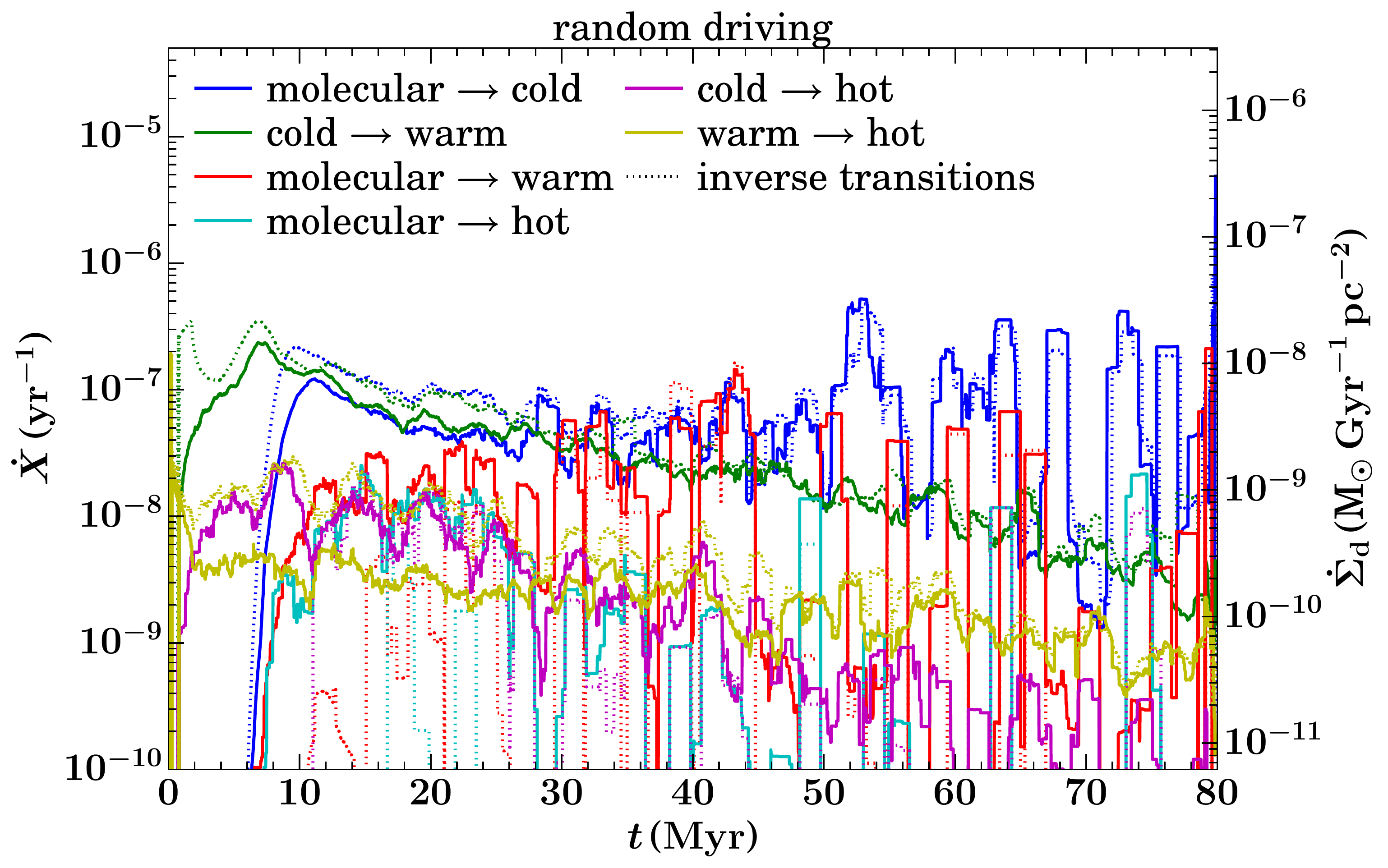}
\includegraphics[width=0.5\linewidth]{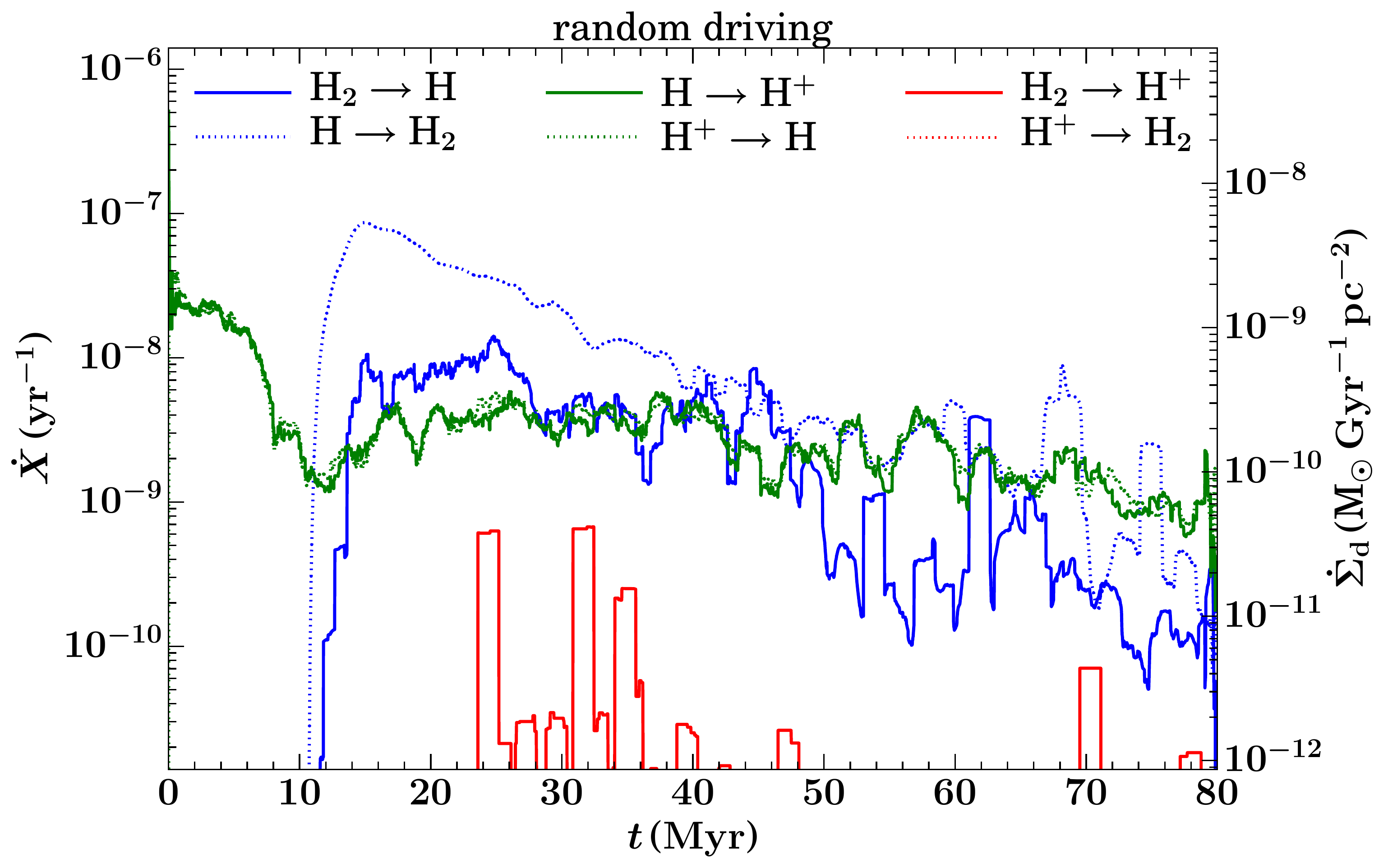}}
\centerline{\includegraphics[width=0.5\linewidth]{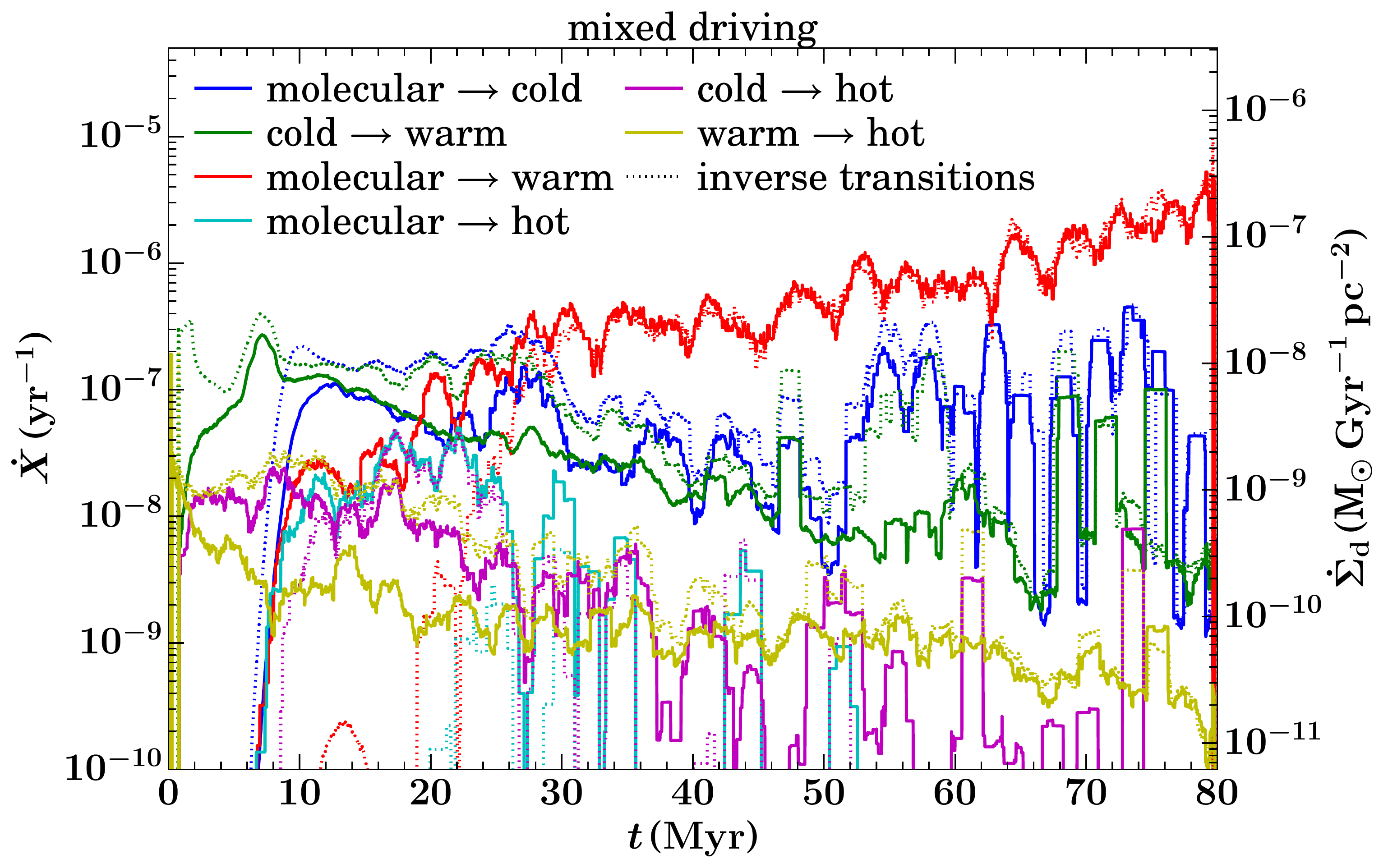}
\includegraphics[width=0.5\linewidth]{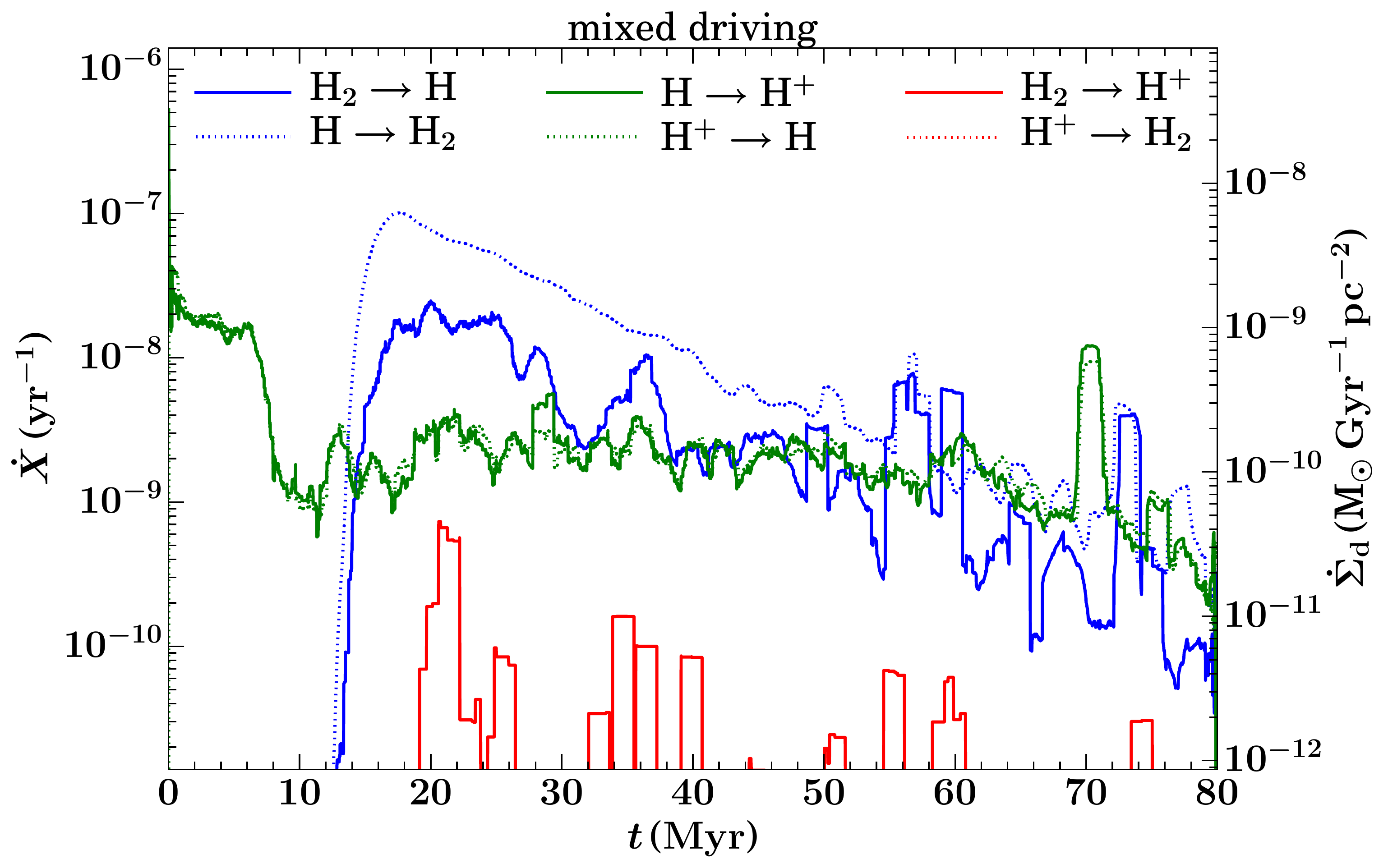}}
\centerline{\includegraphics[width=0.5\linewidth]{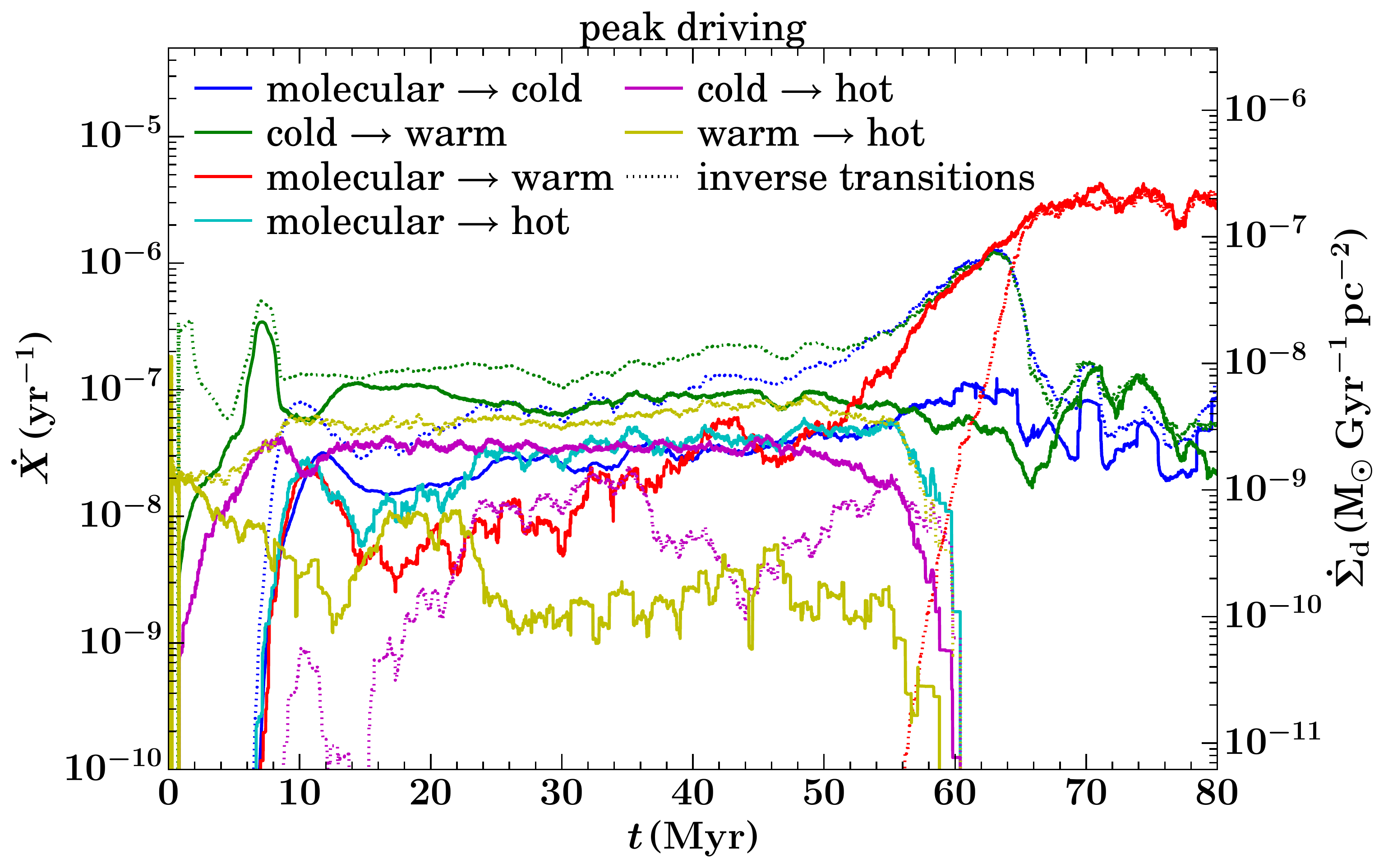}
\includegraphics[width=0.5\linewidth]{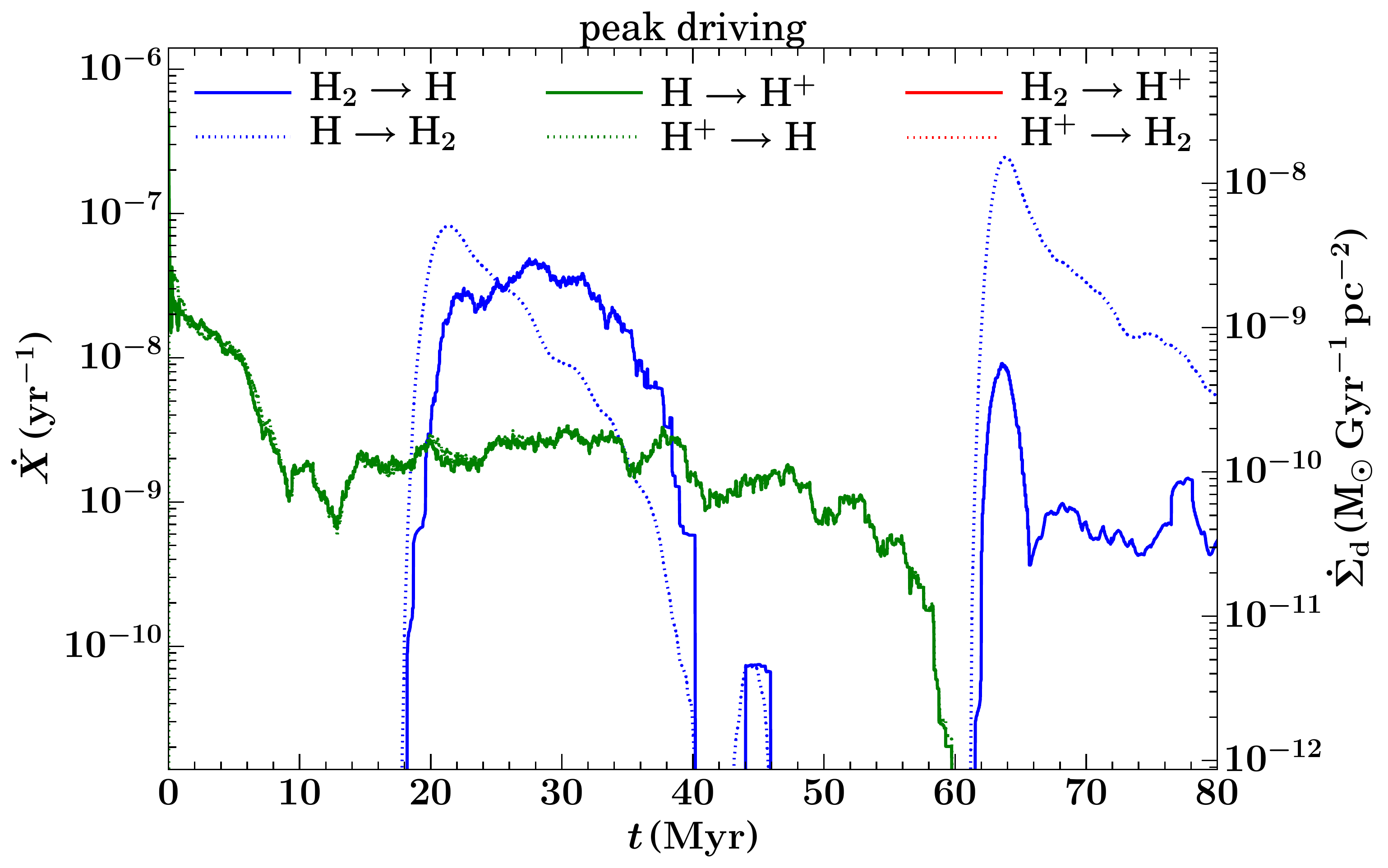}}
\caption{Transition rates of particles moving
between different ISM phases defined by temperature cuts (left) and chemical abundances (right)
as function of time $t$ for random (top), mixed (middle) and peak (bottom) driving.
Transitions from a cooler to a warmer phase are drawn with solid lines, and the corresponding inverse transitions
from the warmer to the cooler phase with dashed lines. The ordinate on the left shows $\dot{X}$ and the right ordinate $\dot{\Sigma}_\mathrm{d}$.}
\label{fig:trall}
\end{figure*}

The transition rates between the different ISM phases for the simulations with random, mixed
and peak driving are shown in Figure~\ref{fig:trall}
as function of time $t$. We express the transition rates
as time derivatives of the particle fractions, $\dot{X}$, and of the dust surface density, $\dot{\Sigma}_\mathrm{d}$.
Let $\dot{N}_{i\to j}(t)$ denote the number of particles transitioning from phase $i$ to phase $j$ at time $t$, then
the corresponding transition rate of the particle fraction is
\begin{equation}
\dot{X}_{i\to j}(t) = \frac{\dot{N}_{i\to j}(t)}{N_\mathrm{part}},
\end{equation}
and the transition rate for the surface density is
\begin{equation}
\dot{\Sigma}_\mathrm{d}(t) = \frac{m_\mathrm{part} \dot{N}_{i\to j}(t)}{A_\mathrm{box}}
\end{equation}
with the surface area of our simulation box $A_\mathrm{box} = (0.5\,$kpc$)^2$.
As already noted in Section~\ref{sec:sam}, it takes 10\,Myr for the molecular phase to form, which
is also reflected in the transition rates. After this initial period, all four ISM phases coexist in the simulations,
with the exception of the disappearance of the hot phase in the peak driving run after 60\,Myr.

A common feature of all transition rates is their high intermittency. The main reason for this is the sudden
heating of the gas by supernova explosions and the subsequent cooling process. Depending on the location of the supernova
explosion (i.e., the local gas density), both the peak temperature reached during an explosion event and the timescale for cooling of the remnant
afterwards varies. Therefore, the dominant phase transitions and their rates depend on the supernova positioning and
are different for each simulation.

The simulation with random driving is dominated by the transitions between the molecular and
the cold phase. This is reasonable since these two phases are directly connected and contain most tracer particles.
The rate itself varies by more than an order of magnitude around a mean value of
$\dot{X} \sim 10^{-7}\,$yr$^{-1}$ or $\dot{\Sigma}_\mathrm{d} \sim 10^{-8}\,$M$_\odot\,$Gyr$^{-1}\,$pc$^{-2}$.
The next most dominant rate, which is roughly one order of magnitude lower,
is the one between the cold and the warm phase, which again can be
understood in terms of the particle fractions. These two transitions are to a very good approximation
in detailed balance, meaning that the number of particles undergoing the transition from the cooler
to the warmer phase equals the number of particles making the transition in the other direction.
Intermittently, the direct transition from the molecular to the warm phase becomes important, too.
This is an example for a transition rate in which an intermediate phase (the
cold phase) is skipped at the time resolution of our sampling rate for the tracer particle data, which is 10\,kyr.
This effect is caused by the supernovae that explode in dense gas.
The supernova injection then quickly heats up the gas and brings
it into the warm phase. Usually, the cooling time after the supernova injection is much longer than 10\,kyr,
which explains why the inverse transition does not occur very often. Instead, cooling proceeds via the cold
phase as an intermediate step.

The simulation with mixed driving shows a slightly different behaviour. Here, the transition rate between
the molecular and the cold phase is comparable to the case of random driving, but the transitions between
the molecular and the warm phase are now dominant.
This effect is caused by the 50\% of all supernovae that explode at density peaks in this run. 
These density peaks are usually found in the molecular phase. A supernova explosion then triggers a
transition into the warm phase, but at the density peaks the cooling time is short enough that tracer
particles directly fall back into the molecular phase \citep[e.g.][]{wana15,haid16}. A small fraction of these particles does make
a transition via the cold phase as an intermediate step, but the molecular and the warm phase are to a good
approximation in detailed balance, as are all transitions between neighbouring phases.

The peak driving simulation shows yet another behaviour.
Here, because of the suppression of the molecular phase, the molecular-cold transition is insignificant.
Instead, the direct transition cold-warm and its inverse transition warm-cold are dominant.
After the molecular phase has formed, the indirect transition molecular-warm and the inverse transition cold-molecular also
become significant since many particles reside in the dense gas. These transitions reflect the heating and cooling cycle after supernova explosions.
The injection of thermal energy brings the gas from the cold into the warm or hot phase. In the hot phase,
the gas cools via the warm phase down to the cold phase again. After 60\,Myr, the hot phase cannot be
maintained anymore, and the supernova explosions only produce a warm phase
since they explode in high-density environments and cool quickly.

The transition rates for the ISM phases defined by chemical abundances behave differently from the
phases defined by temperature cuts. Here, detailed balance only occurs between the H phase and the
H$^+$ phase. The chemical evolution for random and mixed driving is almost identical, despite
the large differences in the temperature evolution. Both simulations show a significantly larger
transition from the H phase into the H$_2$ phase than vice versa, which reflects the net H$_2$ formation
observed in the simulations (compare Figure~\ref{fig:pnabun}). The transition rates for the peak
driving run reveal the formation and destruction of H$_2$ between $20$ and $40\,$Myr, and the
disappearance of the H$^+$ phase and the final formation of an H$_2$ phase after $60\,$Myr.

\subsection{Comparison with existing mass exchange schemes}

In the following, we compare the mass transfer rates directly measured in our simulations 
with the rates required by simple dust evolution models to reproduce the observed differences in 
interstellar element depletion observed in the cold and warm medium. 
\citet{draine90} adopts a three-phase ISM with characteristic temperatures of $30$, $100$ and $6000\,$K
in the molecular, cold and warm phase, respectively.
He considers two schemes for the mass circulation presented in his Table~III, Model~A and Model~B,
which only differ in their mass exchange rates but otherwise have identical parameters.
Model~A has a molecular-cold transition rate of $1.5 \times 10^{-7}\,$yr$^{-1}$ and a cold-warm
transition rate of $8 \times 10^{-9}\,$yr$^{-1}$. Both transitions are assumed to be in detailed balance.
This model is broadly consistent with the time-averaged behaviour we see in our simulation with random driving.
In contrast, his Model~B assumes a molecular-cold transition rate of $3 \times 10^{-8}\,$yr$^{-1}$
and an inverse transition rate of $2 \times 10^{-8}\,$yr$^{-1}$.
The cold-warm transition rate in this model is assumed to be $8 \times 10^{-9}\,$yr$^{-1}$, and
the inverse transition goes directly
from the warm to the molecular phase with the same rate.
This model is not consistent with any of our simulations, for several reasons. The magnitude of his
molecular-cold transition is too small, the relative strengths of the molecular-cold and cold-warm
transitions do not match, and it is difficult to get a warm-molecular transition rate that is of
similar strength as the cold-warm transition rate.

\cite{Weingartner:1999p6573} adopt a different scheme of the mass exchange in the ISM, in which molecular
clouds are formed from the CNM and, upon destruction, are circulated directly to the warm phase. The CNM
is formed by cooling of the WNM and can exchange mass with both the WNM and molecular clouds.
Transitions from molecular clouds to the CNM or from the WNM to the molecular phase are not considered in the model.
Their model
includes dust growth by accretion, dust destruction in the ISM and input from stellar sources or galactic inflows.
The main feature of this model is that it incorporates the enhanced accretion rates due to focusing of ions on small negatively charged grains
in the cold and warm phases (see the following Subsection). The transition rates between phases are then computed from the model parameters
and are given in their Table~3.

Their scheme of mass exchange is also not supported
by our results as we observe intense mass transfer from the warm to the molecular phase in the simulations
with mixed driving and from the molecular to the cold phase in the case of random driving,
which are absent in \citet{Weingartner:1999p6573}. The transition
rates included in their scheme of the order of a few  $10^{-8}\,$yr$^{-1}$ are roughly similar to the values predicted by our simulations,
except the last $20$\,Myr for the mixed and peak driving simulations, when the dominant mass exchange rates exceed $10^{-7}\,$yr$^{-1}$.

Since relative dust abundances are higher in molecular clouds, the pathway of circulation of matter from molecular clouds to warm gas
influences the distribution of interstellar element depletion \citep{ODonnell:1997p683}. If dust-rich gas from
molecular clouds rapidly circulates to the WNM, the average element abundances in dust in the WNM will be higher compared
to the case when dust-rich matter circulates from molecular clouds to the WNM through the CNM. The latter scenario agrees with our
simulation with random driving. The faster, direct circulation between molecular clouds and the warm medium appears in the
simulation with mixed driving.

\subsection{Model predictions for element depletion}

\begin{figure}
\includegraphics[width=\linewidth]{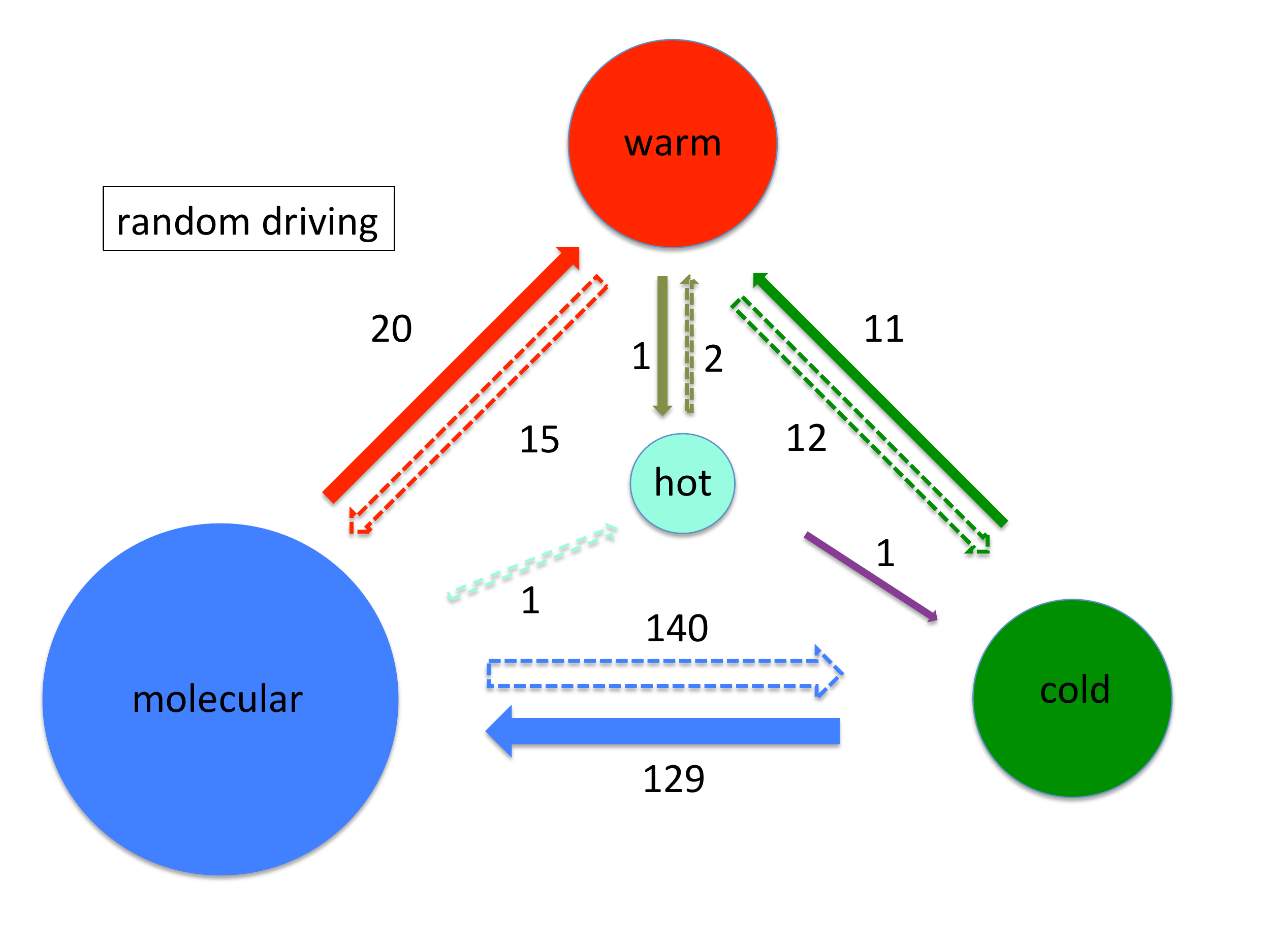}
\vphantom{h}\\
\includegraphics[width=\linewidth]{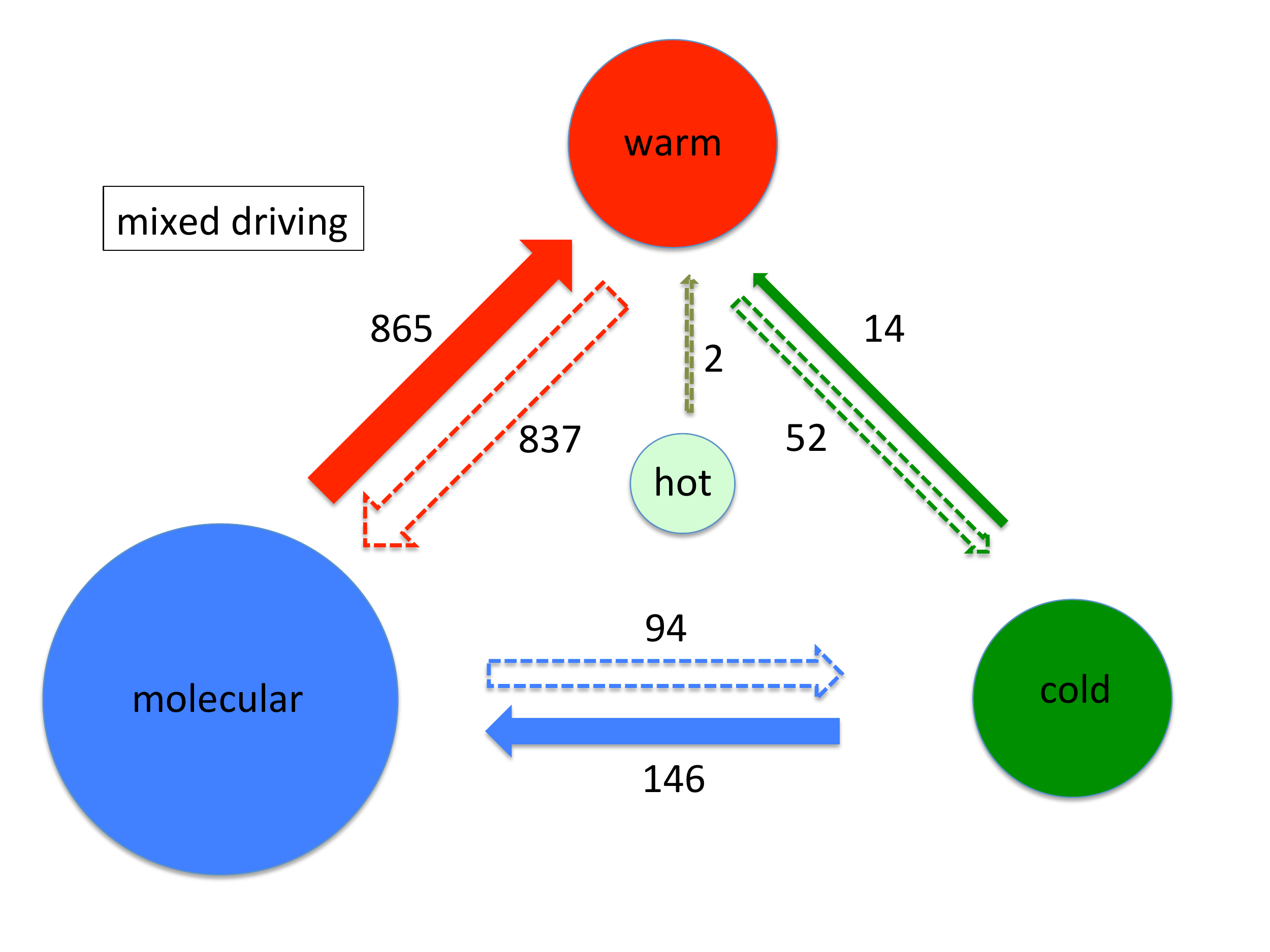}
\caption{Schemes illustrating the dominant phase transitions (in units of $10^{-9}\,$yr$^{-1}$) in the simulation with random (top)
and mixed (bottom) driving averaged around $t = 60\,$Myr $\pm 10\,$Myr. Transitions involving the hot phase are neglected in the depletion models.
The sizes of the arrows indicate the strength of the transitions, and the extent of the circles illustrates the corresponding mass fractions.}
\label{fig:cartoon}
\end{figure}

For random and mixed driving, 
we study the implications of our measured transition rates for predictions of element depletion on dust grains
with the simple model of \cite{Weingartner:1999p6573}.
They used the observed depletion to infer the transition rates between the phases in their
mass transfer scheme. Here, we reverse the procedure and take the transition rates as given to compute the depletion.
For consistency, we leave all other parameters of the model unchanged.
We adopt Model~A and Model~B of \cite{Weingartner:1999p6573}, which use different grain size distributions and destruction timescales.
Below we briefly describe the models and refer to the original publication for 
more details.
We then compare our results with the observed depletion in the warm and cold medium.
Since the transition rates in the simulations are highly fluctuating, we take the
average transition rates from $t = 50\,$Myr to $t = 70\,$Myr.
Figure~\ref{fig:cartoon} shows the two mass-exchange schemes and the measured rates.

\begin{table*}
\caption{Comparison of the predicted Ti depletion $\log \delta_j$ for mass exchange rates from the
simulations with random and mixed driving with observations.}
\label{table:Depl}
\begin{center}
\begin{tabular}{l c c c c c }
\hline
Phase           &  Observed $\log \delta_j$ &  \multicolumn{2}{c}{Random driving} &  \multicolumn{2}{c}{Mixed driving} \\
                &                           & Model A & Model B                   & Model A & Model B \\ 
\hline
warm            &  $-1.3$                   & $-1.6$  & $-1.3$                    & $-2.4$ & $-2.1$ \\
cold            &  $-3.0$                   & $-2.8$  & $-2.7$                    & $-2.8$ & $-2.8$ \\
molecular       &  ---                      & $-2.6$  & $-2.4$                    & $-2.5$ & $-2.2$ \\   
\hline
\end{tabular}
\medskip\\
\end{center}
\end{table*}

The depletion $\delta(\mathrm{X})$ of element X is defined as its gas-phase abundance relative to a reference abundance, for which we take its abundance in the Sun,
\begin{equation}
\delta(\mathrm{X}) = \left(\frac{n_{\rm X}}{n_\mathrm{H}}\right)_{\rm gas} {\Huge /} {\left(\frac{n_{\rm X}}{n_\mathrm{H}} \right)_{\sun} }.
\end{equation} 
Here, $n_\mathrm{H}$ and $n_{\rm X}$ are the number densities of hydrogen and element X, respectively.
The main constituents of interstellar dust are the ``big 5'' elements: Si, Mg, Fe, O and C. However, the evolution of less abundant,
but more depleted elements such as Ti has been considered in the literature, because their very high depletion levels
(only $5\%$ of Ti remains in the gas in the warm and less than $0.1\%$ in the cold diffuse medium) represent a more challenging problem
to explain with dust evolution models \citep{ODonnell:1997p683, Weingartner:1999p6573}.

\cite{Weingartner:1999p6573} assume a steady state of the Ti depletion distribution between the phases 
to derive the mass exchange rates for the observed Ti depletion. We reverse this procedure to compute 
the Ti depletion $\delta_j$ in the molecular, cold and warm phase ($j=\rm{m},\rm{c},\rm{w}$) for our measured mass exchange rates
by solving the equations 
\begin{equation}
f_j \frac{1-\delta_j}{\tau_{\mathrm{d},j}} + \sum_{k \neq j} R_{k,j}(\delta_k - \delta_j) + \frac{q_j}{f_j} \frac{1}{\tau_{\rm in}}
(\delta_{\rm in} - \delta_j) = \frac{\delta_j}{\tau_{\mathrm{a},j}},
\end{equation}
where $f_j$ is the mass fraction of the phase $j$. The first term denotes the release of Ti into the gas phase due to dust destruction
on the timescale $\tau_{\mathrm{d},j}$. The second term is the contribution from other phases,
with $R_{k,j}$ denoting the mass exchange rate from the phase $k$ to the phase $j$ in units of yr$^{-1}$.
The third term denotes the mass input from stars or infall that happens on the timescale $\tau_{\rm in}$, with
the Ti depletion $\delta_{\rm in}$ in this material and $q_j$ the mass fraction that goes into the phase $j$.
It is assumed that $q_\mathrm{m}=0$.
Finally, the term on the right-hand side denotes the decrease of the gas-phase Ti abundance due to accretion onto the
grain surfaces on the timescale $\tau_{\mathrm{a},j}$.

Since this is a steady state model, the net inflow and outflow rates for each phase $j$
should cancel. This condition is not strictly satisfied by our schemes, as depicted in Figure~\ref{fig:cartoon}. However, the deviation is so small that
it would imply changes on $\sim 100\,$Myr timescales, much longer than the time interval over which the transition rates were averaged.

The timescale $\tau_{\mathrm{d},j}$ of destruction of grains in the phase $j$ is related to the 
timescale $\tau_{\mathrm{d}}$ of destruction of grains in the entire ISM by
$f_\mathrm{c} \tau_{\mathrm{d},\mathrm{c}}^{-1} = g_\mathrm{c} \tau_{\mathrm{d}}^{-1}$
and $f_\mathrm{w} \tau_{\mathrm{d},\mathrm{w}}^{-1} = (1-g_\mathrm{c}) \tau_{\mathrm{d}}^{-1}$, where $g_\mathrm{c}$ is the
fraction of destruction that occurs in the cold medium. Additionally, a small amount of dust is destroyed in the molecular phase on the timescale $\tau_{\mathrm{d},\mathrm{m}}$.

We compute the Ti depletion for Model A and B \citep[Table~3 in][]{Weingartner:1999p6573}
that differ by the size distribution of small grains and the destruction timescale $\tau_\mathrm{d}$.
Both models assume extensions of the MRN power law \citep{Mathis:1977p750}
\begin{equation}
\frac{{\rm d}n_{\rm gr}(a)}{{\rm d}a} \sim a^{-K}
\end{equation}
with $K=3.5$. \cite{Weingartner:1999p6573} extend the classical MRN distribution, which runs from $a_{\min}=5$~nm 
to $a_{\rm max}=0.3\, $\textmu m, down to $a_{\min}=0.4$~nm. Model A retains the MRN distribution throughout the extended grain sizes, 
while Model B assumes a steeper power $K=4$ for $a<3$~nm. Due to the higher abundance of small grains, Model B has shorter timescales
of accretion in the warm and cold medium ($\tau_{\mathrm{a},\mathrm{w}}$ and $\tau_{\mathrm{a},\mathrm{c}}$, respectively). In the molecular phase,
both models assume a larger $a_{\min}=15$~nm. To compensate for the shorter accretion timescale in Model B, the timescale
of destruction $\tau_{\mathrm{d}}$ is shorter, $6.3\times 10^8$~yr vs. $1.4\times 10^9$~yr for Model A.
We adopt the same values for the model parameters for the dust model
($\tau_{\mathrm{a},\mathrm{m}}$, $\tau_{\mathrm{a},\mathrm{c}}$, $\tau_{\mathrm{a},\mathrm{w}}$, $\tau_{\mathrm{d},\mathrm{m}}$, $\tau_{\mathrm{d}}$),
the dust input from stars ($\tau_{\rm in}$, $\delta_{\rm in}, q_\mathrm{c}, q_\mathrm{w}$) as in Model A and B.
The mass fractions $f_j$ and mass exchange rates $R_{k,j}$ are taken from the simulations.

As discussed above, the mass exchange scheme inferred from our simulations includes additional transitions in the mass exchange model
compared to that of \cite{Weingartner:1999p6573}: from warm to molecular and from molecular to cold. For simplicity, we neglect
the mass exchange with the hot phase, given its small mass fraction, and consider the mass exchange in the three-phase idealised ISM model.

Table~\ref{table:Depl} compares the Ti depletion values derived for the mass exchange rates from the
simulations with random and mixed driving with those from observations. It is common to use the logarithmic
depletion $\log \delta({\rm{X}}) \equiv \left[ \frac{{\rm{X}}_{\rm gas}}{{\rm{H}}}\right]$ instead of the linear depletion $\delta({\rm{X}})$.
\cite{Weingartner:1999p6573} adopt $\log \delta_\mathrm{w}({\rm Ti})= -1.3$ in the warm medium and
$\log \delta_\mathrm{c}({\rm Ti})= -3$ in the cold medium. This value
for $\log \delta_\mathrm{w}({\rm Ti})$ is consistent with more recent data compiled by \cite{Jenkins:2009p2144}
for $F_*=0.12$ recommended for the WNM in \citet{zhuk16}.
The parameter $F_*$ measures the overall depletion level and varies from $0$ to$1$ for a given data sample.
The adopted value $\log \delta_\mathrm{c}({\rm Ti})= -3$
corresponds to the most depleted lines of sight with mean densities of $10$\,cm$^{-3}$ in the data from \cite{Jenkins:2009p2144}.
\citet{zhuk16} find that the actual local density in a cloud may be up to $100$ times larger than the mean density of the line of sight.
Therefore, the value of $\log \delta=-3$ may correspond to the sight lines with local densities more typical for translucent molecular clouds
than for the CNM. Hence, we consider the value of $-3$ as a lower limit for $\log \delta_\mathrm{c}({\rm Ti})$. There are no
measurements of the gas-phase abundances of Ti in the molecular phase. At lower densities, observations reveal a clear increasing trend of the
depletion with density \citep{Jenkins:2009p2144} that may also continue in molecular clouds.

We find a larger dependence on the mass exchange scheme than on the grain size distribution. The effect of the enhanced accretion in the
cold phase causes low values of $\log \delta_\mathrm{c}$ that are close to the observed value of $-3$ with the exchange rates from both simulations.
However, in the simulation with mixed driving, Ti is too depleted in the warm phase. Because of the efficient mixing
between the molecular and the warm medium in this simulation, the derived value of $\delta_\mathrm{m}$ is actually higher than $\delta_\mathrm{c}$
and is similar to $\delta_\mathrm{w}$, in contrast to expectations from the trend at lower densities. 
The mass exchange scheme from the simulation with random driving yields values of $\log \delta_\mathrm{w}$ which are closer to the observed ones,
$-1.6$ and $-1.3$ for Models A and B, respectively. Thus, the simple model
of element depletion on grains implemented here favours the mass exchange scheme from the simulation with a random positioning of supernovae.

\section{Residence times}
\label{sec:res}

\begin{figure*}
\centerline{\includegraphics[width=0.5\linewidth]{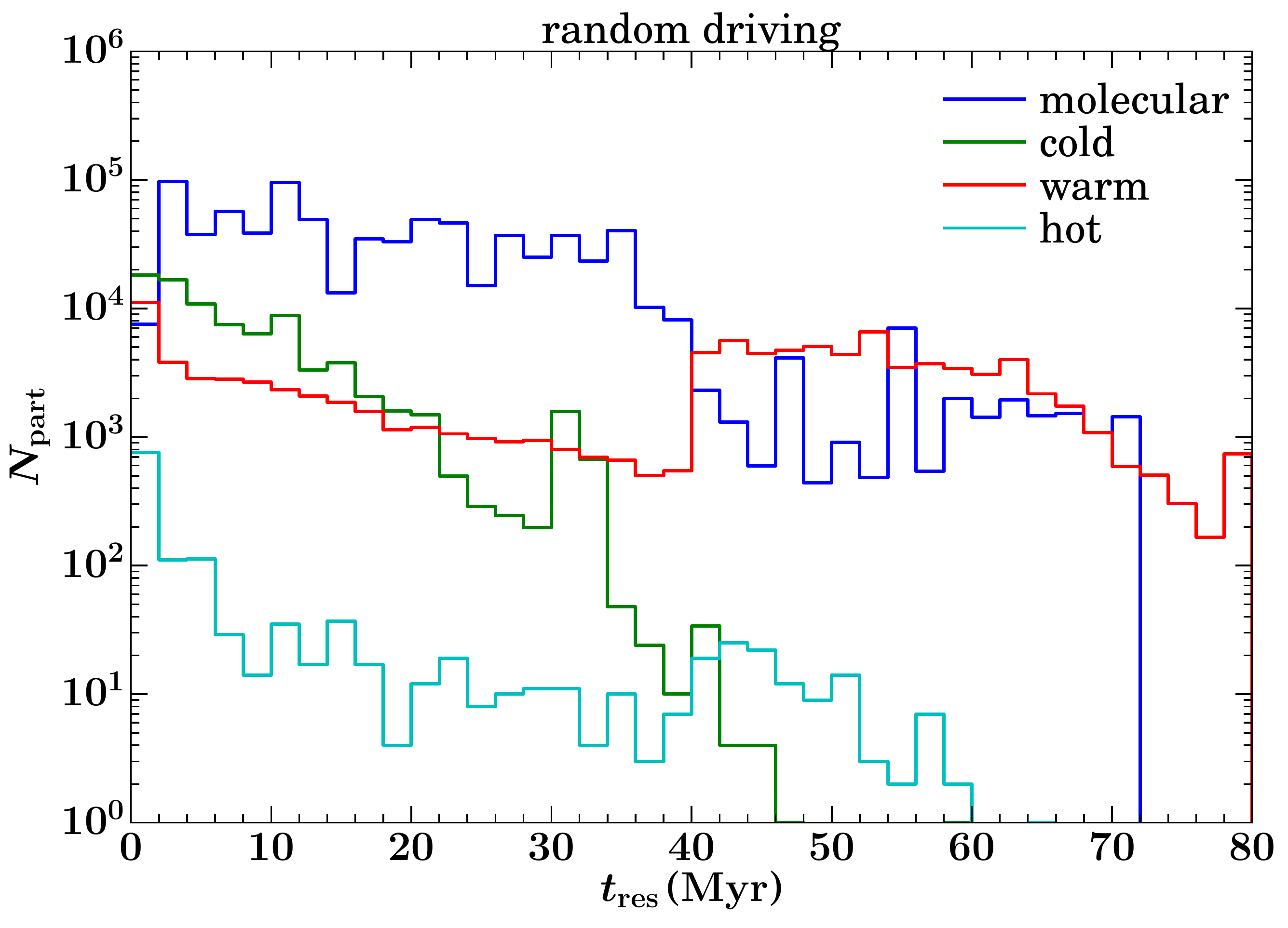}
\includegraphics[width=0.5\linewidth]{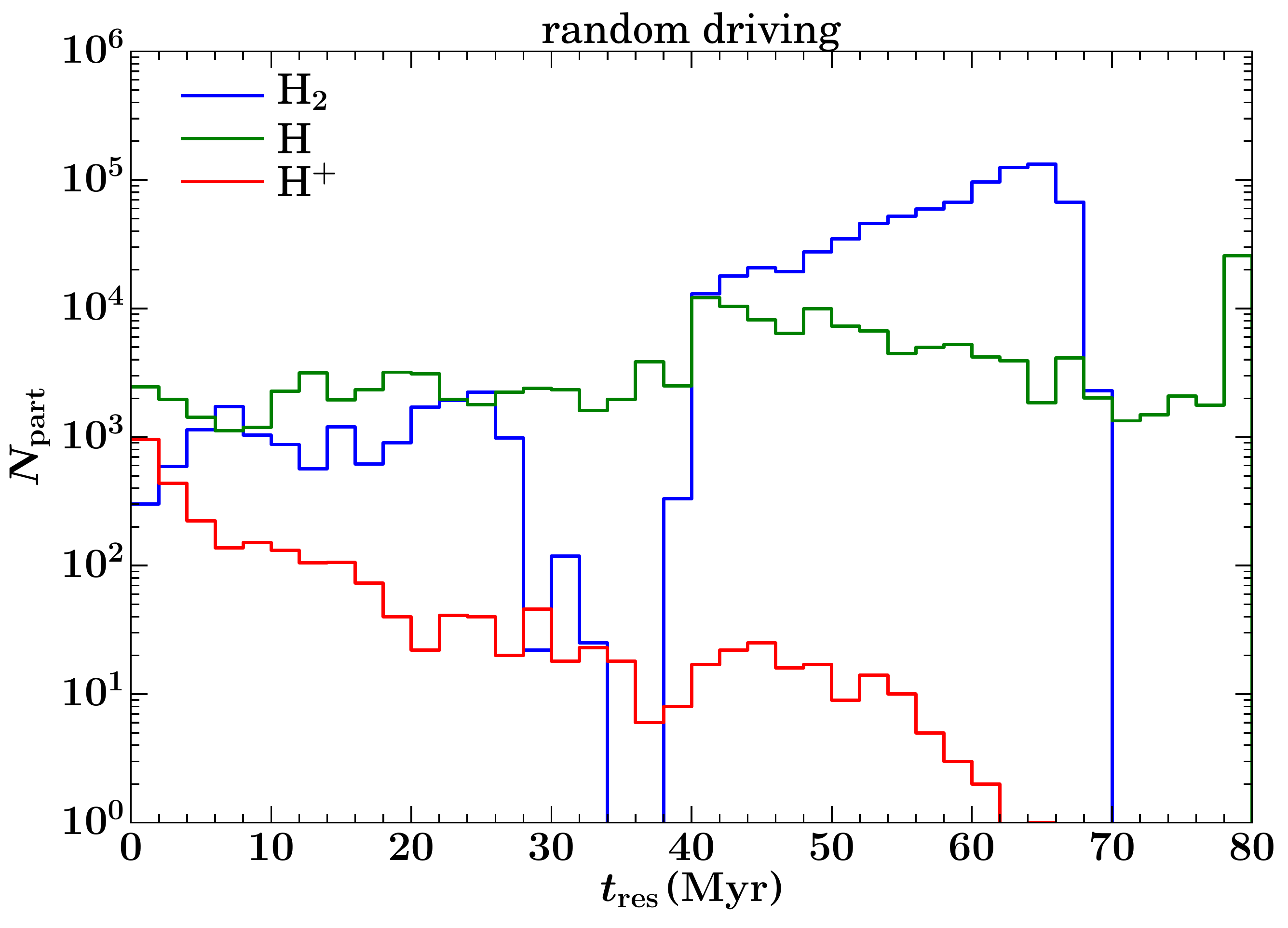}}
\centerline{\includegraphics[width=0.5\linewidth]{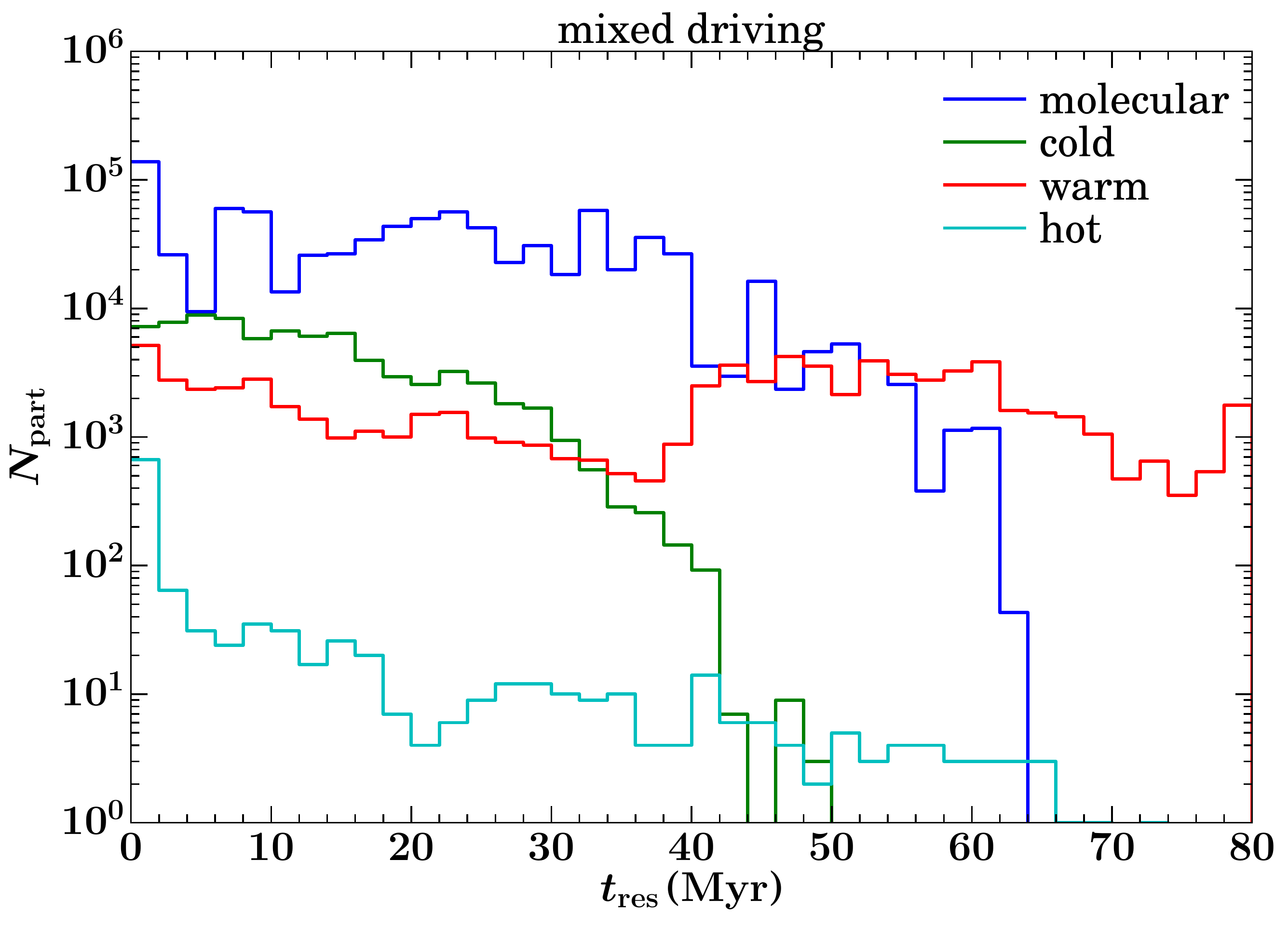}
\includegraphics[width=0.5\linewidth]{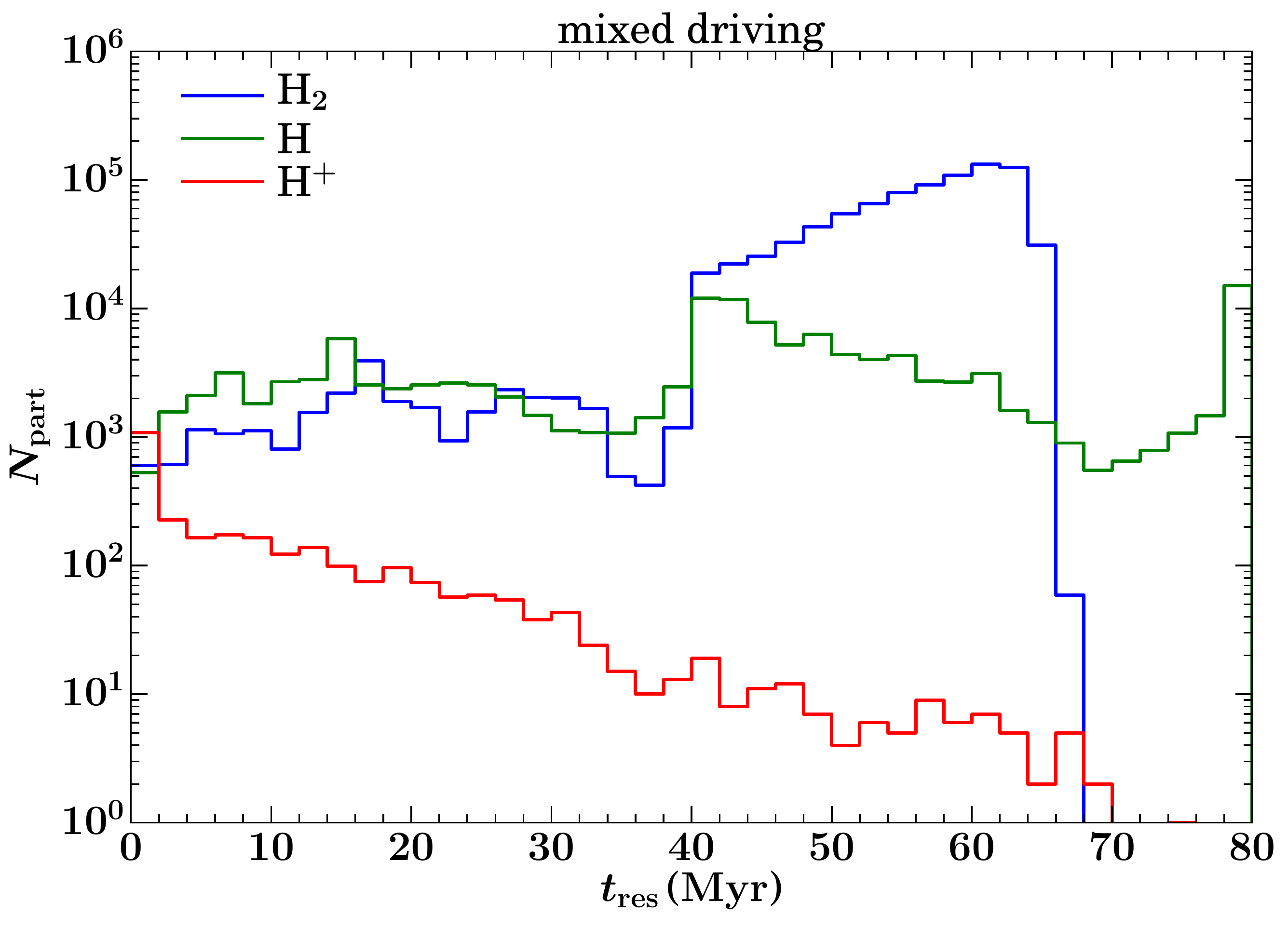}}
\centerline{\includegraphics[width=0.5\linewidth]{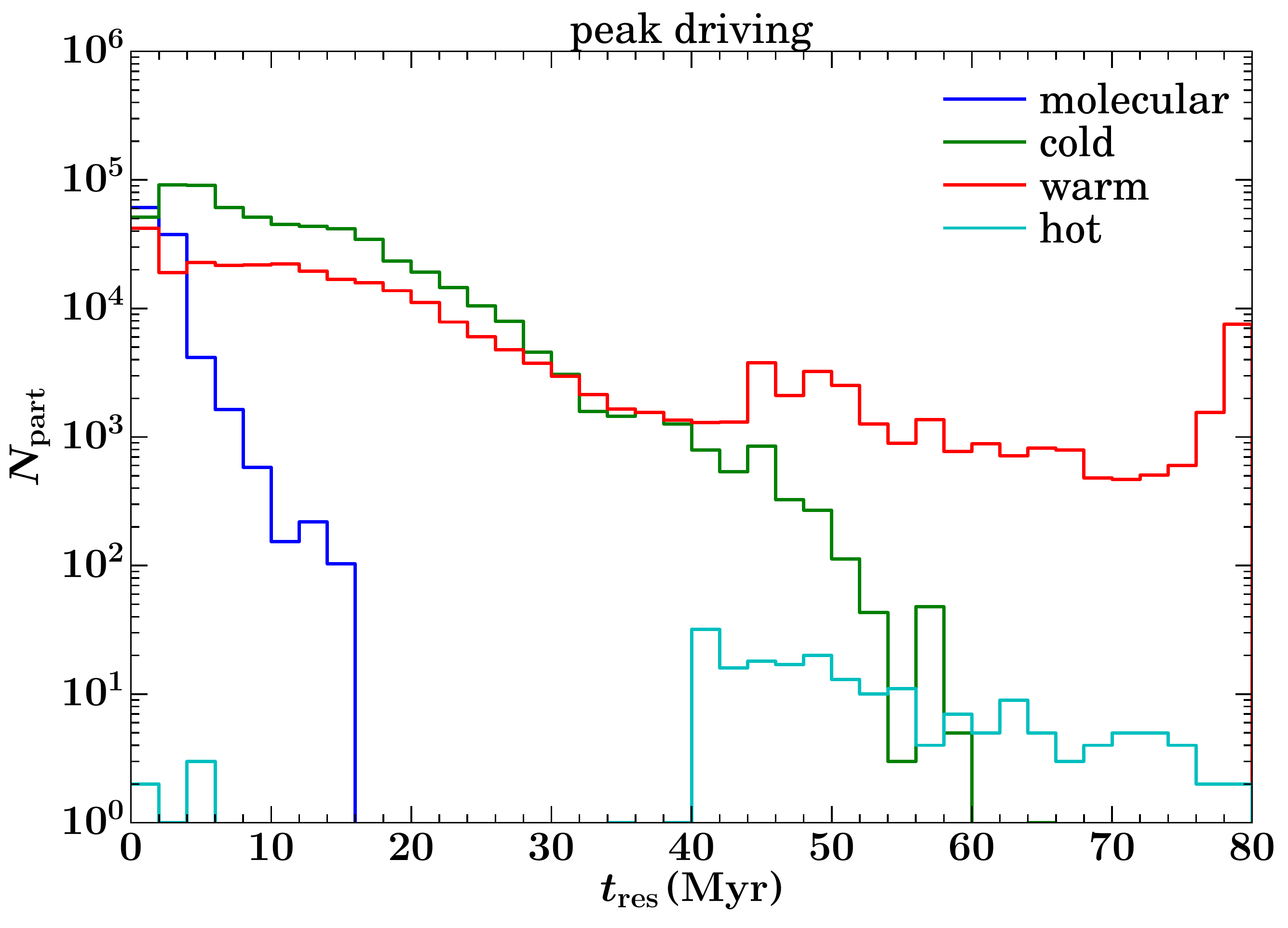}
\includegraphics[width=0.5\linewidth]{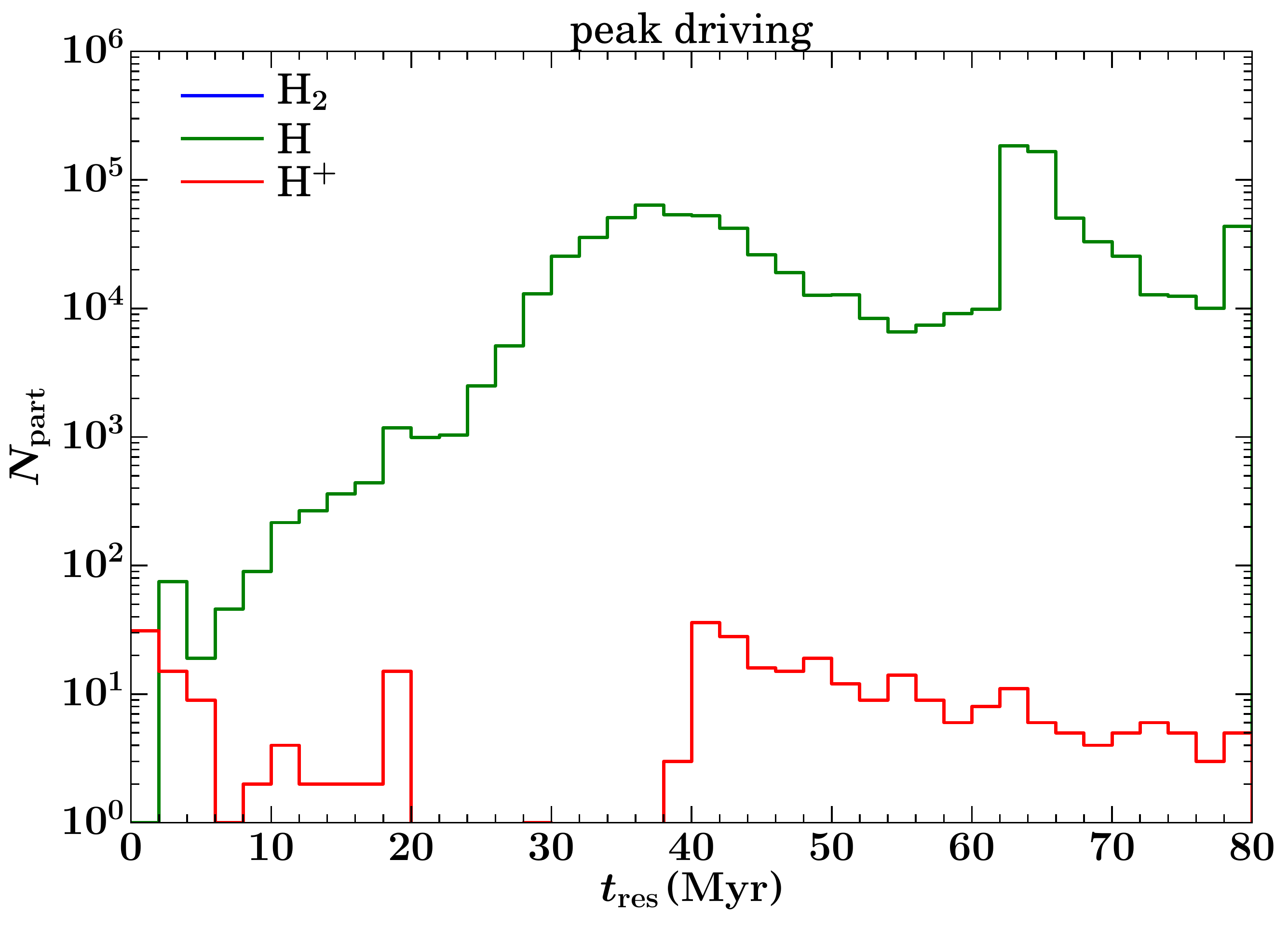}}
\caption{Histograms of the total residence time in the ISM phases defined by temperature cuts (left)
and chemical abundances (right) for the particles at $t = 40\,$Myr
for the simulation with random (left), mixed (middle) and peak (right) driving.}
\label{fig:rthistlin}
\end{figure*}

\begin{figure*}
\centerline{\includegraphics[width=0.5\linewidth]{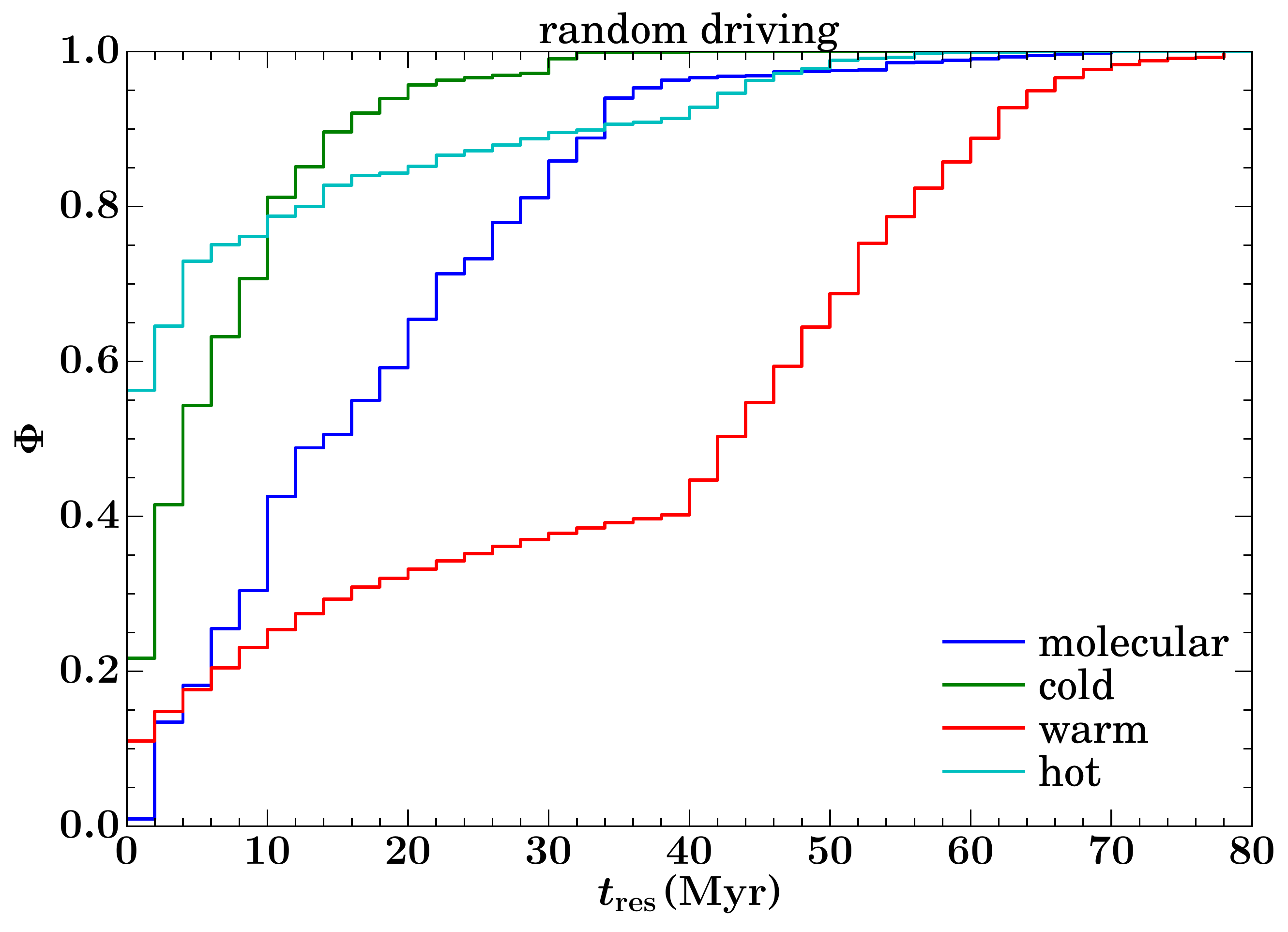}
\includegraphics[width=0.5\linewidth]{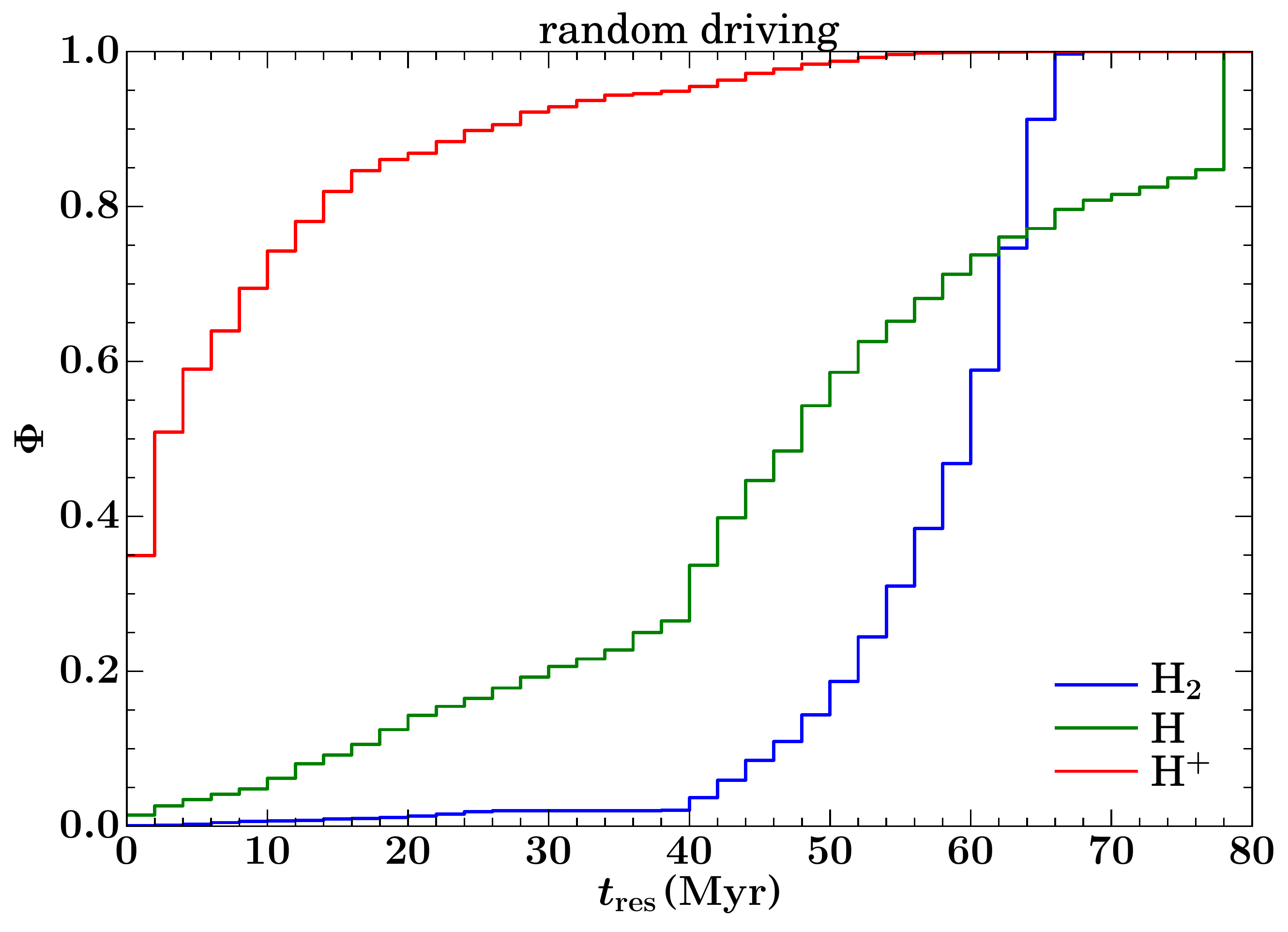}}
\centerline{\includegraphics[width=0.5\linewidth]{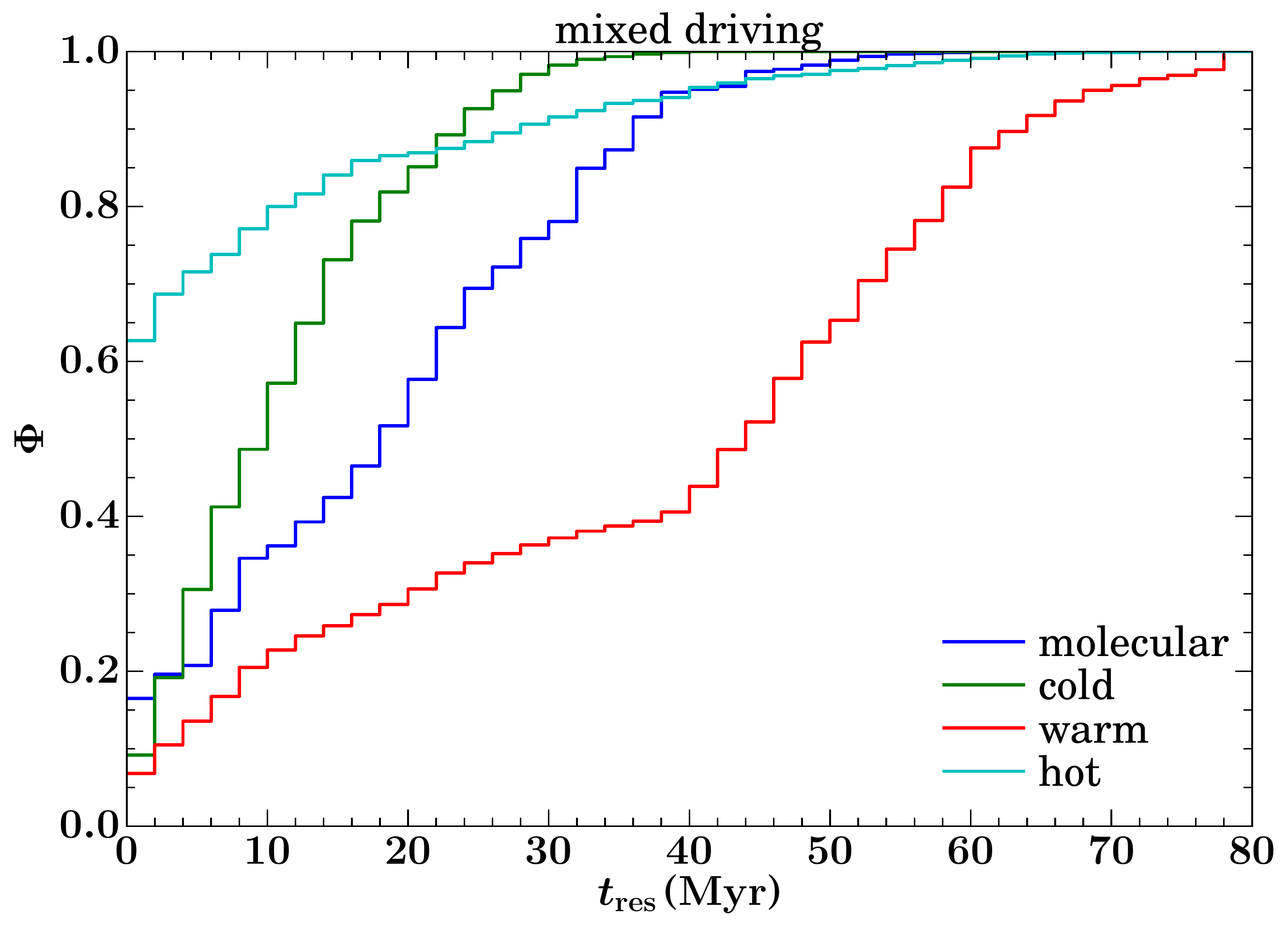}
\includegraphics[width=0.5\linewidth]{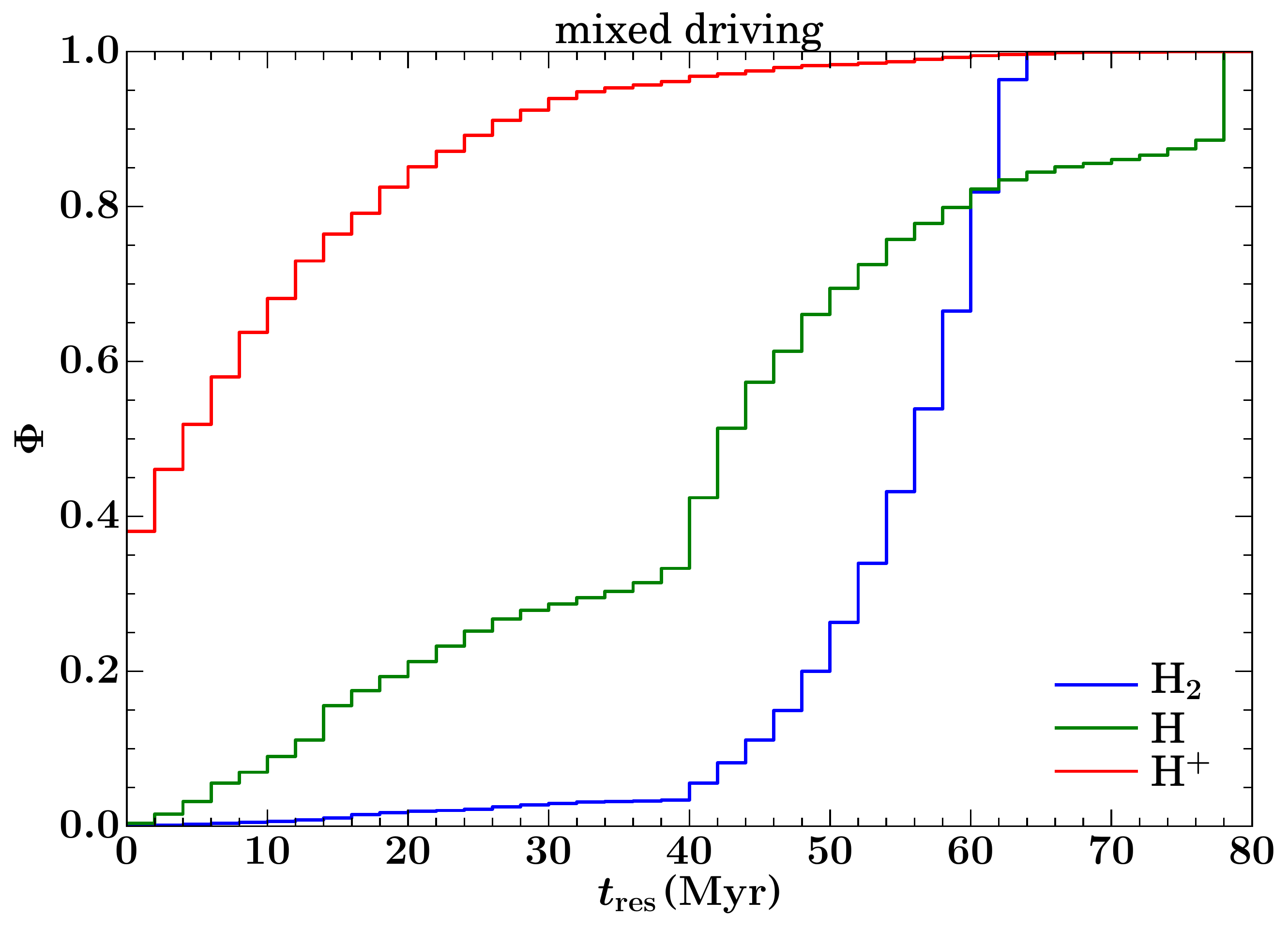}}
\centerline{\includegraphics[width=0.5\linewidth]{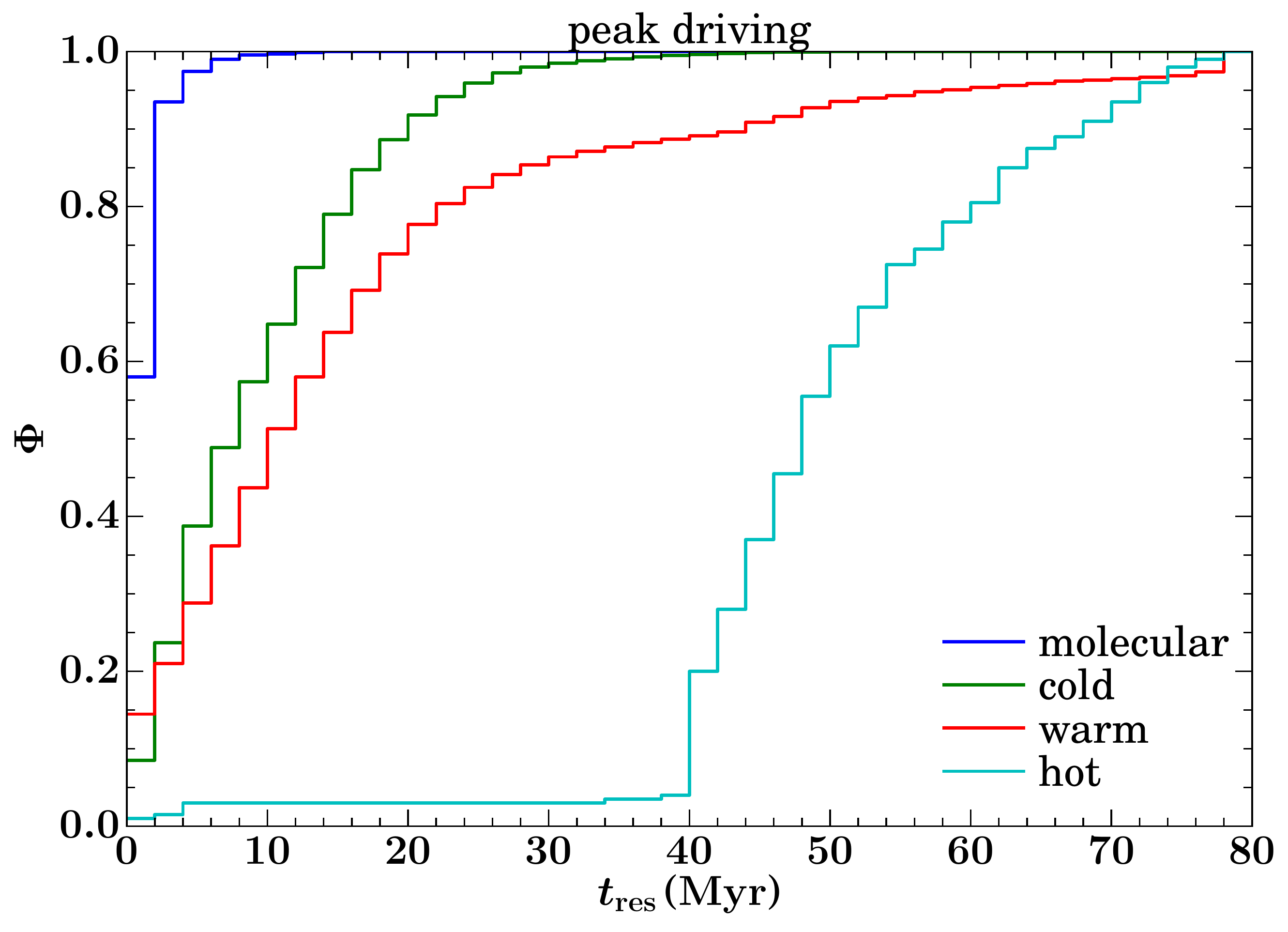}
\includegraphics[width=0.5\linewidth]{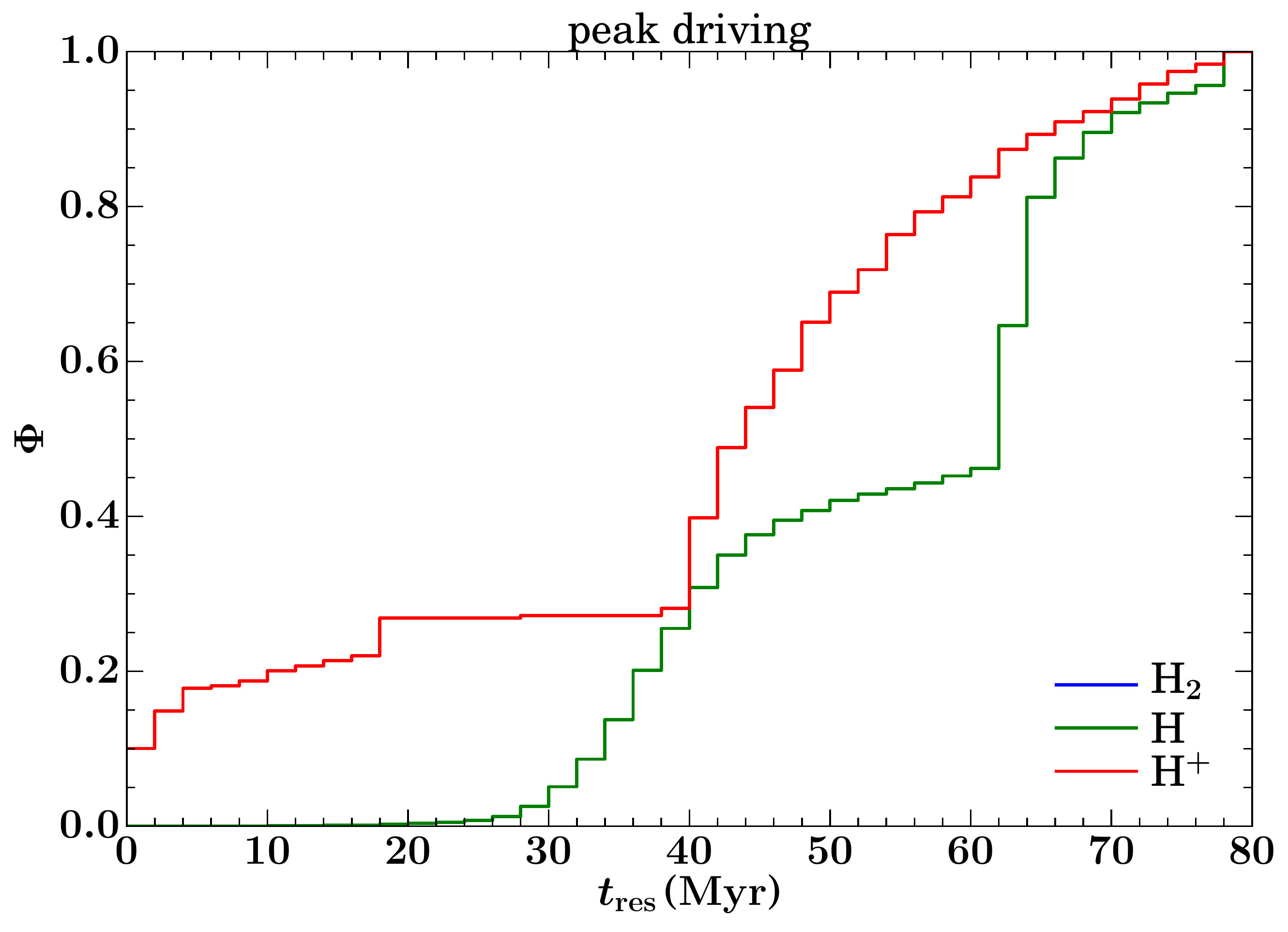}}
\caption{Cumulative probability distribution functions $\Phi$ of the total residence time in the ISM phases defined by temperature cuts (left)
and chemical abundances (right) for the particles at $t = 40\,$Myr
for the simulation with random (left), mixed (middle) and peak (right) driving. The data is normalised to the total particle count per phase.}
\label{fig:rthistlincumnorm}
\end{figure*}

An important quantity for understanding the processing of dust in the ISM is its residence time in the different ISM phases. 
In our simulations, we can directly measure these residence times from the tracer particle histories.
In general, each tracer particle resides in more than one phase during the simulation runtime.
In the beginning of the simulations, the ISM structure and hence the residence times are affected by
our initial conditions. Near the end of the simulations, the residence times are artificially
short because of the finite simulation runtime. We therefore consider the residence times for particles in a snapshot
in the middle of the simulation, at $t = 40\,$Myr. We have checked that we do not obtain
significanly different results at $t = 30\,$Myr or at $t = 50\,$Myr.

At $t = 40\,$Myr, we determine the current ISM phase for all tracer particles. For each particle, we then go back in time
and determine at which point the particle entered this phase, and go forward to determine when it will leave the phase again.
The total duration then gives the residence time for the particle in its current phase. Both the backward
and the forward residence times are limited to $40\,$Myr, so that we cannot measure total residence times longer than
$80\,$Myr. Some of the distributions extend all the way up to this maximum, and in these cases we would likely measure even
larger residence times if we ran the simulations for longer.
Likewise, we are unable to measure residence times shorter than our sampling period of 10\,kyr.

Histograms of the residence time distributions are shown in Figure~\ref{fig:rthistlin} and
cumulative probability distribution functions in Figure~\ref{fig:rthistlincumnorm}.
We list the mean and median
residence times in Table~\ref{table:mrt}. For random driving, the mean residence time in the molecular phase is $18$~Myr.
Half of the particles have a residence time less than $15$~Myr, and 90\% of all particles have a residence time less than $34\,$Myr.
Above $40\,$Myr, the frequency drops markedly. This is likely a result of the
finite simulation runtime, since the molecular phase needs some time to build up. Nevertheless, the characteristic residence time
in the molecular phase is of the order of a few Myr and thus unaffected by this. The cold phase has a very short
typical residence time of only a few Myr, with a mean (median) value of $8$ ($5$) Myr. The longest residence times are found
for the warm phase, with a mean (median) value of $36$ ($44$) Myr. These are the particles between the clouds and in the outflow,
where the dynamical times are long. Once the particles are in the warm phase, they rarely transition to the hot phase, but the dominant
escape route is to the cold or molecular phase (compare Figure~\ref{fig:trall}). These routes are only available for particles
within the disc, not for the majority of particles that are within the outflow.
The jump at $40\,$Myr in the warm phase distribution comes from the fact that this is the initial phase of all particles,
so we see the particles that have not left this phase yet at the time of our measurement. In contrast, particles with shorter residence times
than $40\,$Myr must have entered the warm phase after the start of the simulation.
The particles in the hot phase have a mean (median)
residence time of $9$ ($2$) Myr. The very short median is the result of supernova explosions, which transfer particles into the hot phase,
but when the remnant cools down, they leave the hot phase again. The long tail of residence times comes from particles in the higher disc layers.

The residence time distributions for mixed driving are similar to the case of random driving.
For peak driving, in contrast, the distributions are very different. At the time of the residence time measurement,
$t = 40\,$Myr, only $10\%$ of all particles are in the molecular phase (compare Figure~\ref{fig:pn}).
This molecular gas is efficiently destroyed by supernova explosions at density peaks, resulting in
very short residence times of only a few Myr. The hot phase is almost completely dominated by particles within the higher disc layers. As discussed
previously, for peak driving supernova injections transfer particles to the warm phase, which explains the enhanced frequency of
short residence times in the warm phase compared to the cases of random and mixed driving.
This leads to a much reduced mean (median) value of $17$ ($12$) Myr. The cold phase
contains the majority of particles. They condense out of the warm phase, but because of the peak driving cannot
settle into the molecular phase. Interestingly, the mean and median values of the residence time distribution are
nevertheless very similar to the case of random and mixed driving, although the ISM is very different. We see that the multi-phase ISM is not
appropriately characterised by typical residence times in the temperature regimes.

For the phases defined by the chemical abundances, the residence times in the H$_2$~phase are long.
For random and mixed driving, where this phase exists, they are between $55$ and $60$\,Myr typically.
The distribution jumps at $40$\,Myr since many particles stay in the H$_2$ phase until the end of the simulation at $80$\,Myr,
so we see the particles that have entered the H$_2$ phase at or earlier than the time of our measurement.
Here, the maximum residence time that can be measured is affected by the simulation runtime, as can be
seen in the sharp cutoff at a residence time of $70\,$Myr. Therefore, the true residence times would likely be even
higher. The residence time distribution of the H~phase is relatively flat for random and mixed driving,
but has peaks at $40$ and $80\,$Myr. The former peak is produced by particles that have remained in the H~phase
since the beginning of the simulation, while the latter one is created by particles that enter the wind.
For peak driving, the local maximum in the H~phase distribution shifts from $40$ to $60\,$Myr.
This characteristic time
is created by the beginning conversion of atomic into molecular hydrogen at $t = 60\,$Myr
(compare Figure~\ref{fig:pnabun} and Figure~\ref{fig:trall}).
These are the particles that have resided in the H~phase since the simulation start and then transition into the H$_2$~phase.
The residence times for the H$^+$~phase
are similar to the hot phase in all three simulations.

The characteristic residence times are broadly consistent with the measured transition rates.
For random driving, the dominant transition rates out of the molecular and cold phase are of
the order $10^{-7}$\,yr$^{-1}$ at $t = 40\,$Myr, leading to a $\sim 10$\,Myr residence time.
Likewise, the dominant rate out of the H$_2$ phase is $5 \times 10^{-9}\,$Myr, resulting in a
$\sim 200$\,Myr residence time. These order of magnitude estimates explain why the chemical phases
are more affected by the finite simulation runtime than the phases defined by temperature cuts.

\begin{table*}
\caption{Mean and median residence times in the ISM phases}
\label{table:mrt}
\begin{center}
\begin{tabular}{lccccccccccc}
\hline
supernova       & $t_\mathrm{res}^\mathrm{mol}$ & $t_\mathrm{res}^\mathrm{col}$ & $t_\mathrm{res}^\mathrm{war}$ & $t_\mathrm{res}^\mathrm{hot}$
& $t_\mathrm{res}^{\mathrm{H}_2}$ & $t_\mathrm{res}^{\mathrm{H}}$ & $t_\mathrm{res}^{\mathrm{H}^+}$
& $t_\mathrm{res}^{A_\mathrm{V} \leq 1}$ & $t_\mathrm{res}^{A_\mathrm{V} > 1}$ & $t_\mathrm{res}^{A_\mathrm{V} \leq 3}$ & $t_\mathrm{res}^{A_\mathrm{V} > 3}$
\\
driving         & (Myr)                       & (Myr)                        & (Myr)                        & (Myr)
& (Myr)                        & (Myr)                        & (Myr)
& (Myr)                        & (Myr)                        & (Myr)                        & (Myr)
\\
\hline
random          &    $17.9$                &       $7.51$           & $36.3$               & $8.50$
& $58.0$        &    $48.5$                &       $9.26$
& $61.9$        &    $60.0$                &       $64.6$           &  $52.4$\\
                &    $15.4$                &       $5.24$           &  $43.9$               & $1.55$
& $60.6$        &    $48.4$                &       $3.77$
& $67.6$        &    $61.6$                &       $79.9$           &  $54.4$
\\
mixed           &    $19.1$                &       $11.8$           &  $38.2$               & $7.60$
& $55.1$        &    $43.2$                &       $10.1$
& $55.3$        &    $56.5$                &       $56.7$           &  $47.4$\\
                &    $19.3$                &       $10.3$           &  $44.6$               & $0.765$
& $57.3$        &    $43.8$                &       $5.40$
& $56.2$        &    $58.2$                &       $55.4$           &  $50.4$
\\
peak            &    $2.06$                 &       $10.2$             & $17.0$               & $50.6$
& ---           &    $54.7$                 &       $40.2$
& $50.2$        &    $11.6$                 &       $68.2$             &  ---\\
                &    $1.70$                 &       $8.29$             & $11.7$               & $48.6$
& ---           &    $63.0$                 &       $44.6$
& $54.6$        &    $7.98$                 &       $65.4$             &  ---
\\
\hline
\end{tabular}
\medskip\\
Mean and median residence times of the tracer particle ensemble in the ISM phases for the three simulations at $t = 40\,$Myr.
The first (upper) number is the arithmetic mean, the second (lower) one the median value of the distribution.
\end{center}
\end{table*}

\subsection{Comparison with estimates of molecular cloud lifetimes}

It seems natural to compare the residence times in the molecular and H$_2$ phases with theoretical estimates of molecular cloud
lifetimes. Using smoothed particle hydrodynamics (SPH) simulations of isolated disc galaxies with supernova feedback, \citet{Dobbs:2012in} estimated that
giant molecular clouds with masses above $10^5\,$M$_\odot$ live for at least $50\,$Myr before they return their material
back to the diffuse ISM. In this work, a simple density criterion was used to decide whether an SPH particle was part of
a molecular cloud or in the diffuse phase. In contrast, \citet{dobbs13}, using similar simulations, employed a clump-finding algorithm
to identify SPH particles belonging to giant molecular clouds and determined the cloud lifetime through the dispersal of the selected particles.
This criterion led to much shorter timescales of only $4$ to $25$\,Myr for giant molecular clouds.

With a semi-analytic model of cloud destruction by photoionization feedback, \citet{krmamc06} obtained cloud lifetimes of
$20$ to $30\,$Myr. Our characteristic residence times in the molecular phase are shorter, even without early stellar feedback by winds and radiation.
In fact, our peak driving simulation may give better values for the molecular phase residence times in this respect than
the runs with random and mixed driving, although it otherwise does not produce a realistic ISM. The small values for the molecular
phase residence time are remarkable since they represent
upper limits to the cloud lifetimes, given that they only measure the time spent in molecular material, during which the particles
could circulate between several clouds. This is even more likely the case for the H$_2$ phase, which has much longer residence times
than the molecular phase.

\subsection{Implications for dust evolution models}

Our study of the residence times has implications for simple models of dust evolution that treat the multi-phase ISM
and the mass exchange between phases in a parametrised way \citep{zhuketal08, Zhukovska:2009p7232, Zhukovska:2014ey}. 
In a model proposed by \cite{zhuketal08}, the residence time of grains in molecular clouds has a fixed value and is equivalent to 
the lifetime of clouds $\tau_{\rm cl}$. It is an important parameter that has a two-fold effect on the rate of dust growth by accretion in the molecular clouds. 
If the lifetime of molecular clouds is $\gtrsim 10$ times longer than the timescale of accretion of gas-phase species on dust, which has 
a value of 1~Myr for the solar metallicity,
then refractory elements are almost completely condensed on dust grains during their residence in molecular clouds. The matter 
is thus returned to the diffuse phase with the maximum dust abundances upon disruption of the clouds. 
However, a further increase of the cloud lifetime delays the return of the dust-rich matter to the diffuse phase and 
decreases the total rate of dust production in the ISM. Because the latter is the dominant mechanism of dust production in the 
present-day Milky Way, a longer residence time results in a lower dust-to-gas ratio. 

\begin{figure}
\centerline{\includegraphics[width=\linewidth]{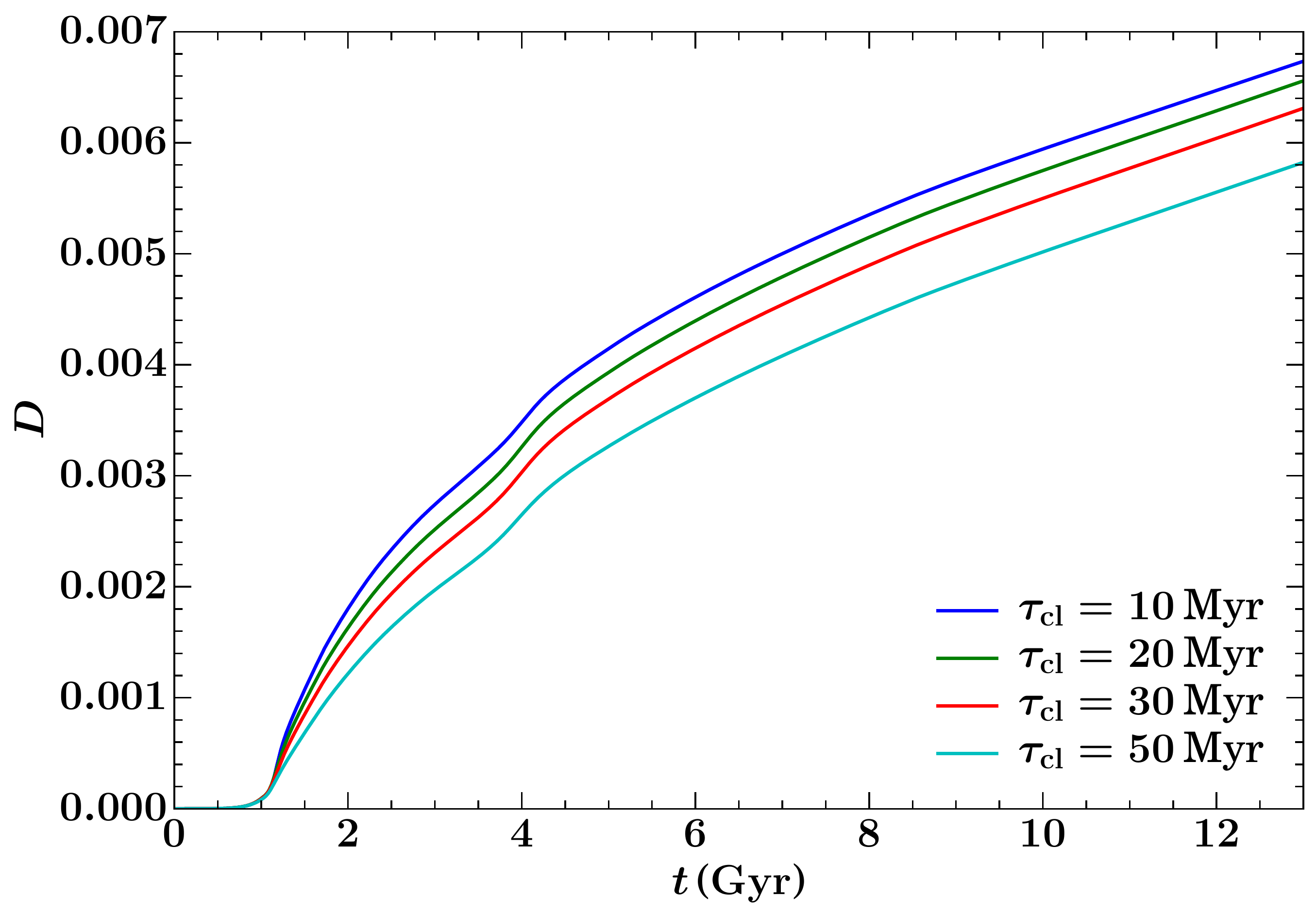}}
\caption{Evolution of the dust-to-gas ratio $D$ in the local Milky Way for the different values of the lifetime of molecular clouds
$\tau_{\rm cl} = 10$, $20$, $30$ and $50$\,Myr (from top to bottom) predicted by the model of dust evolution from \citet{zhuketal08}.}
\label{fig:DGR}
\end{figure}

\cite{zhuketal08} adopted a value of $10$\,Myr for the molecular cloud lifetime. It is $1.5$ to $2$ times shorter than 
the mean residence time in the molecular phase derived from our simulations with random and mixed driving. 
The distribution of the residence times is flat from a few up to $40$\,Myr, a large portion of 
grains thus reside up to $4$ times longer than $10$\,Myr. A longer value of the cloud lifetime of $30$\,Myr has also been adopted by 
\cite{Tielens:1998p7054} in his model of the dust cycle between the diffuse medium and clouds. We investigate the impact of a longer
residence of grains in the molecular phase on the dust-to-gas ratio using the model of dust evolution 
at the local solar galactocentric radius from \cite{zhuketal08}. 
Figure~\ref{fig:DGR} shows the time evolution of the average dust-to-gas ratio for different values of the $\tau_{\rm cl}=10$, $20$, $30$, and $50\,$Myr.
For the $\tau_{\rm cl}=20$ and $30$\,Myr, the present dust-to-gas ratio (for $t=13$\,Gyr) is, respectively, only 3\% and 7\% lower than
for the reference value. With a longer $\tau_{\rm cl}=50$\,Myr, the final dust-to-gas ratio is decreased by 15\%. Therefore, 
the broad distributions of the residence times in the molecular phase from our simulations imply a decrease in the average 
dust abundances by less than 10\%. 
 
The derived distributions of the residence times in the ISM phases have larger consequences for the grain size and
gas-phase element abundance distributions.
In molecular clouds, grains grow their sizes by coagulation and accretion. \cite{Hirashita:2014bu} demonstrated that, 
because of these processes, the residence time in the dense gas is imprinted in such observable dust characteristics as
the extinction curve and element depletion. 
Broad distributions of the residence times in molecular clouds as found in this work imply that matter that is transferred to the diffuse
phase upon disruption of clouds should exhibit large variations in the grain size and element depletion distributions because of 
the different dynamical histories of the gas. Indeed, large dispersion of Si gas-phase element abundances for a given gas volume density 
was recently found by \citet{zhuk16} with models of dust evolution in an inhomogeneous ISM. 
We conclude that the observed scatter in the interstellar element depletion may arise 
not only because of local conditions, but may be caused by different dynamical histories of the gas. 
Similar arguments are valid for the observed variations of interstellar extinction curves \citep{Fitzpatrick:2007p6352}. 

The residence time in the warm phase determines the degree of grain processing by turbulence, which shatters
large grains into smaller fragments. Our simulations with random and mixed driving yield a mean residence time in 
the warm medium of about $40$\,Myr, with a large dispersion. Longer residence times in the warm medium ($\sim 100$\,Myr)
are required by models of the evolution of interstellar grain sizes to create a population of small grains 
detected by observations \citep{Hirashita:2009p663, Hirashita:2010p6587}. 
Such a longer timescale is necessary, because \cite{Hirashita:2010p6587} adopt an initial size distribution dominated 
by large $0.1\,$\textmu m grains, assuming that small grains are removed by coagulation in molecular clouds. 
However, as we discussed above, not all grains spend equally long times in molecular clouds (see Figure~\ref{fig:rthistlin}), meaning that not all gas is equally 
processed by coagulation. Moreover, coagulation works efficiently at gas densities of $10^4\,\rm cm^{-3}$ and above. 
A large portion of molecular gas in our simulations resides at lower densities and should retain small grains. 

\subsection{Residence time for presolar dust grains}

In addition to the processing of interstellar dust discussed above, grains are irradiated by Galactic cosmic rays.
As a result, grains that remain in the ISM for a certain time have an excess of cosmic-ray-produced nuclides, which increases
with their residence time. Some interstellar grains that survived the formation of the Solar System can be identified in primitive meteorites.
Laboratory measurements of cosmic-ray-produced nuclides, in particular noble gas isotopes, in these presolar grains are
used to estimate the time period the grains were exposed to cosmic rays \citep{Tang:1988ho, Ott:2005jd, Heck:2009eg}.
Information on the distribution of grain residence times directly measured in the laboratory provides an invaluable
test for theoretical models of dust evolution.

A common assumption in this method is that the cosmic ray flux is on average constant in space. The values of the
residence times in the ISM (i.e., the exposure times) derived by \cite{Heck:2009eg} range from $3$ to $1100$\,Myr
and from $40$ to $1000$\,Myr in work by \cite{Gyngard:2009ch}. Since the majority of the studied presolar grains
have ages longer than $100$\,Myr, it means that they cycled a few times between the warm medium and the clouds before
they got trapped during the solar system formation. It has been shown recently that the cosmic ray ionisation rate decreases
in the dense molecular clouds compared to the diffuse lines of sight
because low-energy cosmic rays, which are most efficient at ionisation, lose their energy before they reach the interiors
of dense clouds \citep[see][and references therein]{Indriolo:2012ip}.
Using
hydrodynamical simulations of the ISM evolution, we find that on average the grains spend two times longer in the warm
gas compared to the molecular cloud (about $20$ and $40$\,Myr, respectively), and, additionally, $5$--$15$\,Myr in the cold diffuse medium.
Thus, about one third of their residence in the ISM, grains spend at the environment with a lower ionisation rate.
Taking this into account may lead to longer residence times in the ISM in presolar grain studies.
ISM simulations including cosmic rays \citep{petersetal15,giricr,simps16} will help to understand this effect.

\section{Shielding}
\label{sec:shield}

Dust grains undergo different processing when they are exposed to the interstellar radiation field
or shielded from it.
For every grid cell, we determine $A_\mathrm{V}$ by the arithmetic mean of
$4\pi$ steradian maps of the total hydrogen column density
using the \texttt{TreeCol} algorithm \citep{clarketal12}.
We restrict the column density calculation to a radius of $50\,$pc and take the periodic boundaries in the disc plane
into account \citep{walchetal15}.
Since we thus know the strength of the radiation field at the location of the tracer particles,
we can compute the residence times in regions of high and low $A_\mathrm{V}$.
We consider two different threshold
values $A_\mathrm{V, thresh}$, namely $A_\mathrm{V, thresh} = 1$ and $A_\mathrm{V, thresh} = 3$.
The value $A_\mathrm{V} = 3$ to $4$ is the threshold extinction for the detection of ice mantles
around dust grains derived in infrared spectroscopic observations \citep{Murakawa:2000hg, Whittet:2001ch}.

\begin{figure}
\centerline{\includegraphics[width=\linewidth]{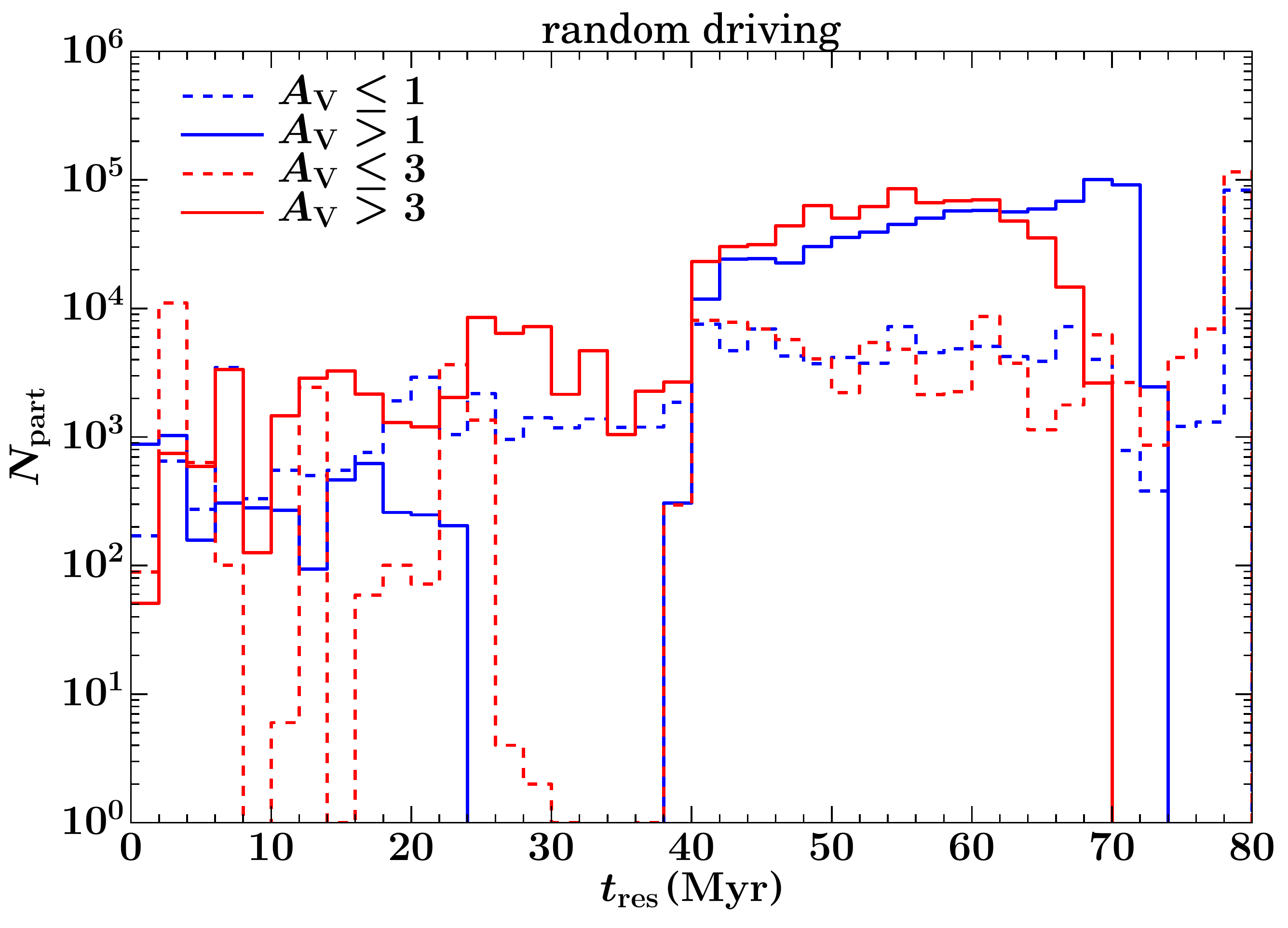}}
\centerline{\includegraphics[width=\linewidth]{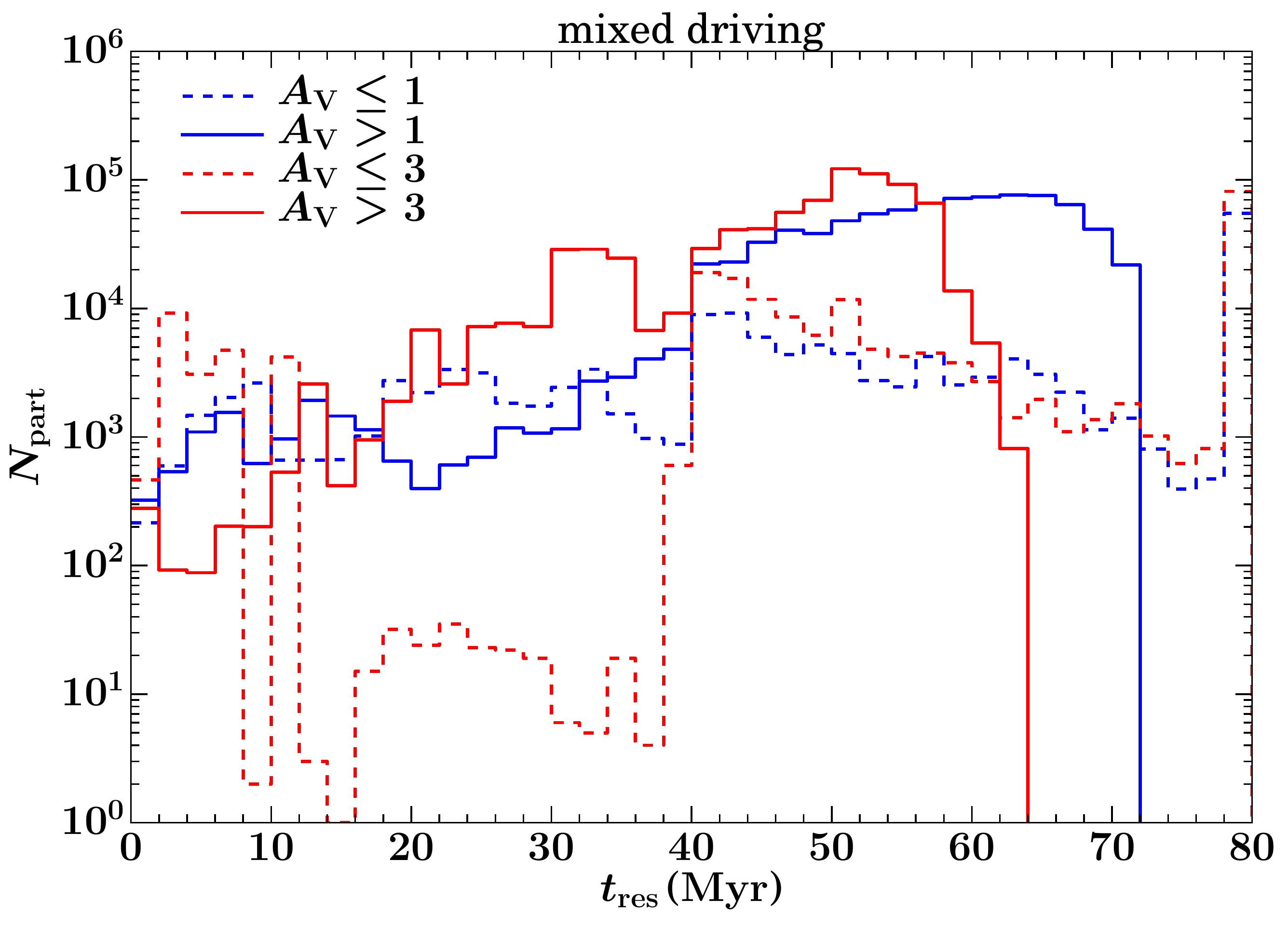}}
\centerline{\includegraphics[width=\linewidth]{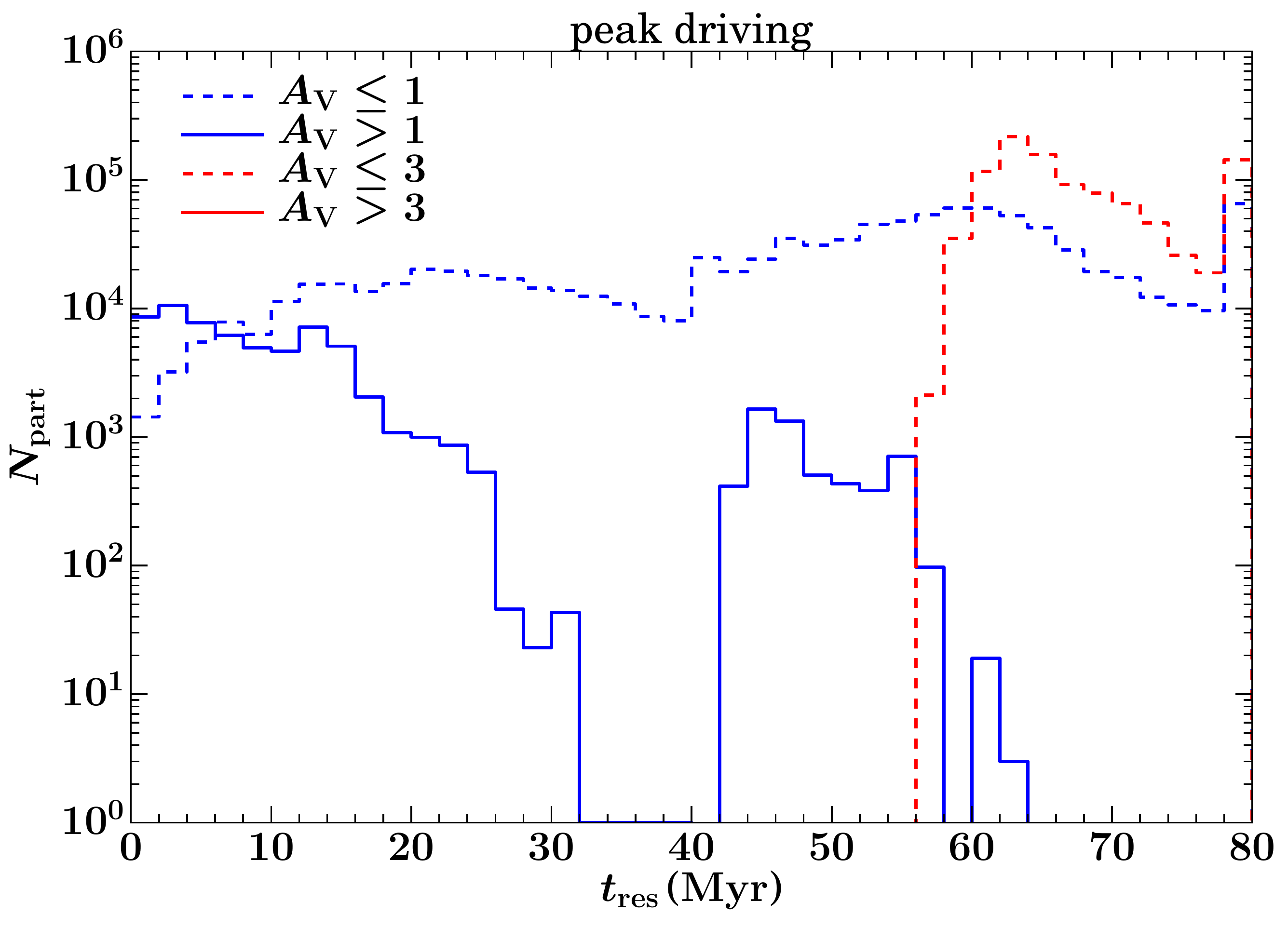}}
\caption{Histograms of total residence times at $t = 40\,$Myr 
in regions with $A_\mathrm{V} \leq A_\mathrm{V, thresh}$ and $A_\mathrm{V} > A_\mathrm{V, thresh}$ for $A_\mathrm{V, thresh} = 1$ and $A_\mathrm{V, thresh} = 3$
for the simulation with random (top), mixed (middle) and peak (bottom) driving.}
\label{fig:rthistAV}
\end{figure}

\begin{figure}
\centerline{\includegraphics[width=\linewidth]{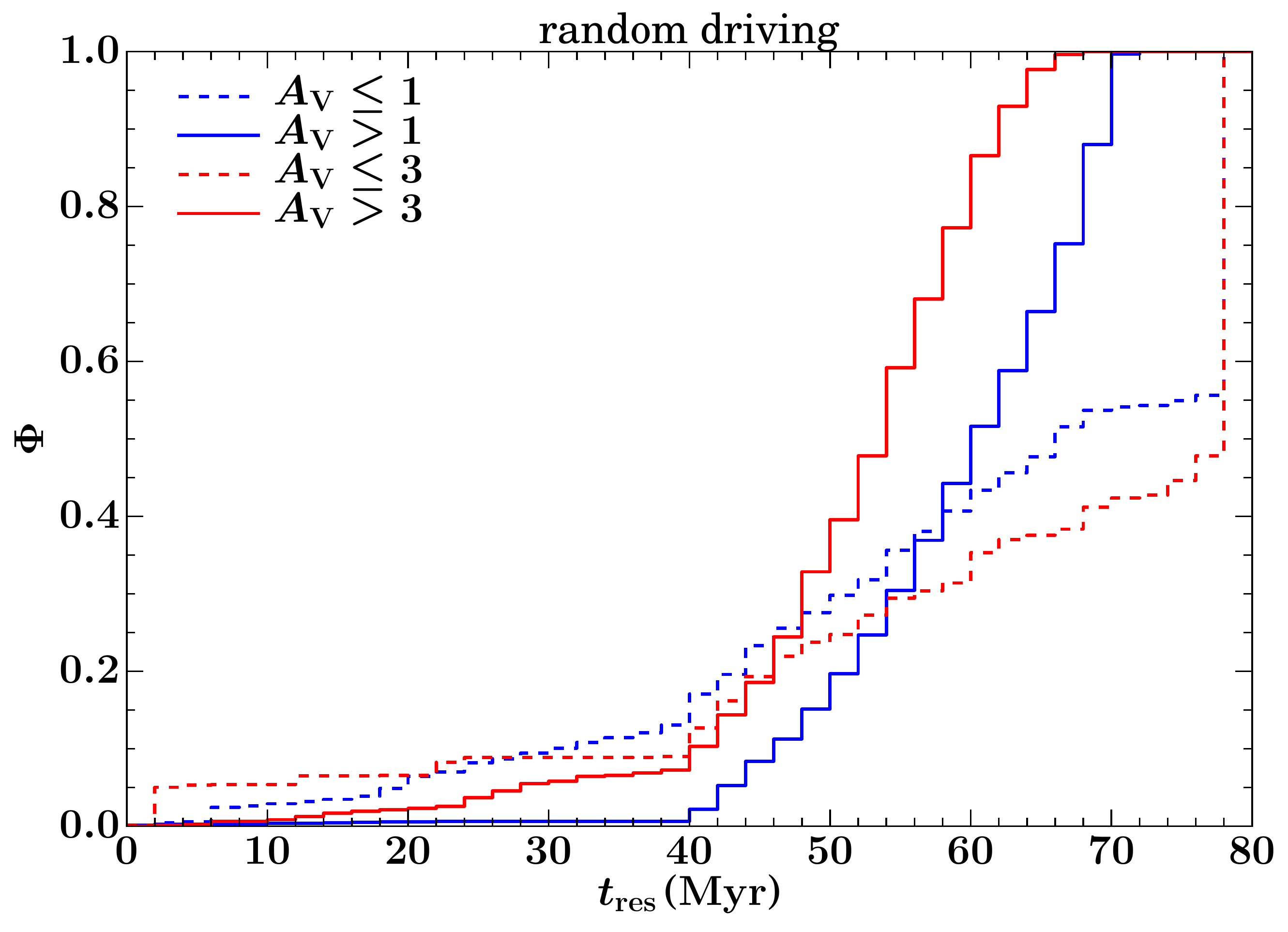}}
\centerline{\includegraphics[width=\linewidth]{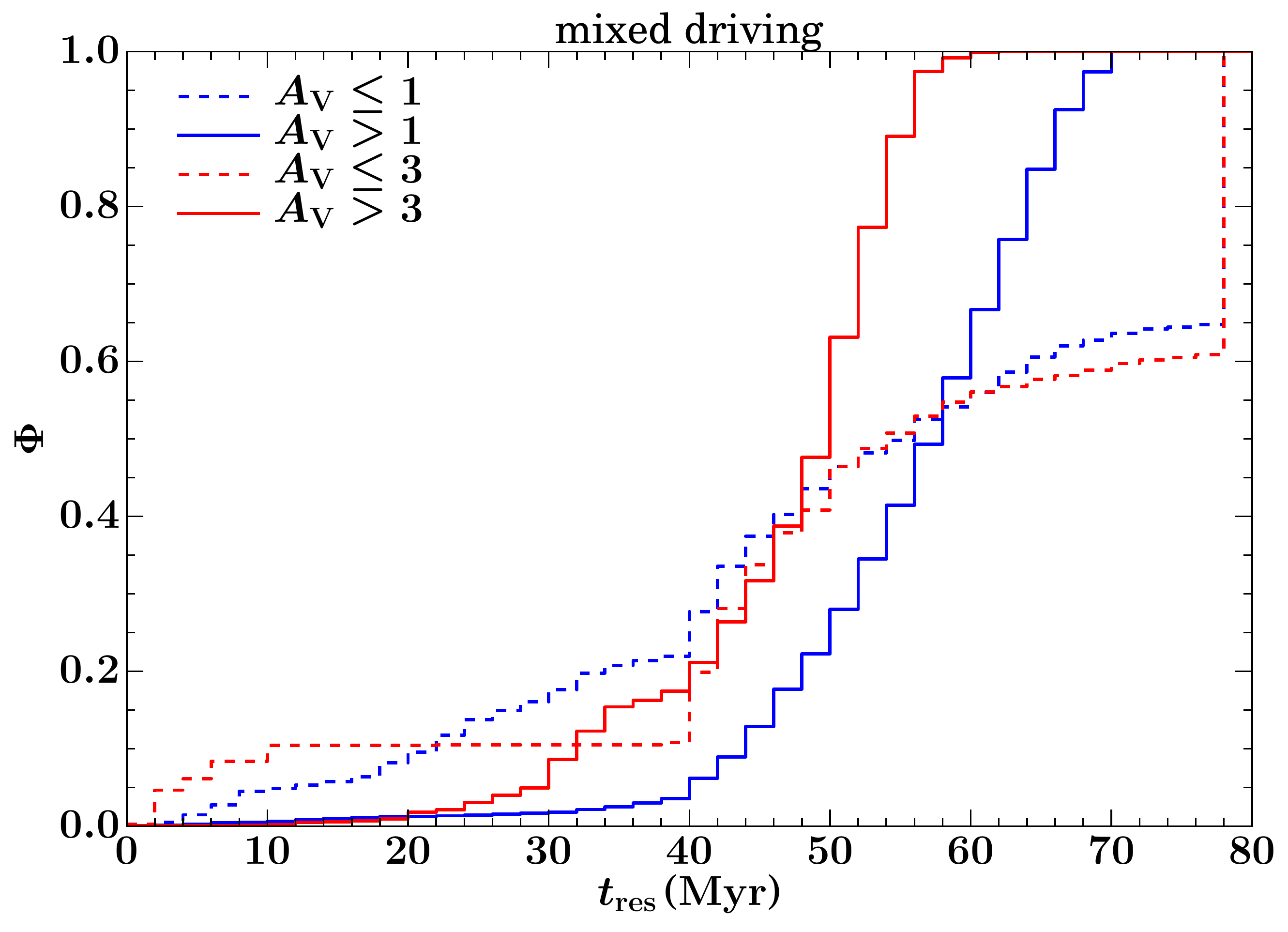}}
\centerline{\includegraphics[width=\linewidth]{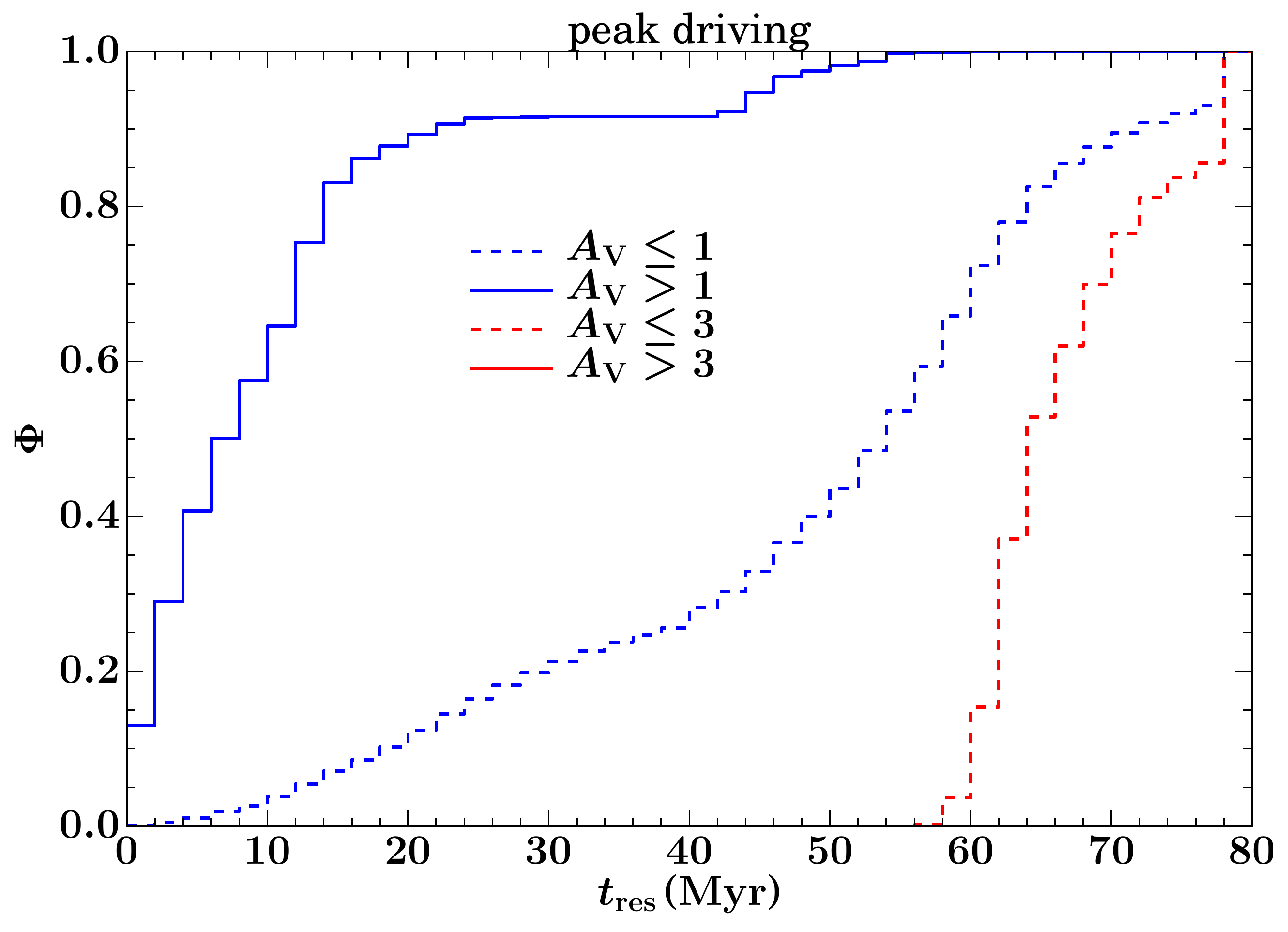}}
\caption{Cumulative probability distribution functions $\Phi$ of total residence times at $t = 40\,$Myr 
in regions with $A_\mathrm{V} \leq A_\mathrm{V, thresh}$ and $A_\mathrm{V} > A_\mathrm{V, thresh}$ for $A_\mathrm{V, thresh} = 1$ and $A_\mathrm{V, thresh} = 3$
for the simulation with random (top), mixed (middle) and peak (bottom) driving. The data is normalised to the total particle count per phase.}
\label{fig:rthistAVnormcum}
\end{figure}

We show the histograms of residence times at $t = 40\,$Myr for the three simulations in Figure~\ref{fig:rthistAV}
and the cumulative probability distribution functions in Figure~\ref{fig:rthistAVnormcum}.
We have used exactly the same procedure to produce the statistics as for the ISM phases.
Just as with the previous histograms, the results for random and mixed driving are very similar. The
statistics for $A_\mathrm{V, thresh} = 1$ and $A_\mathrm{V, thresh} = 3$ is not very different at the highest particle counts,
which implies that gas at these extinctions collapses into molecular clouds and is dispersed on very similar time scales.
Qualitatively, the histrograms resemble the distributions for the H$_2$ phase in Figure~\ref{fig:rthistlin}
for random and mixed driving. Quantitatively, there are more particles in the H$_2$ phase than with
$A_\mathrm{V} > 1$ since $A_\mathrm{V} \sim 0.1$ to $0.2$ is sufficient to from H$_2$ due to self-shielding.
The large residence times imply that both shielding from
as well as exposure to the interstellar radiation field are typically long-term processes that
last more than $50\,$Myr, at least in the absence of early feedback.
For peak driving, with the constant supernova injections in dense gas, particles with $A_\mathrm{V} > 3$
do not exist, and most particles reside in regions with $A_\mathrm{V} \leq 1$ for a long time.
Some particles do have $A_\mathrm{V} > 1$, although there is no H$_2$ phase at this time of the simulation.
This again illustrates that it is not possible to properly characterise the multi-phase ISM
with a single quantity like $A_\mathrm{V}$.

\section{Caveats}
\label{sec:cav}

The analysis presented in this work is only a first step.
Here we aim at a systematic investigation of different methods of supernova positioning in simulations where supernovae
explode at a constant rate. Clearly, some of our results will be different for more realistic simulations.
Self-consistent and early feedback by stellar winds 
\citep{gatto16} and radiation \citep{peters16b} will likely reduce the residence time of grains in the molecular and H$_2$ phase.
Early feedback will also mitigate the overcooling problem for supernova explosions in dense gas by creating cavities
inside clouds before the first supernovae explode \citep{wana15}. This may significantly
alter the transition rates. Self-consistent feedback also allows for self-regulation of star formation, so that
our simulations can reach some form of steady state in the ISM properties.
Our measured transition rates and residence times would then be independent of the stage of the simulation.

Further caveats are related to our simulation setup. Including galactic rotation through shearing box boundary conditions
may contribute to cloud disruption and further reduce the molecular and H$_2$ residence times.
Furthermore, our limited numerical resolution in the dense gas may also affect
our measured residence times in the molecular and H$_2$ phase and in regions with $A_\mathrm{V} > 3$.
Finally, the non-uniform sampling of the ISM by the Lagrangian tracer particles leads to discrepancies between gas mass fractions
and particle fractions. This disagreement should be eliminated to better represent the simulated ISM with the tracer particle technique.
We plan to address these issues in future work.

\section{Conclusions}
\label{sec:conc}

We present a pilot study in which we measure, for the first time, the residence times of interstellar dust in the
different ISM phases, and the transition rates between the phases.
We use hydrodynamical simulations from the SILCC project that are similar to previous work \citep{walchetal15,girietal16},
but with a smaller box height and with Lagrangian tracer particles to probe the physical conditions for interstellar dust grains.
We find complex evolutionary histories of the dust
grains, which in general reside in multiple ISM phases during the simulation runtime. The development of the ISM phases,
the dominant transition rates between phases, and the residence times are a strong function of supernova positioning,
that is they differ depending on whether supernovae
explode at random positions, at density peaks, or as a mixture of the two.

In simulations with random and mixed driving, the resulting ISM is similar. For random driving, transitions between the molecular and
cold phase dominate, followed by transitions between the cold and warm and intermittently between the molecular and warm phases.
For mixed driving, molecular-warm transitions outweigh the other transitions because of the 50\% fraction of peak supernovae.
Most of the transitions are in detailed balance. For peak driving, the formation of a molecular phase is suppressed initially,
and the hot phase eventually disappears. Hence, the transition rates change significantly at different stages of the simulation.
For the most realistic simulations with random and mixed driving, we model the Ti gas-phase depletion with a simple model of dust evolution using the measured
transition rates. We find the best agreement with observations for random driving, although we under-predict the Ti depletion
in the molecular phase in all models.

Residence times generally have broad distributions, in contrast to the single values used in idealised models.
The residence time distribution in the runs with random and mixed driving is similar, with median residence times in the molecular,
cold, warm and hot phase around $17$, $7$, $44$ and $1\,$Myr, respectively. Peak driving, in contrast, leads to median residence times of
$2$, $8$, $12$ and $49\,$Myr in these phases. The residence times in the cold phase agree well in all simulations, although the ISM
structure is very different. Therefore, characteristic residence times are not a good description of the multi-phase ISM.
The broad distribution of residence times in the molecular phase implies a reduction of the dust-to-gas ratio
at the percent level in the dust and chemical evolution model by \cite{zhuketal08}, and may contribute to the scatter in the
observed gas-phase element depletion.

ISM phases defined by chemical abundance rather than temperature cuts are more stable with respect to perturbations
by shocks and supernova explosions. They therefore generally show smaller transition
rates and have longer residence times. For random and mixed driving, the median residence times in the H$_2$ and H~phase
are around $59$ and $45$ Myr, respectively. The residence time in H$_2$-dominated gas is therefore significantly longer than the residence
time in gas with $T \leq 50\,$K, which stresses the importance of including chemistry in simulations of the multi-phase ISM. In contrast,
the residence time distribution in the H$^+$ phase coincides with the hot phase distribution.
The $A_\mathrm{V}$ residence time statistics agree well with the statistics for the H$_2$ phase.

Our results indicate the great potential of hydrodynamical simulations of
the multi-phase ISM to provide detailed information on variable physical
conditions for interstellar grains. They can further the development of more physical models of the dust life cycle in the ISM.
Future investigations with a more self-consistent and complete treatment
of star formation and stellar feedback as well as simulations over longer
time intervals will overcome some of our current limitations.

\section*{Acknowledgements}

We thank Fabian Heitsch for helpful discussions.
All simulations have been performed on the Odin and Hydra clusters hosted by the Max Planck Computing \& Data Facility (http://www.mpcdf.mpg.de/).
TP, ZS, TN, PG, SW, SCOG and RSK acknowledge the {\em Deutsche Forschungsgemeinschaft (DFG)} for funding through the SPP 1573
``The Physics of the Interstellar Medium''.
TN acknowledges support by the DFG cluster of excellence ``Origin and structure of the Universe''. 
SW acknowledges funding by the Bonn-Cologne-Graduate School,
by SFB 956 ``The conditions and impact of star formation'', and from the European Research Council under the European Community's Framework
Programme FP8 via the ERC Starting Grant RADFEEDBACK (project number 679852).
SCOG and RSK acknowledge support from the DFG via SFB 881 ``The Milky Way System'' (sub-projects B1, B2 and B8).
RSK acknowledges support from the European Research Council under the European Community's Seventh Framework Programme
(FP7/2007-2013) via the ERC Advanced Grant STARLIGHT (project number 339177).
The software used in this work was developed in part by the DOE NNSA ASC- and DOE Office of Science ASCR-supported Flash Center for
Computational Science at the University of Chicago.
The data analysis was partially carried out with the \texttt{yt} software \citep{turketal11}.

\end{document}